\documentclass[a4paper,11pt]{article}
\usepackage{jheppub} 
\usepackage{lineno}
\usepackage{hyperref}
\usepackage{graphicx}
\usepackage{subcaption}
\usepackage{adjustbox}
\usepackage{xcolor}
\usepackage{placeins}
\usepackage{bm}
\usepackage{tikz}
\usepackage{tikz-feynman}
\usepackage{multirow}
\usepackage{float}
\usepackage{placeins}

\tikzset{
    fermion/.style={line width=0.8pt, postaction={decorate}, decoration={markings, mark=at position .55 with {\arrow[scale=1.5]{>}}}},
    noarrowfermion/.style={line width=0.4pt},
    boson/.style={line width=0.8pt, decorate, decoration={snake, segment length=4pt, amplitude=3pt}},
    scalar/.style={dashed, line width=0.8pt},
}

\newcommand{\mailto}[1]{\href{mailto:#1}{#1}}

\definecolor{colorDG}{HTML}{008000}

\newcommand{\cp}{$\mathcal{CP}$}

\newcommand{\sss}{\scriptscriptstyle}
\newcommand{\pdp}{\ensuremath{\varphi^\dagger\varphi}}

\title{\boldmath Sensitivity to \cp-violating effective couplings in the top-Higgs sector}

\author[1]{V\'ictor Miralles,\note{\mailto{victor.miralles@manchester.ac.uk}}}
\author[2]{Yvonne Peters,\note{\mailto{yvonne.peters@manchester.ac.uk}}}
\author[3]{Eleni Vryonidou,\note{\mailto{eleni.vryonidou@manchester.ac.uk}}}
\author[4]{Joshua K. Winter,\note{\mailto{joshua.winter@manchester.ac.uk}}}
\affiliation{Department of Physics and Astronomy, University of Manchester, Oxford Road, Manchester M13 9PL, United
Kingdom}

\abstract{The observed baryon asymmetry of the Universe requires new sources of charge-parity ($\mathcal{CP}$) violation beyond those in the Standard Model. In this work, we investigate $\mathcal{CP}$-violating effects in the top-Higgs sector using the Standard Model Effective Field Theory (SMEFT) framework. Focusing on top-pair production in association with a Higgs boson and single top-Higgs associated production at the LHC, we study $\mathcal{CP}$ violation in the top-Higgs Yukawa coupling and other Higgs and top interactions entering these processes.
By analysing $\mathcal{CP}$-sensitive differential observables and asymmetries, we provide direct constraints on $\mathcal{CP}$-violating interactions in the top-Higgs sector. Our analysis demonstrates how combining $t\bar{t}h$ and $thj$ production can disentangle the real and imaginary components of the top-Yukawa coupling, offering valuable insights into potential sources of $\mathcal{CP}$ violation. The sensitivity of these observables to SMEFT operators provides model-independent constraints on the parameter space, advancing the search for new physics in the top-Higgs sector.}

\begin{document}
\maketitle

\section{Introduction}
\label{sec:intro}

Despite the extraordinary success of the Standard Model (SM) in describing the fundamental interactions, it fails to account for the baryon asymmetry of the Universe. In order to describe baryogenesis, the Sakharov conditions \cite{Sakharov:1967dj} must be satisfied, which imply the existence of new sources of charge-parity ($\mathcal{CP}$) violation \cite{Zhang:1994fb,Trodden:1998ym}. This has motivated a campaign of searches for $\mathcal{CP}$ violation both at low energy experiments and at high-energy colliders, such as the Large Hadron Collider (LHC). 

Searches for $\mathcal{CP}$ violation at the LHC rely on the construction of suitable observables, sensitive to the presence of $\mathcal{CP}$-violating effects. As direct searches for new particles have not yielded any observations, attention has shifted towards indirect searches. The absence of discovery of any additional beyond the Standard Model (BSM) particle by the LHC suggests the existence of an energy gap among the electroweak (EW) scale and the new physics (NP) scale. In this scenario, the Standard Model Effective Field Theory (SMEFT) constitutes a robust theoretical framework to parametrise deviations from the SM predictions, including those arising from $\mathcal{CP}$-violating interactions.   When the validity of the SMEFT is guaranteed through a sufficiently large mass gap, we can study the effects of NP in a model-independent way, just assuming that the BSM particles, with a mass well above the EW scale, satisfy the gauge symmetries of the SM.

In this work we investigate the effects of \cp-violating couplings in the top-Higgs sector, where the direct measurements of the couplings  are only accessible at the LHC. The top-Higgs Yukawa coupling can be probed directly through the associated production of a Higgs boson with either a top-antitop quark pair or a single top quark. This class of processes has become accessible at the LHC for the first time, with $t\bar{t}h$ observed in 2018 \cite{CMS:2018uxb,ATLAS:2018mme} and current analyses leading to a bound on the single top-Higgs production \cite{ATLAS:2022vkf,CMS:2022dbt}.

The top-Higgs sector, and in particular its \cp-violating interactions, can also be accessed indirectly. For instance, the top-Yukawa coupling affects the gluon-fusion Higgs boson production, providing the dominant contribution to this process in the SM. Gluon fusion in association with two jets has been proven to be \cp~sensitive \cite{Klamke:2007cu,Freitas:2012kw,Djouadi:2013qya,Demartin:2014fia,Englert:2019xhk,Bhardwaj:2023ufl,Bahl:2023qwk}. 
The top-Higgs \cp-violating component also generates huge contributions to the chromo-electric and electric dipole moment (EDM) of the nucleons --- through its contribution to the Weinberg operator \cite{Weinberg:1989dx} and to the Barr-Zee diagrams \cite{Barr:1990vd} --- as well as to the EDM of the leptons \cite{Brod:2013cka,Brod:2018pli,Fuchs:2020uoc,Bahl:2022yrs,Brod:2022bww,Brod:2023wsh}. For the former, the uncertainties of the hadronic matrix elements are typically too large to provide competitive bounds. For the latter, the extremely good individual constraints become much looser in a global analysis once several operators affect the same observable, generating blind directions \cite{Brod:2022bww}. As such, the more  direct LHC searches for the top-Higgs interactions provide crucial and complementary information.

Top-pair production in association with a Higgs boson has been widely explored in the context of extended Higgs sectors and the top-Yukawa $\mathcal{CP}$ structure \cite{Gunion:1996xu,Demartin:2014fia,He:2014xla,Chien:2015xha,Boudjema:2015nda,Buckley:2015vsa,Rindani:2016scj,Gritsan:2016hjl,Mileo:2016mxg,Kobakhidze:2016mfx,Cao:2016wib,Azevedo:2017qiz,Goncalves:2018agy,Hou:2018uvr,Cao:2019ygh,Faroughy:2019ird,Bortolato:2020zcg,Cao:2020hhb,Bahl:2020wee,Barman:2021yfh,Goncalves:2021dcu,Martini:2021uey,Barman:2022pip,Hermann:2022vit,Azevedo:2022jnd,Bahl:2024tjy,Maltoni:2024wyh}. Whilst suffering from a significantly smaller cross section, single top-Higgs associated production has been proven to be of particular interest in the search for new physics, due to its sensitivity to an anomalous top-Yukawa coupling and in particular to its sign \cite{Agrawal:2012ga,Ellis:2013yxa,Kobakhidze:2014gqa,Yue:2014tya,Chang:2014rfa,Demartin:2015uha,Gritsan:2016hjl,Demartin:2016axk,Kobakhidze:2016mfx,Kraus:2019myc,Patrick:2019nhv,Faroughy:2019ird,Barger:2019ccj,Bahl:2020wee,Martini:2021uey,Bhattacharya:2022kje}. In this work we aim to comprehensively study both processes, focusing on the classes of differential observables and asymmetries, which can be sensitive to \cp\,violation.

In this work, we employ the SMEFT framework since it allows us to systematically capture potential \cp-violating effects in the top-Higgs sector arising from new physics at a heavy mass scale. By parametrising these effects through dimension-six operators, the SMEFT approach allows us to remain agnostic about the details of any specific UV-complete model while still accounting for all possible deviations from the SM. Although many past analyses have especially focused on the effective top-quark Yukawa coupling, working within the SMEFT provides a more comprehensive exploration of the interplay between this interaction and other \cp-violating couplings. The \cp-violating top-Higgs interactions have been studied previously in the context of the SMEFT \cite{Chien:2015xha,Englert:2016ljt,Cirigliano:2016njn,Cirigliano:2016nyn,Haisch:2019xyi,Cirigliano:2019vfc,Kley:2021yhn,Degrande:2021zpv,Bhattacharya:2022kje,Brod:2022bww}. Distinct from previous studies, our work proposes and analyses a wide array of \cp-sensitive observables and asymmetries that could be measured at the LHC. In particular, we focus on both top-pair production in association with a Higgs boson and single top-Higgs associated production, incorporating realistic projections for Run~3 and the HL-LHC. This comprehensive approach aims to highlight future ways to improve current constraints on the \cp-violating parameter space, focusing on the complementarity of multiple channels and the importance of differential measurements.

This work is organised as follows. In Sec.~\ref{sec:setup} we introduce the operators and describe the computational setup for the processes we consider. In Sec.~\ref{sec:diffdist} we present results for various differential observables and in Sec.~\ref{sec:constWC} we perform an analysis to constrain the Wilson coefficients using Run 3 LHC and HL-LHC expected measurements. We conclude in Sec.~\ref{sec:con}.

\section{Theoretical Framework}
\label{sec:setup}
\subsection{Operator basis}
In this work we parameterise the effects of BSM physics in terms of effective interactions among the SM particles. The SMEFT Lagrangian takes the form
\begin{equation}
    \mathcal{L}_{\rm{SMEFT}} = \mathcal{L}_{\rm{SM}} + \sum_{d>4} \sum^{N_d}_{i} \frac{C_i O^{(d)}_i}{\Lambda^{d-4}}\quad,
\end{equation}
where the coefficients $C_i$, known as Wilson coefficients (WC), can be related to the physical parameters of BSM extensions, while the operators, $O_i$, contain only SM fields.

The first contribution relevant for our work appears at dimension six, $d=6$, i.e. 
\begin{equation}
    \mathcal{L}_{\rm{SMEFT}} = \mathcal{L}_{\rm{SM}} +  \sum_{i} \frac{C_i O_i}{\Lambda^{2}} +  \sum_{i} \frac{\hat{C}_i \hat{O}_i + \rm{h.c.}}{\Lambda^{2}} + \mathcal{O}(\Lambda^{-4}) \quad,
\end{equation}
where the non-hermitian operators are marked with a hat. At leading order, the imaginary and real part of the complex WC will generate \cp-odd and \cp-even interactions, respectively.  

The SMEFT contribution to a physical observable can therefore be written as
\begin{equation}
    X_{\text{SMEFT}}=X_{\text{SM}}+\sum_i \frac{C_i}{\Lambda^2}X^{\rm{int}}_i+\sum_{ij}\frac{C_i C_j}{\Lambda^4} X^{\rm{quad}}_{ij}+\mathcal{O}(\Lambda^{-4}) \quad.
\end{equation}
The  contribution at linear order in the WC originates from the interference of the dimension-six operators with the SM, while the contribution at quadratic order arises from the square of dimension-six operators.\footnote{There would be an additional contribution from double insertions of dimension six operators which will not be considered in this work.} The latter contribution is suppressed by the NP scale at the same order as the interference of the SM with dimension-eight operators, which are neglected in this analysis. However, we include results with quadratic terms to examine their impact.

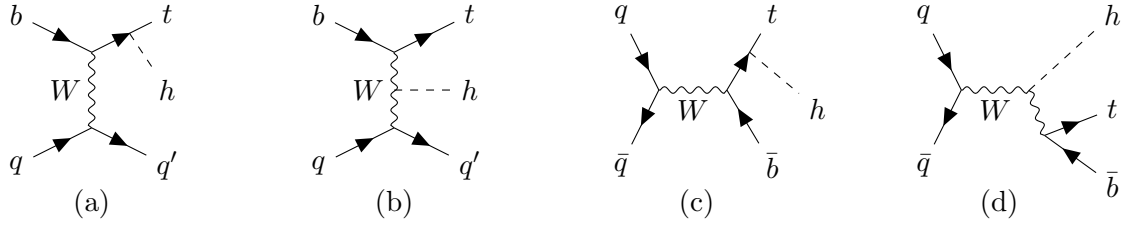
\begin{figure}[h!]
    \centering
    \begin{tikzpicture}
        \begin{scope}[shift={(-8,2)}]
            \node at (1, -1.5) {(a)};
            \begin{feynman}
                \vertex (a) at (0, 1) {\(b\)};
                \vertex (ab) at (1., 0.5) ;
                \vertex (b) at (2, 1) {\(t\)};
                \vertex (c) at (0, -1) {\(q\)};
                \vertex (cd) at (1, -0.5);
                \vertex (d) at (2, -1) {\(q'\)};
                \vertex (abd) at (1.5, 0.75);
                \vertex (e) at (2, 0.) {\(h\)};
                
                \diagram*{
                    (a) -- [fermion] (ab),
                    (ab) -- [fermion] (b),
                    (c) -- [fermion] (cd),
                    (cd) -- [fermion] (d),
                    (ab) -- [boson, edge label'=\(W\)] (cd),
                    (abd) -- [scalar] (e)
                    
                };
            \end{feynman}
        \end{scope}

        \begin{scope}[shift={(-4,2)}]
            \node at (1, -1.5) {(b)};
            \begin{feynman}
                \vertex (a) at (0, 1) {\(b\)};
                \vertex (ab) at (1., 0.5) ;
                \vertex (b) at (2, 1) {\(t\)};
                \vertex (c) at (0, -1) {\(q\)};
                \vertex (cd) at (1, -0.5);
                \vertex (d) at (2, -1) {\(q'\)};
                \vertex (abcd) at (1., 0.) ;
                \vertex (e) at (2, 0.) {\(h\)};
                
                \diagram*{
                    (a) -- [fermion] (ab),
                    (ab) -- [fermion] (b),
                    (c) -- [fermion] (cd),
                    (cd) -- [fermion] (d),
                    (ab) -- [boson, edge label'=\(W\)] (cd),
                    (abcd) -- [scalar] (e)
                    
                };
            \end{feynman}
        \end{scope}

        \begin{scope}[shift={(0,2)}]
            \node at (1, -1.5) {(c)};
            \begin{feynman}
                \vertex (a) at (0, 1.) {\(q\)};
                \vertex (ab) at (0.5, 0.) ;
                \vertex (b) at (0, -1) {\(\bar{q}\)};
                \vertex (c) at (2, 1) {\(t\)};
                \vertex (cd) at (1.4, 0.);
                \vertex (d) at (2, -1) {\(\bar{b}\)};
                \vertex (cdc) at (1.7, 0.5) ;
                \vertex (e) at (2.6, -0.25) {\(h\)};
                
                \diagram*{
                    (a) -- [fermion] (ab),
                    (ab) -- [fermion] (b),
                    (cd) -- [fermion] (c),
                    (d) -- [fermion] (cd),
                    (ab) -- [boson, edge label'=\(W\)] (cd),
                    (cdc) -- [scalar] (e)
                    
                };
            \end{feynman}
        \end{scope}

        \begin{scope}[shift={(4,2)}]
            \node at (1, -1.5) {(d)};
            \begin{feynman}
                \vertex (a) at (0, 1.) {\(q\)};
                \vertex (ab) at (0.5, 0.) ;
                \vertex (b) at (0, -1) {\(\bar{q}\)};
                \vertex (c) at (2.5, 1) {\(h\)};
                \vertex (cd) at (1.4, 0.);
                \vertex (cdd) at (1.6, -0.6);
                
                \vertex (e1) at (2.5, -0.25) {\(t\)};
                \vertex (e2) at (2.5, -1.25) {\(\bar{b}\)};
                
                \diagram*{
                    (a) -- [fermion] (ab),
                    (ab) -- [fermion] (b),
                    (c) -- [scalar] (cd),
                    (cd) -- [boson] (cdd),
                    (ab) -- [boson, edge label'=\(W\)] (cd),
                    (cdd) -- [fermion] (e1),
                    (e2) -- [fermion] (cdd)
                    
                };
            \end{feynman}
        \end{scope}
    \end{tikzpicture}
    \caption{Leading-order Feynman diagrams contributing to $thj$ production at the LHC in the SM for the five flavour scheme.}
    \label{fig:diag_thj}
\end{figure}

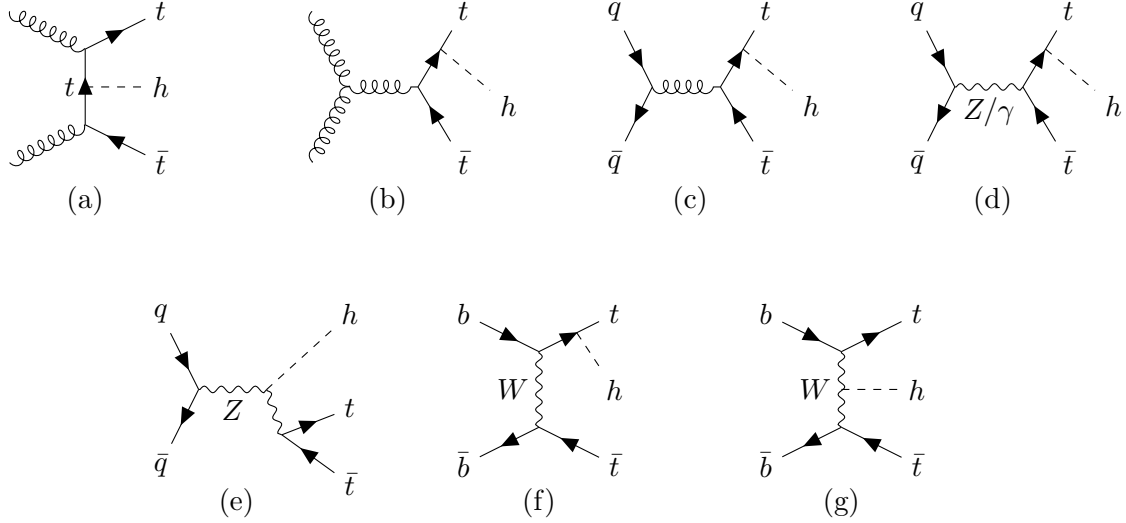
\begin{figure}[h!]
    \centering
    \begin{tikzpicture}
        \begin{scope}[shift={(-8,2)}]
            \node at (1, -1.5) {(a)};
            \begin{feynman}
                \vertex (a) at (0, 1) ;
                \vertex (ab) at (1., 0.5) ;
                \vertex (b) at (2, 1) {\(t\)};
                \vertex (c) at (0, -1) ;
                \vertex (cd) at (1, -0.5);
                \vertex (d) at (2, -1) {\(\bar{t}\)};
                \vertex (abcd) at (1., 0.) ;
                \vertex (e) at (2, 0.) {\(h\)};
                \diagram*{
                    (a) -- [gluon] (ab),
                    (ab) -- [fermion] (b),
                    (c) -- [gluon] (cd),
                    (d) -- [fermion] (cd),
                    (cd) -- [fermion, edge label=\(t\)] (ab),
                    (abcd) -- [scalar] (e)
                };
            \end{feynman}
        \end{scope}

        \begin{scope}[shift={(-4,2)}]
            \node at (1, -1.5) {(b)};
            \begin{feynman}
                \vertex (a) at (0, 1.) ;
                \vertex (ab) at (0.5, 0.) ;
                \vertex (b) at (0, -1) ;
                \vertex (c) at (2, 1) {\(t\)};
                \vertex (cd) at (1.4, 0.);
                \vertex (d) at (2, -1) {\(\bar{t}\)};
                \vertex (cdc) at (1.7, 0.5) ;
                \vertex (e) at (2.6, -0.25) {\(h\)};
                
                \diagram*{
                    (a) -- [gluon] (ab),
                    (ab) -- [gluon] (b),
                    (cd) -- [fermion] (c),
                    (d) -- [fermion] (cd),
                    (ab) -- [gluon] (cd),
                    (cdc) -- [scalar] (e)
                    
                };
            \end{feynman} 
        \end{scope}

        \begin{scope}[shift={(0,2)}]
            \node at (1, -1.5) {(c)};
            \begin{feynman}
                \vertex (a) at (0, 1.) {\(q\)};
                \vertex (ab) at (0.5, 0.) ;
                \vertex (b) at (0, -1) {\(\bar{q}\)};
                \vertex (c) at (2, 1) {\(t\)};
                \vertex (cd) at (1.4, 0.);
                \vertex (d) at (2, -1) {\(\bar{t}\)};
                \vertex (cdc) at (1.7, 0.5) ;
                \vertex (e) at (2.6, -0.25) {\(h\)};
                
                \diagram*{
                    (a) -- [fermion] (ab),
                    (ab) -- [fermion] (b),
                    (cd) -- [fermion] (c),
                    (d) -- [fermion] (cd),
                    (ab) -- [gluon] (cd),
                    (cdc) -- [scalar] (e)
                    
                };
            \end{feynman}      
        \end{scope}

        \begin{scope}[shift={(4,2)}]
            \node at (1, -1.5) {(d)};
            \begin{feynman}
                \vertex (a) at (0, 1.) {\(q\)};
                \vertex (ab) at (0.5, 0.) ;
                \vertex (b) at (0, -1) {\(\bar{q}\)};
                \vertex (c) at (2, 1) {\(t\)};
                \vertex (cd) at (1.4, 0.);
                \vertex (d) at (2, -1) {\(\bar{t}\)};
                \vertex (cdc) at (1.7, 0.5) ;
                \vertex (e) at (2.6, -0.25) {\(h\)};
                
                \diagram*{
                    (a) -- [fermion] (ab),
                    (ab) -- [fermion] (b),
                    (cd) -- [fermion] (c),
                    (d) -- [fermion] (cd),
                    (ab) -- [boson, edge label'=\(Z/\gamma\)] (cd),
                    (cdc) -- [scalar] (e)
                    
                };
            \end{feynman} 
        \end{scope}

        \begin{scope}[shift={(-6,-2)}]
            \node at (1, -1.5) {(e)};
            \begin{feynman}
                \vertex (a) at (0, 1.) {\(q\)};
                \vertex (ab) at (0.5, 0.) ;
                \vertex (b) at (0, -1) {\(\bar{q}\)};
                \vertex (c) at (2.5, 1) {\(h\)};
                \vertex (cd) at (1.4, 0.);
                \vertex (cdd) at (1.6, -0.6);
                
                \vertex (e1) at (2.5, -0.25) {\(t\)};
                \vertex (e2) at (2.5, -1.25) {\(\bar{t}\)};
                
                \diagram*{
                    (a) -- [fermion] (ab),
                    (ab) -- [fermion] (b),
                    (c) -- [scalar] (cd),
                    (cd) -- [boson] (cdd),
                    (ab) -- [boson, edge label'=\(Z\)] (cd),
                    (cdd) -- [fermion] (e1),
                    (e2) -- [fermion] (cdd)
                    
                };
            \end{feynman}
        \end{scope}

        \begin{scope}[shift={(-2,-2)}]
            \node at (1, -1.5) {(f)};
            \begin{feynman}
                \vertex (a) at (0, 1) {\(b\)};
                \vertex (ab) at (1., 0.5) ;
                \vertex (b) at (2, 1) {\(t\)};
                \vertex (c) at (0, -1) {\(\bar{b}\)};
                \vertex (cd) at (1, -0.5);
                \vertex (d) at (2, -1) {\(\bar{t}\)};
                \vertex (abd) at (1.5, 0.75);
                \vertex (e) at (2, 0.) {\(h\)};
                \diagram*{
                    (a) -- [fermion] (ab),
                    (ab) -- [fermion] (b),
                    (cd) -- [fermion] (c),
                    (d) -- [fermion] (cd),
                    (ab) -- [boson, edge label'=\(W\)] (cd),
                    (abd) -- [scalar] (e)
                    
                };
            \end{feynman}        
        \end{scope}

        \begin{scope}[shift={(2,-2)}]
            \node at (1, -1.5) {(g)};
            \begin{feynman}
                \vertex (a) at (0, 1) {\(b\)};
                \vertex (ab) at (1., 0.5) ;
                \vertex (b) at (2, 1) {\(t\)};
                \vertex (c) at (0, -1) {\(\bar{b}\)};
                \vertex (cd) at (1, -0.5);
                \vertex (d) at (2, -1) {\(\bar{t}\)};
                \vertex (abcd) at (1., 0.) ;
                \vertex (e) at (2, 0.) {\(h\)};
                \diagram*{
                    (a) -- [fermion] (ab),
                    (ab) -- [fermion] (b),
                    (cd) -- [fermion] (c),
                    (d) -- [fermion] (cd),
                    (ab) -- [boson, edge label'=\(W\)] (cd),
                    (abcd) -- [scalar] (e)
                };
            \end{feynman}
        \end{scope}
        
    \end{tikzpicture}
    \caption{Leading-order  Feynman diagrams contributing to $t\bar{t}h$ production at the LHC in the SM for the five flavour scheme.}
    \label{fig:diag_tth}
\end{figure}

In this work we follow the flavour assumptions of Ref.~\cite{Degrande:2020evl}, relaxing the assumption of \cp-conservation in the NP. We follow the conventions of the LHC Top WC \cite{Aguilar-Saavedra:2018ksv}, defining the covariant derivative as in the \texttt{dim6top} model described in the same reference. Our study focuses on the top-Higgs sector, considering only the $pp\rightarrow thj$ and $pp\rightarrow t\bar{t}h$ processes. The leading-order diagrams of these processes in the SM are shown in Figs.~\ref{fig:diag_thj} and \ref{fig:diag_tth}, respectively.  Different classes of top-quark operators enter our processes of interest, as will be discussed.

The $pp\rightarrow t\bar{t}h$ process will be modified by the same four-fermion operators as those entering the more precisely measured top-pair production process. With a plethora of measurements of differential distributions as well as asymmetries and spin correlation observables in the $t\bar{t}$ process, we expect top-pair production to sufficiently constrain these operators and thus we do not consider them further. 

Both processes considered are affected by the top-Yukawa coupling,  $\hat{O}_{t\varphi}$, whilst $\hat{O}_{tG}$ enters the $t\bar{t}h$ process, and $\hat{O}_{tW}$ and $O_{\varphi Q}^{(3)}$  affect the $thj$ process. 
Since the $O_{\varphi Q}^{(3)}$ coefficient receives much stronger constraints from single-top production processes without the production of an additional Higgs boson and is \cp-even by construction, it is not the focus of this work. 
Therefore, the relevant two-quark operators for this analysis are \{$\hat{O}_{t\varphi},\,\, \hat{O}_{tG},\,\hat{O}_{tW}$\}.

Finally, the purely bosonic operators can, in general, have some effect on LHC processes involving top quarks. We have 3 scalar operators
$$
O_\varphi,\,\, O_{\varphi\Box},\,\, O_{\varphi D},
$$
and 12 involving gauge bosons for which we have 6 \cp-even operators
$$
O_G,\,\, O_W,\,\, O_{\varphi G},\,\, O_{\varphi W},\,\, O_{\varphi B},\,\, O_{\varphi WB},
$$
plus their anti-symmetric counterparts
$$
O_{\widetilde{G}},\,\, O_{\widetilde{W}},\,\, O_{\varphi \widetilde{G}},\,\, O_{\varphi \widetilde{W}},\,\, O_{\varphi \widetilde{B}},\,\, O_{\varphi \widetilde{W}B},
$$
which are \cp-odd by construction.

The operators $O_{G}$ and $O_{\widetilde{G}}$, besides generating a relevant contribution for $t\bar{t}h$, can be stringently constrained by multi-jet data at quadratic order \cite{Krauss:2016ely,Hirschi:2018etq}. The three scalar operators, while generating some subleading effects, are all \cp-even and are expected to be constrained by other Higgs processes and electroweak precision observables, given that they modify the Higgs kinetic terms (hence affecting its decays), the Higgs self-coupling as well as the effective weak mixing angle. Therefore, we do not consider them further, leaving the relevant bosonic operators for our processes as \{$O_{\varphi G}$, $O_{\varphi \widetilde{G}}$, $O_{\varphi W}$, $O_{\varphi \widetilde{W}}$\}.

In summary, the set of operators considered will be 
$$
\{O_{\varphi G},\, O_{\varphi \widetilde{G}},\, O_{\varphi W},\, O_{\varphi \widetilde{W}},\, \hat{O}_{t\varphi},\, \hat{O}_{tG},\, \hat{O}_{tW}\},
$$
whose definition can be found in Appendix~\ref{app:operators}. This set of operators contains four bosonic operators that are Hermitian and three two-fermion operators that are non-Hermitian. Hence, we have in total 10 degrees of freedom
$$
\{C_{\varphi G},\, C_{\varphi \widetilde{G}},\, C_{\varphi W},\, C_{\varphi \widetilde{W}},\, C_{t\varphi},\, C^I_{t\varphi},\, C_{tG},\, C^I_{tG},\, C_{tW},\,\, C^I_{tW}\},
$$
where all the coefficients are real and the ones corresponding to the imaginary part of a WC are represented with a super index $I$. Therefore, our analysis will contain a total of five WC introducing \cp-violating effects in the Higgs sector
$$
 \{C_{\varphi \widetilde{G}},\, C_{\varphi \widetilde{W}},\, C^I_{t\varphi},\, C^I_{tG},\, C^I_{tW}\},
$$
and five \cp-conserving WC
$$
 \{C_{\varphi {G}},\, C_{\varphi {W}},\, C_{t\varphi},\, C_{tG},\, C_{tW}\}.
$$

The effects of $\hat{O}_{t\varphi}$ can also be interpreted using an effective parametrisation of the top-Higgs coupling.  The relation between the effective Yukawa couplings and the WC of $\hat{O}_{t\varphi}$ is given by
\begin{gather}
   \nonumber \mathcal{L}_{h\bar{t}t}=-\frac{m_t}{v}\bar{t}(\kappa\cos\alpha+i\gamma_5\kappa\sin\alpha)th,\\
    \kappa\cos\alpha=1-\frac{ v^3}{\sqrt{2}\,m_t}\frac{C_{t\varphi}}{\Lambda^2},\quad
    \kappa\sin\alpha=-\frac{ v^3}{\sqrt{2}\,m_t}\frac{C_{t\varphi}^I}{\Lambda^2}.
    \label{eq:eff_yuk_lag}
\end{gather}

\subsection{Computational setup}
\label{sec:methodology}

The dependence of the observables considered on the WCs is obtained at leading-order using \texttt{MadGraph5\_aMC@NLO}
 \cite{Alwall:2014hca}.
  We use the model \texttt{SMEFTsim3.0}\footnote{In particular the  \texttt{SMEFTsim\_top\_MwScheme} model, using the convention for the covariant derivative as in \texttt{dim6top} (different from the usual \texttt{SMEFTsim} convention).} \cite{Brivio:2020onw} that implements the SMEFT in \texttt{MadGraph5\_aMC@NLO} through the Universal FeynRules Output (UFO) format \cite{Degrande:2011ua,Darme:2023jdn}. Therefore, our parametrisation will be obtained at LO for all the NP effects. However, we have scaled the EFT contributions by the inclusive SM $k-$factor of the relevant process. 
  As benchmark SM predictions we take the values quoted in Ref.~\cite{ATLAS:2022vkf} ($\sigma_{thj}^{\rm{SM}}=84.8$ fb and $\sigma_{t\bar{t}h}^{\rm{SM}}=499.8$ fb), from which we obtain a $k$-factor of 1.51 and 1.24 for the single top-Higgs associated production and the top-pair production in association with a Higgs boson, respectively. Details about the values of the parameters and cuts used can be found in Tab.~\ref{tab:eventparams}.\footnote{We have also checked that applying more aggressive cuts on the pseudorapidity compared to the ones shown in Tab.~\ref{tab:eventparams}, e.g. $|\eta|\le3$, would not change the results of our exploratory study by more than 30\%.}
 
Incoming protons are modelled with the \texttt{nn23lo1} parton distribution function, with dynamic renormalisation and factorisation scales chosen as the default of \texttt{MadEvent}, which uses the central transverse mass after applying a $k_T$-clustering algorithm on the event \cite{Maltoni:2002qb}, in a five flavour scheme. In the single top-Higgs production process, events with top- or antitop-quarks in the final state are both considered, as well as the s- and t-channel, though the s-channel provides only subleading contribution to the cross section.

Outgoing top-quarks are decayed in \texttt{MadSpin} \cite{Artoisenet:2012st} and the Higgs boson is left undecayed. 
Since leptons from top-quark decays are most sensitive in their kinematics to top-quark polarisation and top-pair spin correlations, $W^\pm$ bosons from top-quarks are decayed to electrons, $t \rightarrow b W^+, W^+ \rightarrow e^+ \nu_e$.

\begin{table}[ht]
    \centering
    \begin{small}
    \begin{tabular}{c|c}
        \hline \hline
        Parameter & Value \\ \hline \hline
        $\Lambda$  & $1000$ GeV \\
        \hline
        top quark mass & $172$ GeV \\
        Higgs boson mass & $125$ GeV \\
        top resonance width & $1.508336$ GeV \\
        Higgs resonance width & $4.07\cdot 10^{-3}$ GeV \\
        \hline
        Minimum jet $p^T$ & $20$ GeV \\
        Minimum charged lepton $p^T$ & $4$ GeV \\
        Maximum jet $|\eta|$ & $5$ \\
        Maximum charged lepton $|\eta|$ & $5$ \\
        Maximum jet PDG flavour & $5$ \\ 
    \end{tabular}
    \caption{Parameters required for event generation in \texttt{MadGraph5\_aMC@NLO}, using the \texttt{SMEFTsim3.0} model. The lepton cuts are applied to the electrons/positrons from the $W^\pm$ decay, while the jet cuts are applied to the b-jets originating from top-quark decay as well as to the final-state jets in the case of single-top production.}
    
    \label{tab:eventparams}
    \end{small}
\end{table}

The contributions from the SMEFT operators have only been considered in the production, $pp\rightarrow thj$ or $pp\rightarrow t\bar{t}h$, with top-quark decays assumed to obey the SM predictions. While the impact of higher-dimensional operators in the decay has been shown to be relevant in other channels \cite{deBeurs:2018pvs}, these effects are found to be irrelevant in our case. To validate this approach, we have explored the effects of inserting $C_{tW}$ and $C_{tW}^I$ in the decay, obtaining effects smaller than 5\% for values of $C_{tW}^{(I)}/\Lambda^2$ as large as 10 TeV$^{-2}$. Given this minimal impact, we  focus on the contributions of SMEFT operators only in the production. Note that there are also other production channels that can be relevant as a probe of the \cp-violating interactions in the top-Higgs sector, e.g. $pp\to thW$. However, significant interference of this channel with $t\bar{t}h$ \cite{Demartin:2016axk} complicates both the experimental extraction of the signal and the theoretical interpretation in a SMEFT framework. Therefore, we decided not to include this process in our exploratory analysis.

\section{\cp-odd effects in differential distributions}
\label{sec:diffdist}

The total cross sections are \cp-even observables by construction. Although these observables are still sensitive to \cp-odd couplings at quadratic order, the sensitivity to them would benefit from a differential analysis.
In the following, we propose and describe some manifestly \cp-odd observables which are sensitive to \cp-odd couplings at linear order for single and pair top-Higgs associated production.

\subsection{Single top-Higgs associated production}
\label{sec:cpodd_th}

\begin{figure}[h!]
  \centering
  \vspace{-0.4cm}
  \begin{subfigure}[b]{0.38\textwidth}
    \includegraphics[width=1.\linewidth]{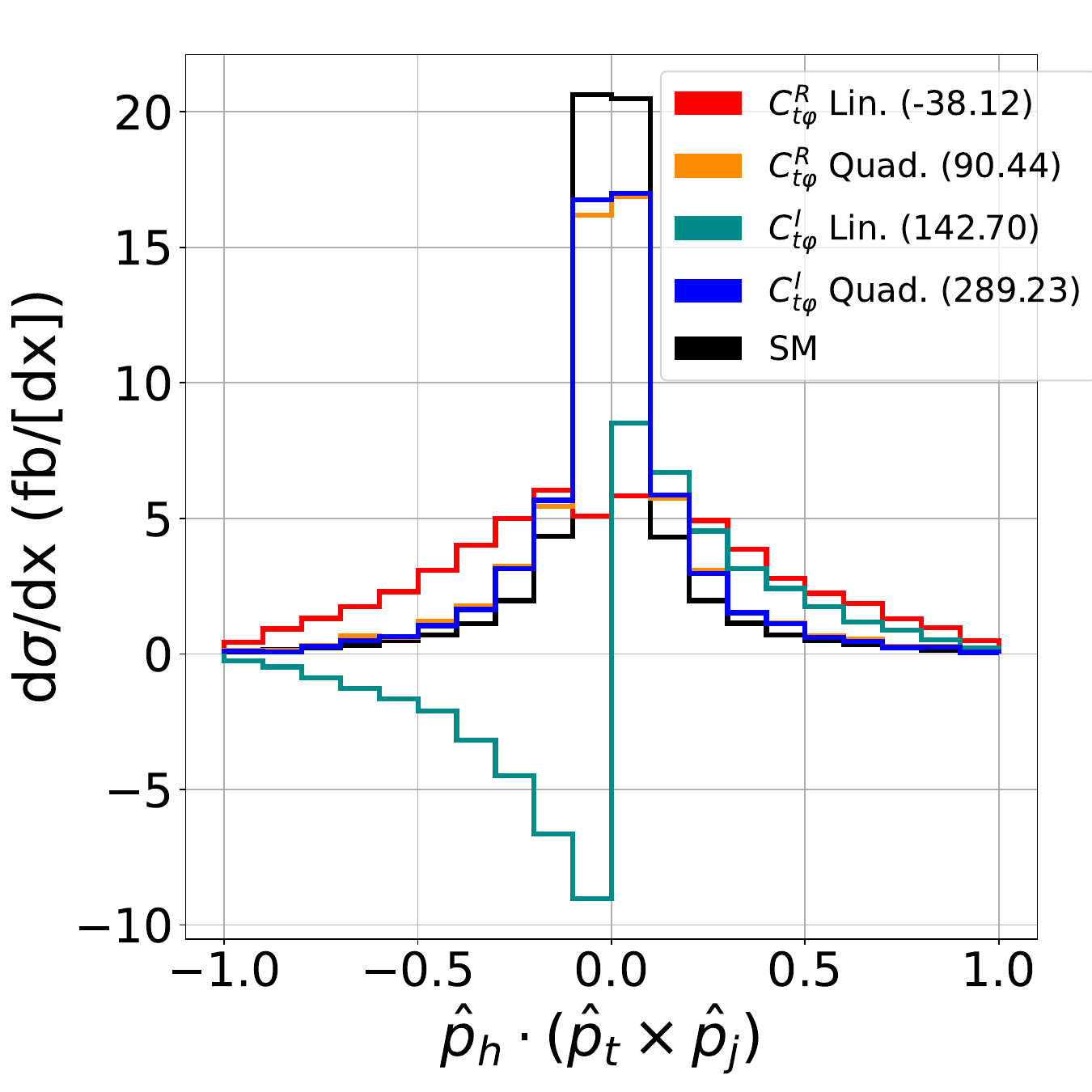}
    \caption{}
    \label{fig:sub1}
  \end{subfigure}%
  \begin{subfigure}[b]{0.38\textwidth}
    \includegraphics[width=1.\linewidth]{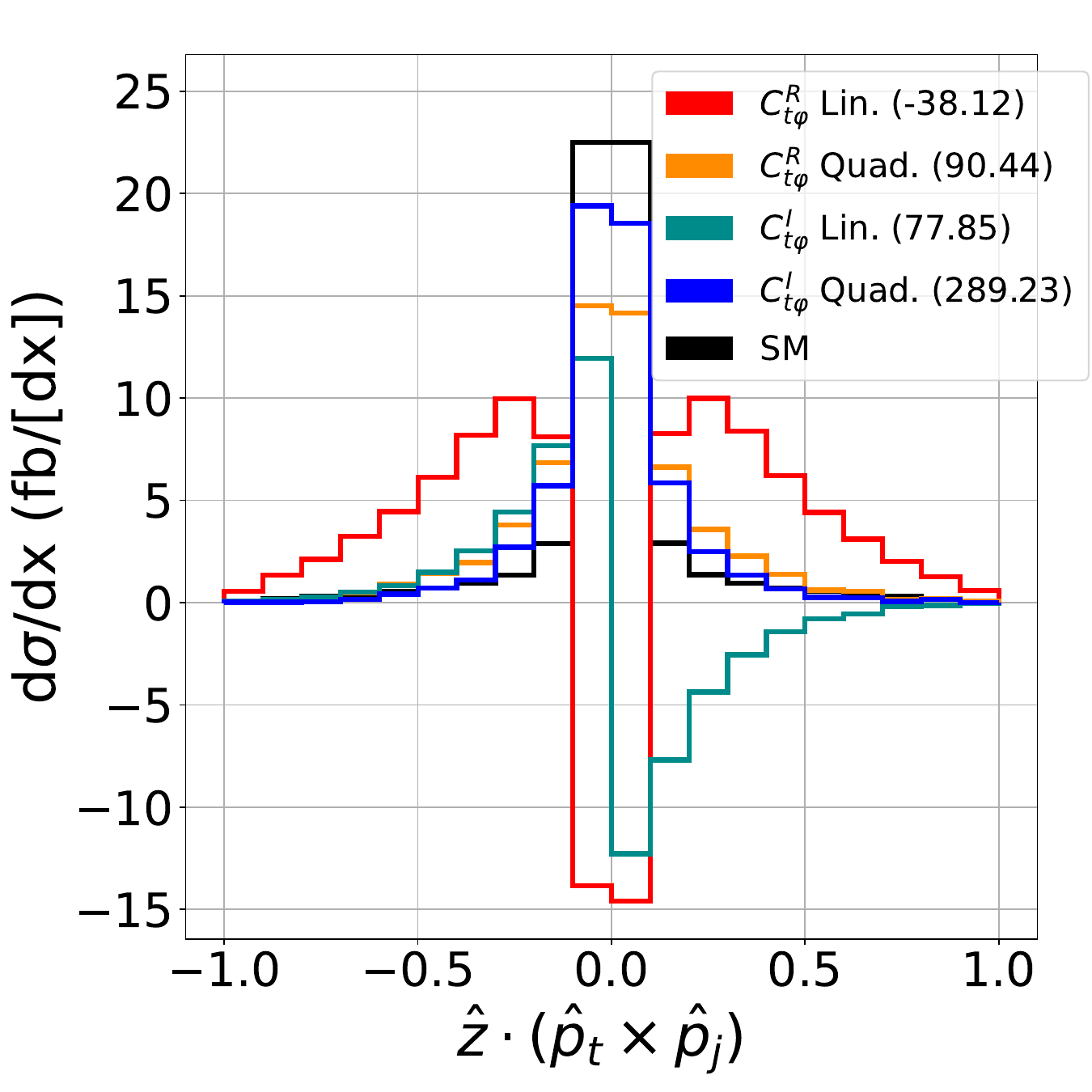}
    \caption{}
    \label{fig:sub2}
  \end{subfigure}
    \begin{subfigure}[b]{0.38\textwidth}
    \includegraphics[width=1.\linewidth]{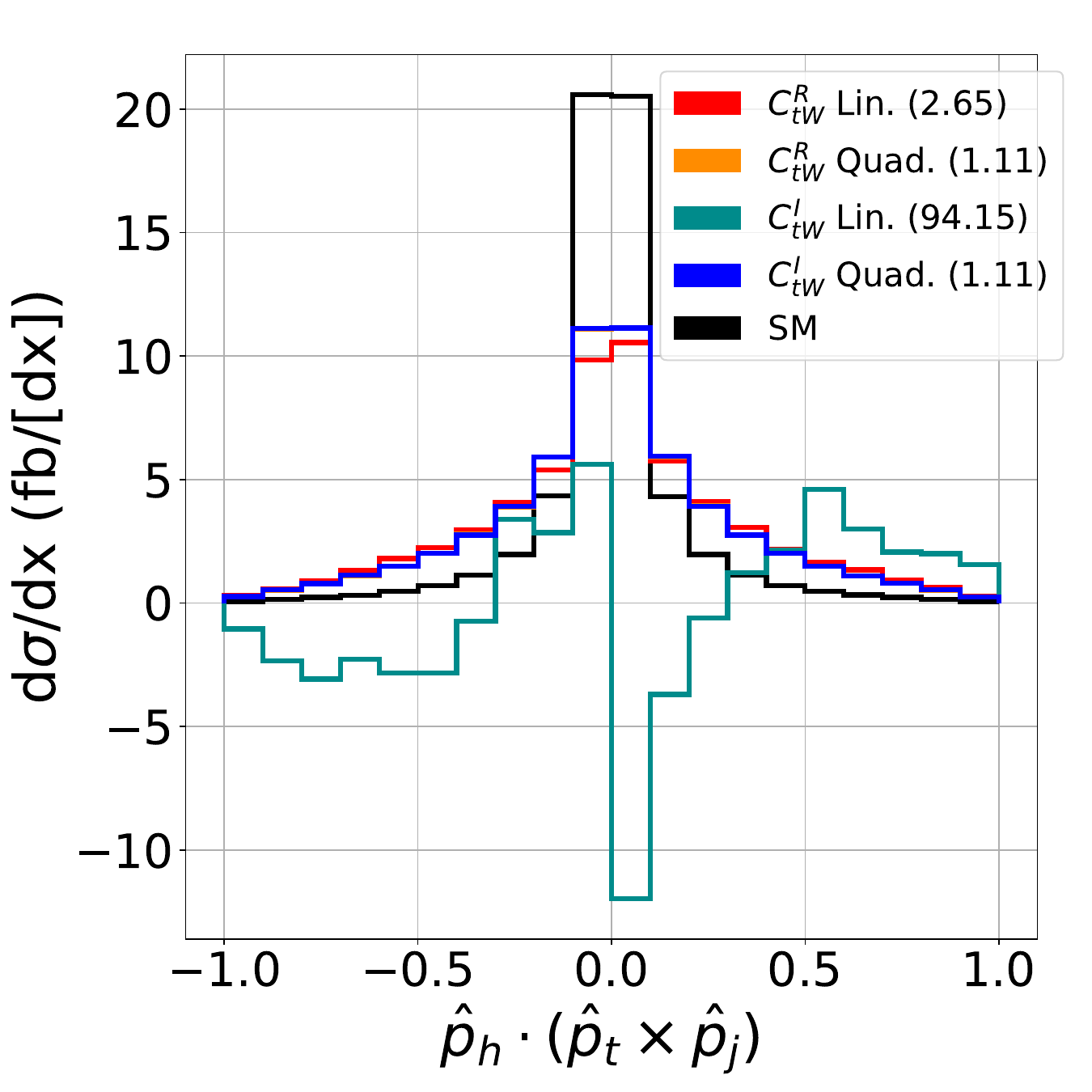}
    \caption{}
    \label{fig:sub2}
  \end{subfigure}%
  \begin{subfigure}[b]{0.38\textwidth}
    \includegraphics[width=1.\linewidth]{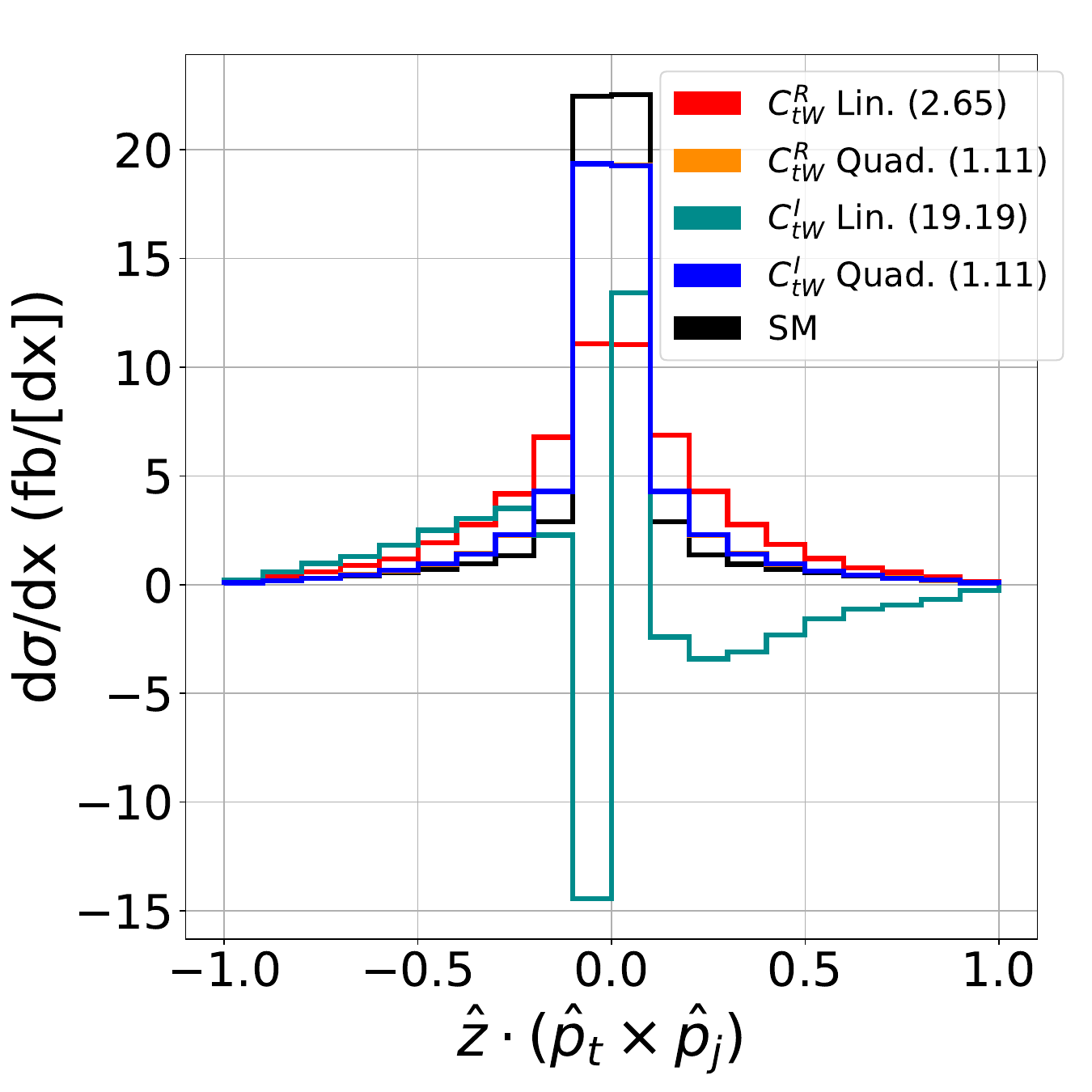}
    \caption{}
    \label{fig:sub1}
  \end{subfigure}
  \begin{subfigure}[b]{0.38\textwidth}
    \includegraphics[width=1.\linewidth]{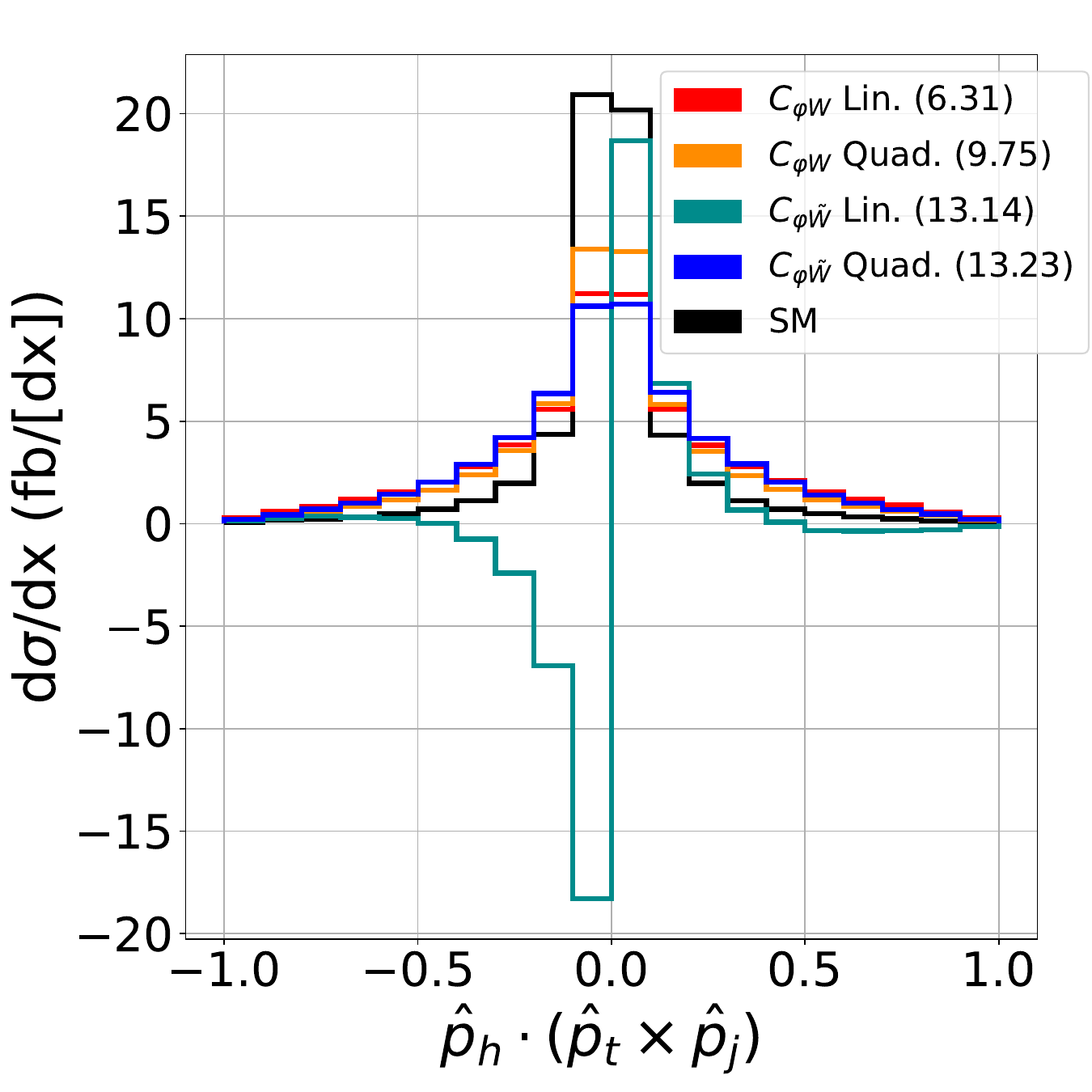}
    \caption{}
    \label{fig:sub2}
  \end{subfigure}%
    \begin{subfigure}[b]{0.38\textwidth}
    \includegraphics[width=1.\linewidth]{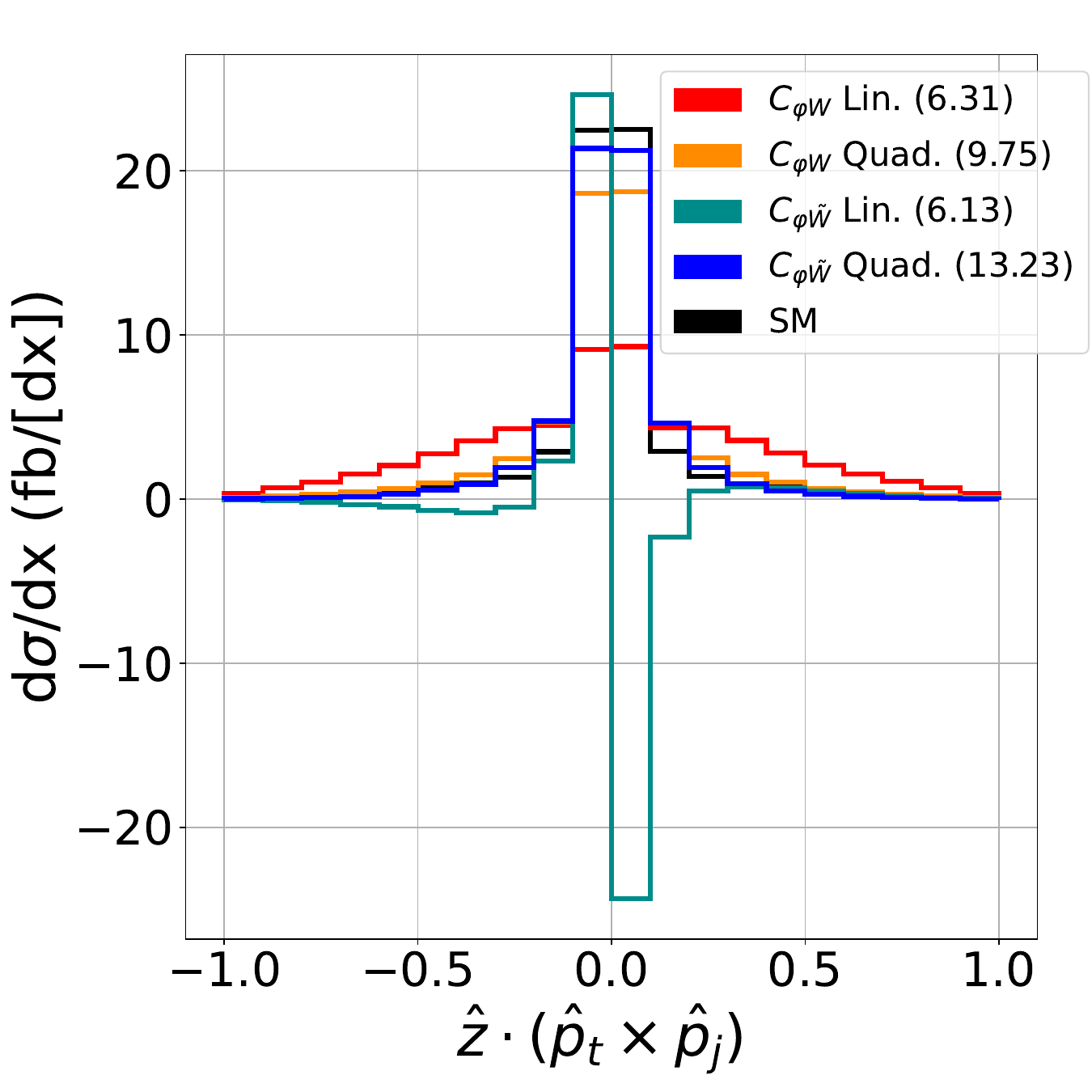}
    \caption{}
    \label{fig:sub2}
  \end{subfigure}
  \caption{\label{fig:thj_tripleprods}Triple product of final state particles (left) and of the beam, top and jet directions in the Higgs rest frame (right), for the operators $\hat{O}_{t\varphi}$ (top), $\hat{O}_{tW}$ (middle) and $O_{\varphi W/\widetilde{W}}$ (bottom). SMEFT predictions are scaled to match the SM curve area, and we show the multiplicative factor used in parenthesis in the label of each curve.
  }
\end{figure}

We start by considering the $thj$ process. Since $thj$ is not sensitive to the gluonic operators, it will  only be sensitive to four out of the seven operators considered in this work:
$\left\{\hat{O}_{t\varphi},\hat{O}_{tW},O_{\varphi W},O_{\varphi\widetilde{W}}\right\}$.
\paragraph{Triple products}
There are several angular observables that have been proven to be sensitive to the \cp-odd couplings at the linear level. A promising set of observables are the triple products of the momenta of the final particles and/or the initial partons \cite{Degrande:2021zpv}. In this work, we have considered the triple product of the 3-momenta of the final particles in the laboratory frame, as well as the triple product of the 3-momenta of the beam axis, the (anti-)top quark, and the final jet in the Higgs rest frame,
\begin{equation}
    \hat{p}_h \cdot (\hat{p}_t \times \hat{p}_j),\qquad \hat{z} \cdot (\hat{p}_t \times \hat{p}_j),
\end{equation}
where $\hat{z}$, $\hat{p}_h$, $\hat{p}_t$, and $\hat{p}_j$ are unit vectors in the direction of beam axis, outgoing Higgs boson, (anti-)top quark, and jet, respectively. The former observable measures the cosine of the angle between the Higgs boson and the perpendicular direction to the plane formed by the (anti-)top quark and the jet in the laboratory frame, while the latter measures the cosine of the angle between the beam axis and that same plane in the Higgs rest frame. The differential distributions for these triple products are presented in Fig.~\ref{fig:thj_tripleprods} for all relevant degrees of freedom at linear and quadratic order in comparison with the SM prediction. As we aim to compare the shape of these distributions, we normalise them in such a way that the absolute value of the area of the distributions is equal to the one of the SM. 
As shown in Fig.~\ref{fig:thj_tripleprods}, these triple products are symmetric for all \cp-even contributions (i.e., the SM and the interference of the \cp-even degrees of freedom with the SM) and demonstrate a clear separation of positive and negative weights for \( C_{t\varphi}^I \), \( C_{tW}^I \), and \( C_{\varphi\widetilde{W}} \). This behaviour allows us to use the triple products to define asymmetries sensitive to the \cp-odd couplings, as demonstrated in Sec.~\ref{sec:asym}.

Although asymmetries are clearly the best option for $C_{t\varphi}^I$ and $C_{\varphi\widetilde{W}}$, the situation differs for $C_{tW}^I$, where the distribution changes sign three times, reducing the sensitivity of the asymmetries. In this case, a binned distribution would achieve better sensitivity, provided the binning is chosen such as to avoid cancellations between positive and negative contributions. 

Finally, in Fig.~\ref{fig:thj_tripleprods} we can also see how the sensitivity of $C_{\varphi\widetilde{W}}$ is around a factor 10 higher than the one of the Yukawa, $C_{t\varphi}^I$, which will be discussed in more detail in Sec.~\ref{sec:asym}.

\begin{figure}[t]
  \centering
  \begin{subfigure}[b]{0.33\textwidth}
    \includegraphics[width=1.\linewidth]{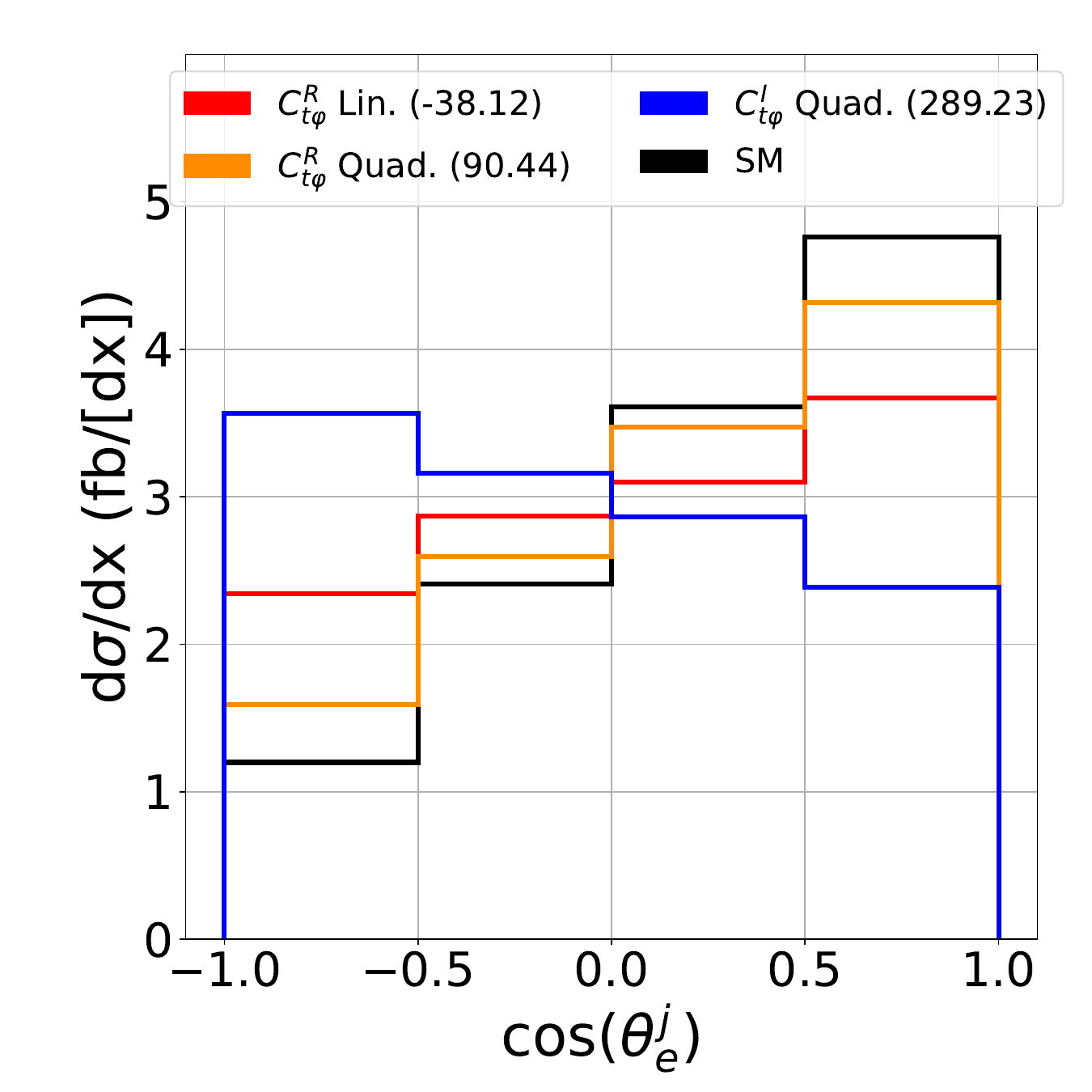}
    \caption{}
    \label{fig:sub1}
  \end{subfigure}%
  \begin{subfigure}[b]{0.33\textwidth}
    \includegraphics[width=1.\linewidth]{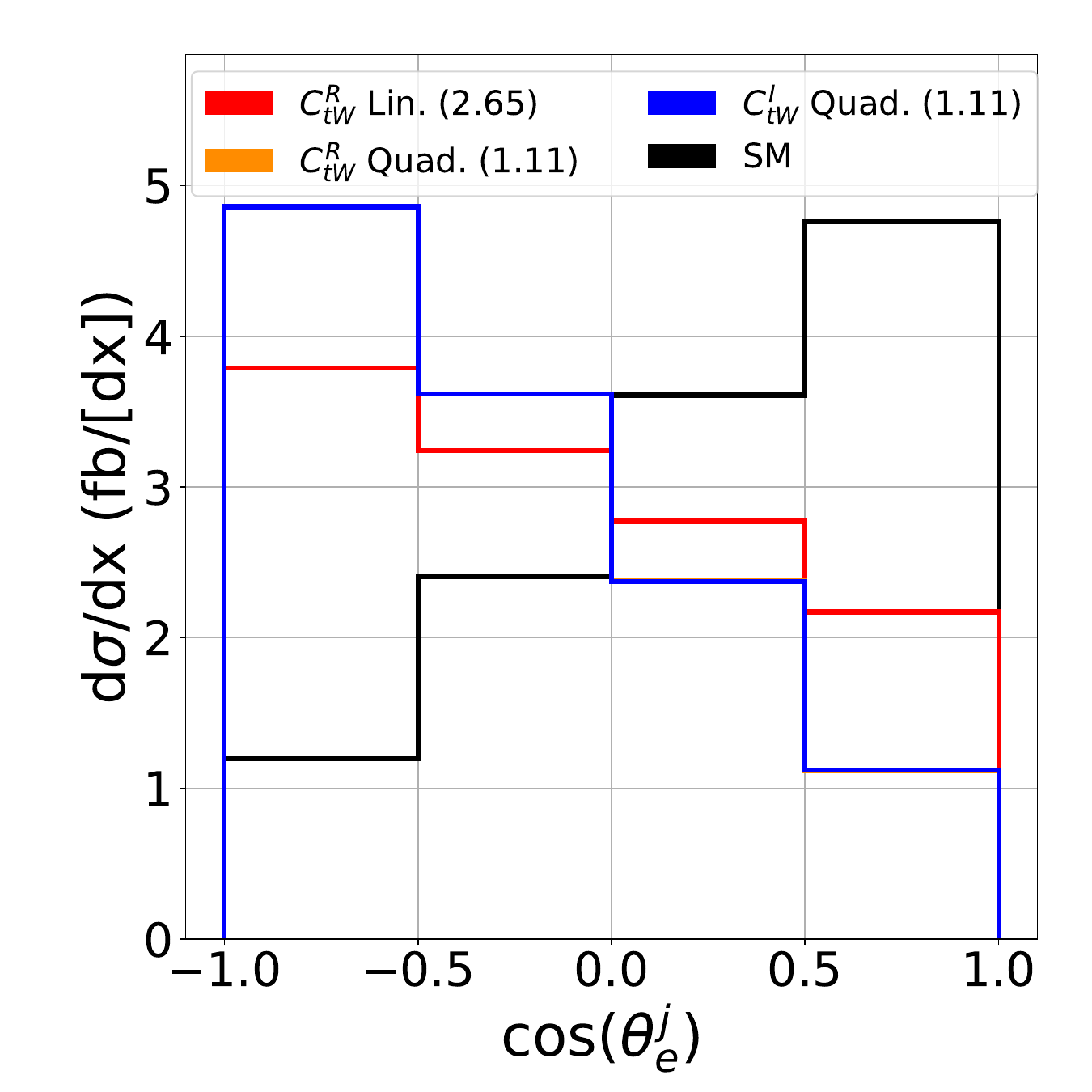}
    \caption{}
    \label{fig:sub2}
  \end{subfigure}
    \begin{subfigure}[b]{0.33\textwidth}
    \includegraphics[width=1.\linewidth]{ 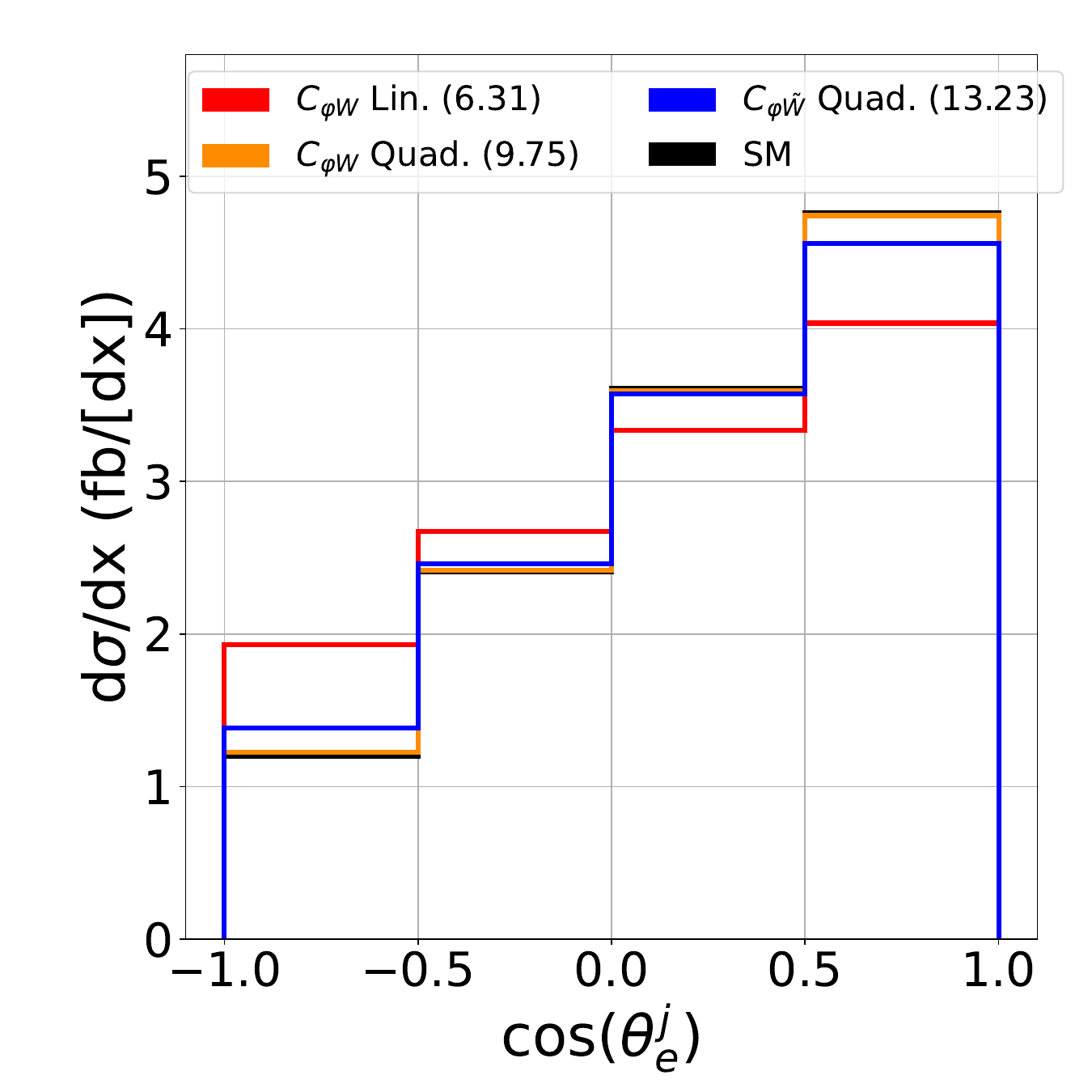}
    \caption{}
    \label{fig:sub2}
  \end{subfigure}
  \begin{subfigure}[b]{0.33\textwidth}
    \includegraphics[width=1.\linewidth]{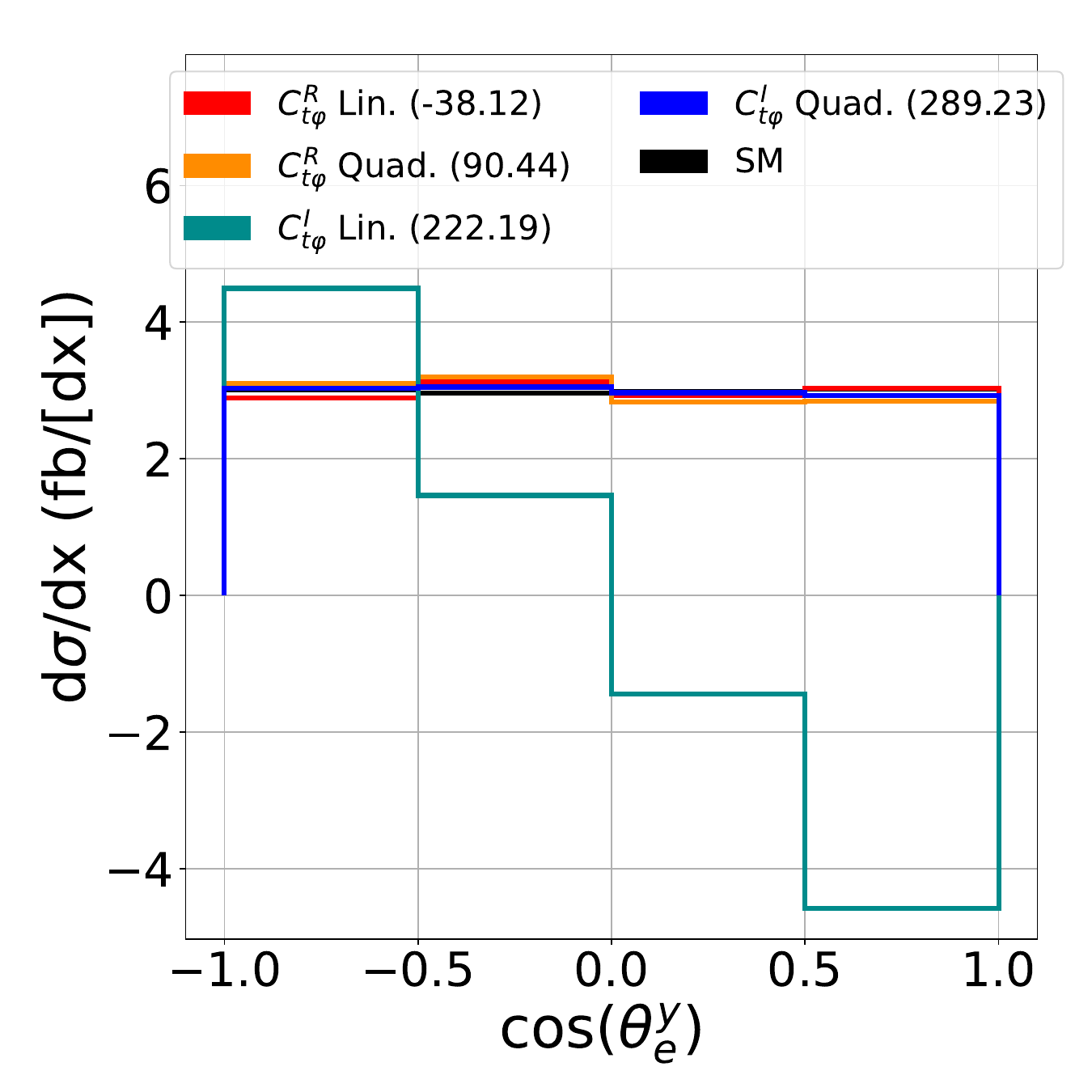}
    \caption{}
    \label{fig:sub1}
  \end{subfigure}%
  \begin{subfigure}[b]{0.33\textwidth}
    \includegraphics[width=1.\linewidth]{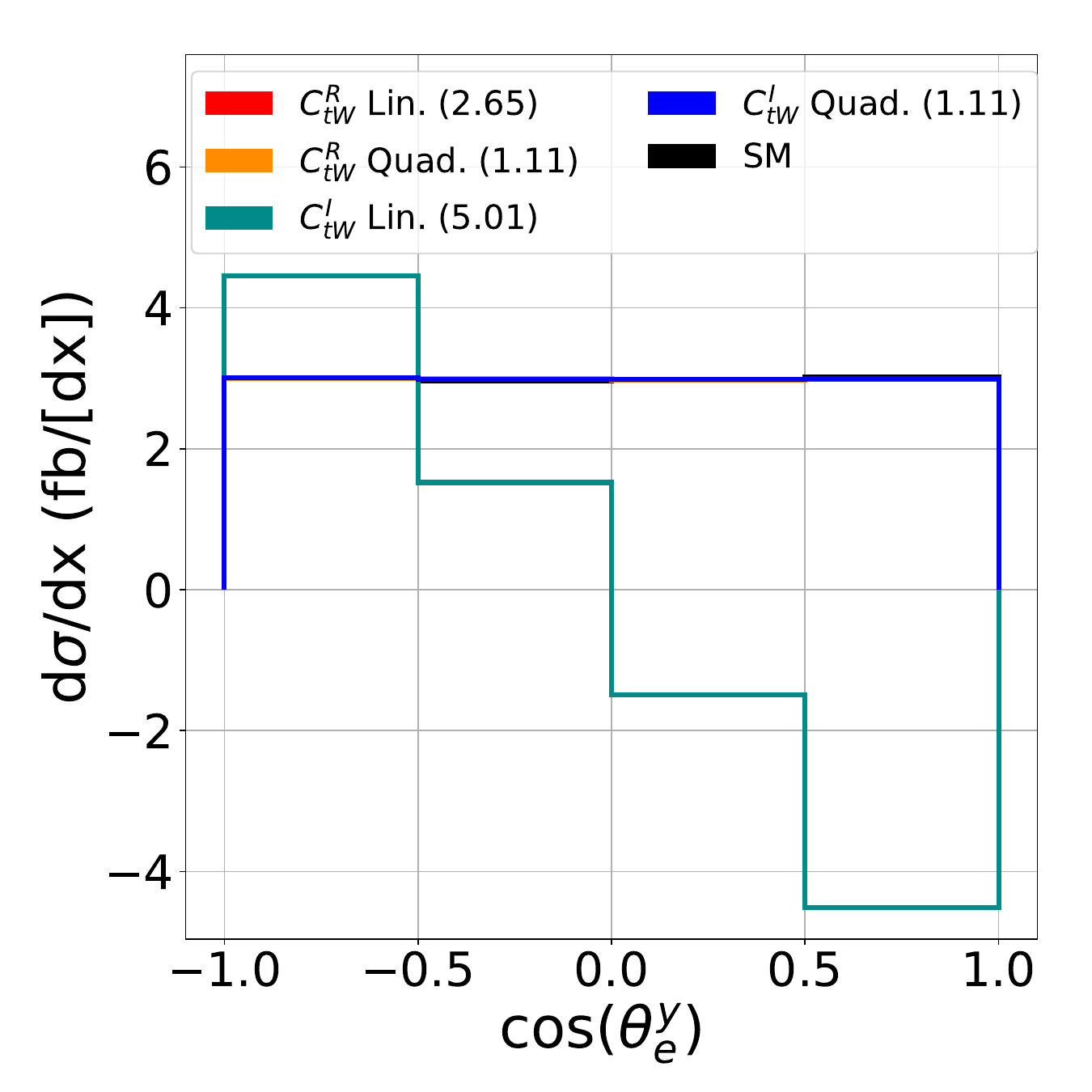}
    \caption{}
    \label{fig:sub2}
  \end{subfigure}
    \begin{subfigure}[b]{0.33\textwidth}
    \includegraphics[width=1.\linewidth]{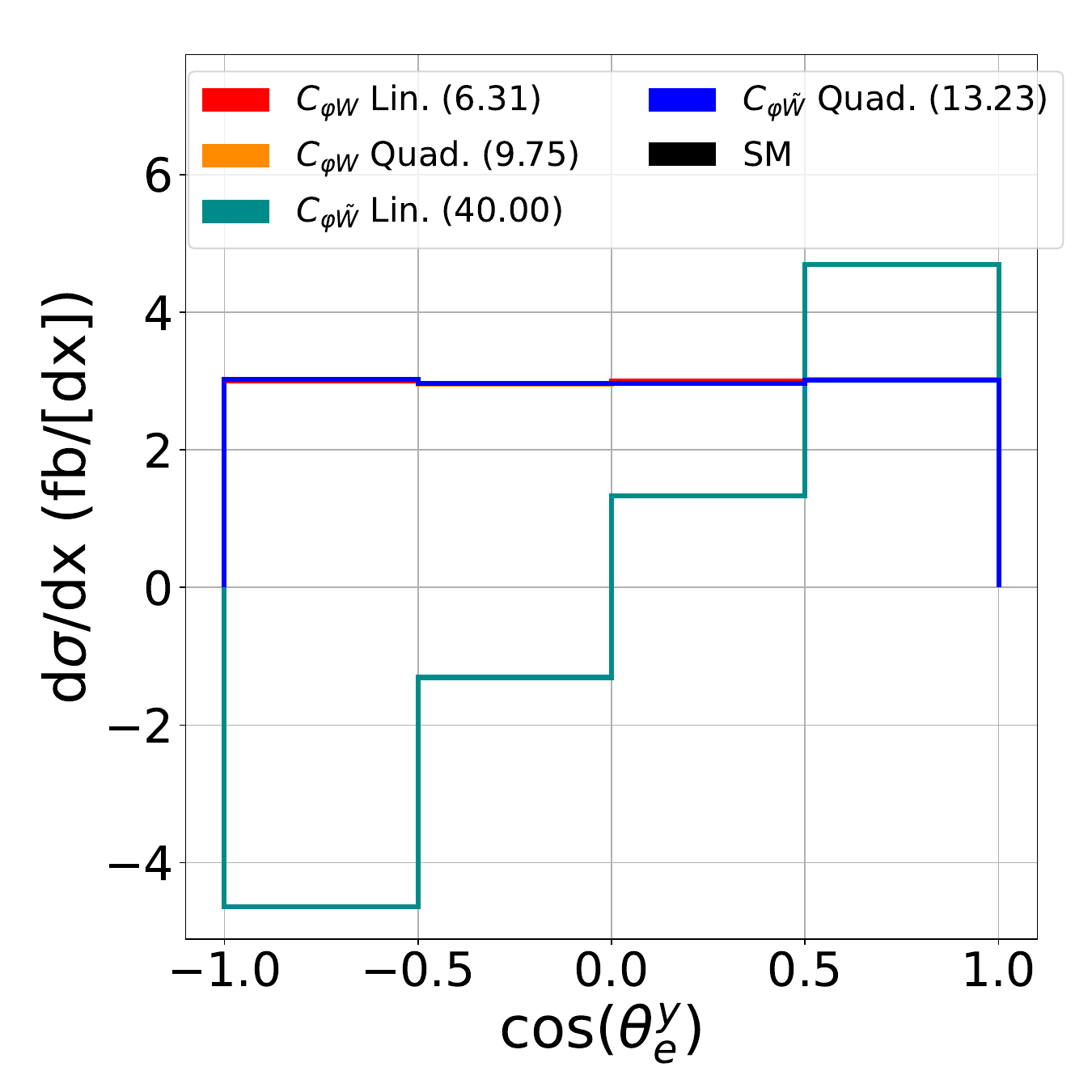}
    \caption{}
    \label{fig:sub2}
  \end{subfigure}
  \begin{subfigure}[b]{0.33\textwidth}
    \includegraphics[width=1.\linewidth]{ 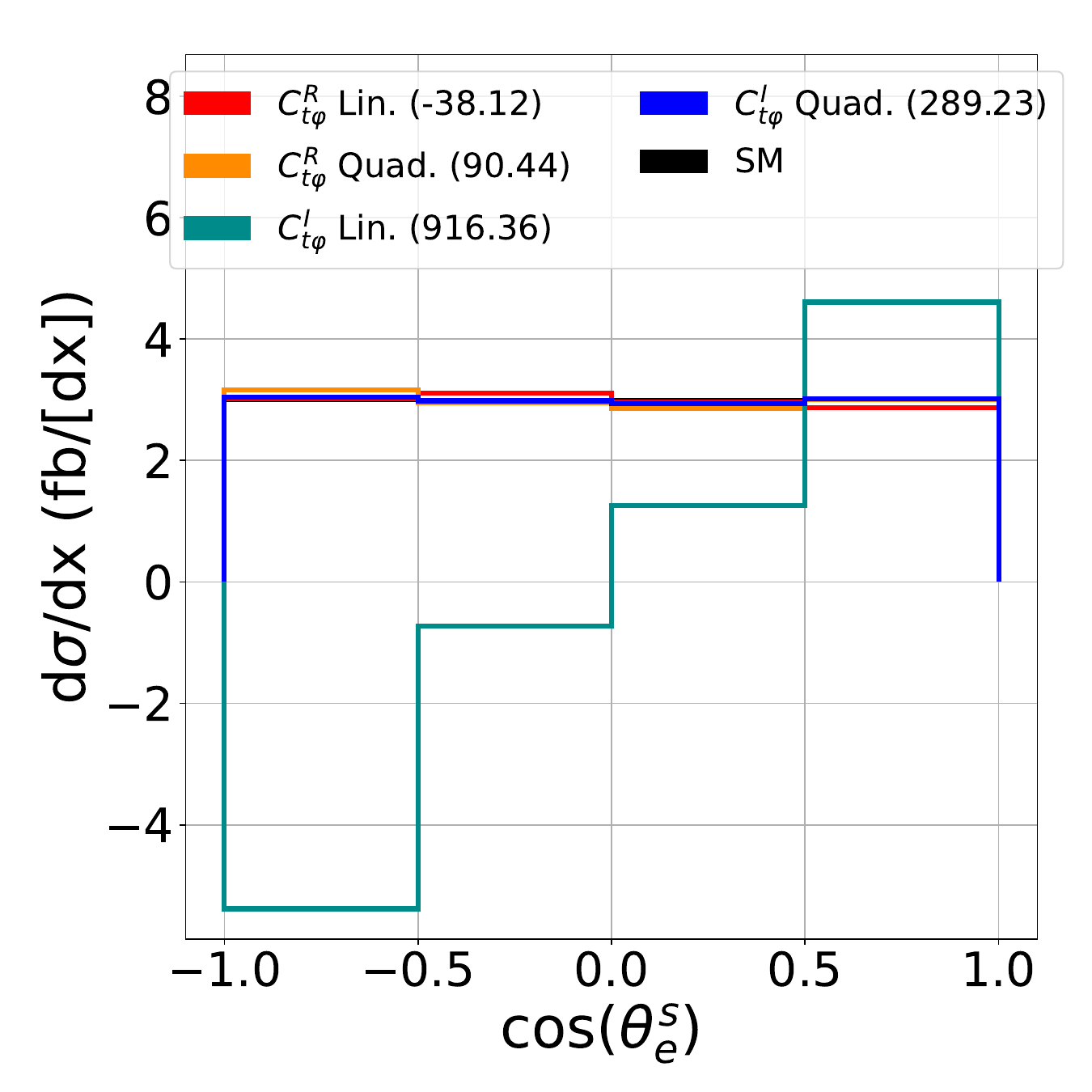}
    \caption{}
    \label{fig:sub1}
  \end{subfigure}%
  \begin{subfigure}[b]{0.33\textwidth}
    \includegraphics[width=1.\linewidth]{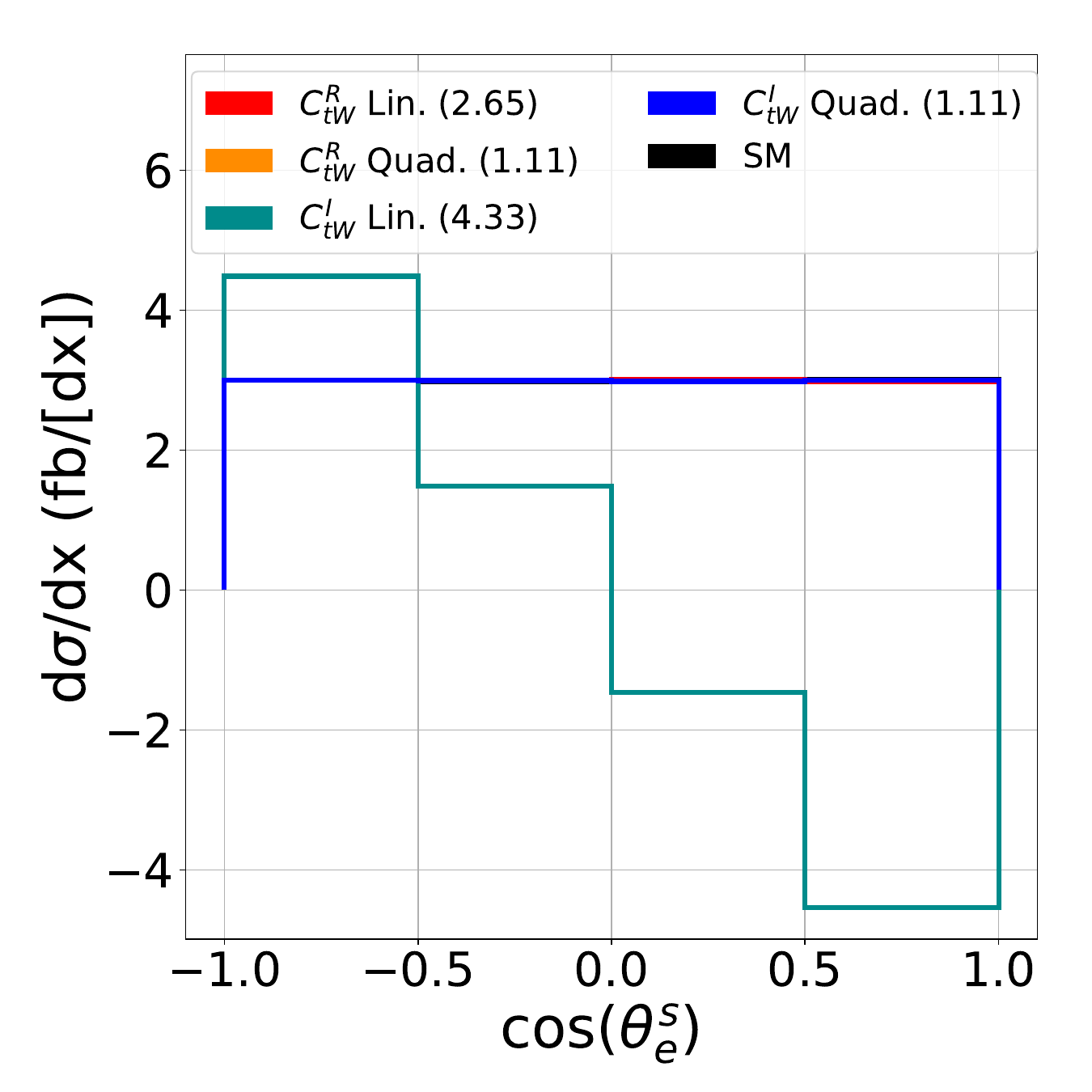}
    \caption{}
    \label{fig:sub2}
  \end{subfigure}
    \begin{subfigure}[b]{0.33\textwidth}
    \includegraphics[width=1.\linewidth]{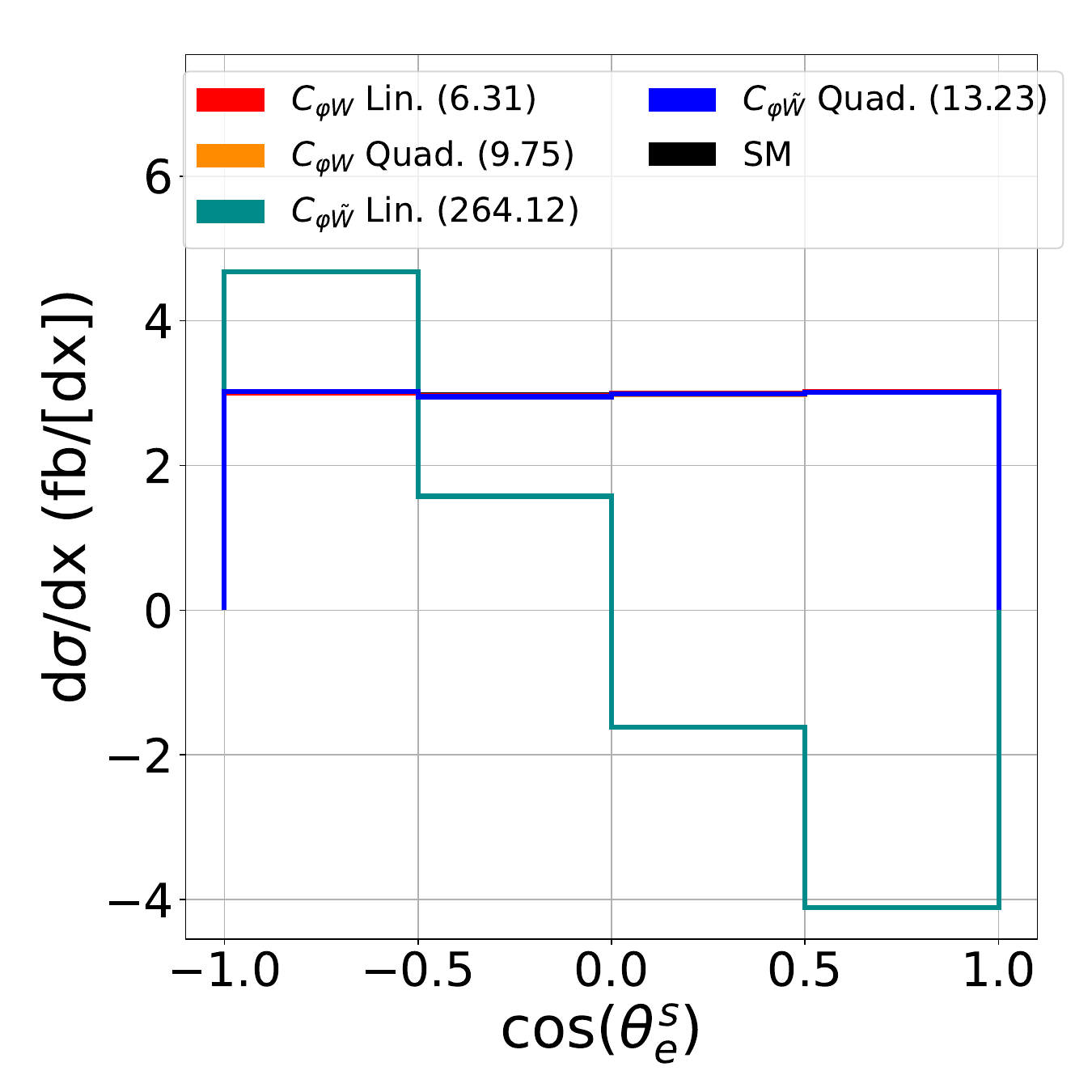}
    \caption{}
    \label{fig:sub2}
  \end{subfigure}
  \caption{\label{fig:thj_polarangs}Differential cross sections of $\cos{\theta_e}$ for the polarisation angles $\theta^j_e,\,\theta^y_e,\,\theta^s_e$ (from top to bottom), for the operators $\hat{O}_{t\varphi}$ (left), $\hat{O}_{tW}$ (middle) and $O_{\varphi W/\widetilde{W}}$ (right). SMEFT predictions are scaled to match the SM curve area, and we show the multiplicative factor used in parenthesis in the label of each curve.
  }
\end{figure}

\paragraph{Polarisation angles}
Another set of observables that are sensitive to the \cp-odd couplings are those accessing  the top-quark polarisation. In single-top production, the top-quark polarisation is accessible via the angular distribution of its decay products. In the top-quark rest frame, the differential cross section can be written as
\begin{equation}
\frac{1}{\sigma}\frac{d\sigma}{d\cos{\theta^x_i}} = \frac{1}{2}\left(1+a_i P\cos{\theta^x_i}\right),
\end{equation}
where $P$ is the top-quark polarisation; $a_i$ is the so-called spin analysing power, which quantifies the effect of the top-quark spin on a given decay product, with $a_e=1$ at leading order; $\theta^x_i$ is the angle of the decay product $i$ relative to the axis $x$. Following the work of Refs.~\cite{Ellis:2013yxa,deBeurs:2018pvs}, we define the electron polarisation angles, $\theta^x_i$, relative to the axes
\begin{equation}
    \hat{j} = \hat{p}_j,\quad
    \hat{y} = \frac{\hat{p}_j \times \hat{z}}{|\hat{p}_j \times \hat{z}|},\quad
    \hat{s} = \frac{\hat{p}_j \times \hat{p}_h}{|\hat{p}_j \times \hat{p}_h|},
\end{equation} in the top-quark rest frame. The polarisation angles can therefore be simply extracted by taking the scalar product of the electron unit 3-momentum with these axes. Given that we consider both top- and antitop-quark production, $\cos\theta_e^j$ is a \cp-even observable. However, $\cos\theta_e^y$ and $\cos\theta_e^s$ are, by construction, \cp-odd. 

The impact of the higher-dimensional operators on the defined polarisation angles is shown in Fig.~\ref{fig:thj_polarangs}. We first note that for the \cp-even angle the \cp-odd operators only contribute at the quadratic level, whilst for the \cp-odd observables, as expected, only the interference of the SM with the \cp-odd operators gives an odd distribution.

The distributions of $\cos{\theta_e^j}$ (Fig.~\ref{fig:thj_polarangs} (a) - (c)) indicate that, in the SM, the lepton from the top-quark decay is preferentially produced in the direction of the spectator quark --- opposing the top quark momentum ($P<0$).
While the differential interference is consistent with zero for $\mathcal{CP}$-odd contributions, the quadratic amplitude distribution of $\hat{O}_{tW}$ shows an inversion of the top-quark polarisation, relative to the SM. 
Additionally, the introduction of modified top-Higgs vertices reduces the degree of top-quark polarisation, which also experiences a sign inversion for a $\mathcal{CP}$-odd top-Yukawa coupling.

The $\mathcal{CP}$-even modified $hWW$ vertex introduced by $O_{\varphi W}$ leaves the top-quark polarisation close to unchanged. Small differential asymmetries in $\cos{\theta_e^y}$ and $ \cos{\theta_e^{s}}$ are seen for the analogous $\mathcal{CP}$-odd operator, but these are not as pronounced as those observed with the triple productions at parton-level.

In the SM, and for modified $\mathcal{CP}$-even amplitudes (including all quadratic amplitudes from $\mathcal{CP}$-odd operators), the lepton direction is invariant along the $\hat{y}$- and $\hat{s}$-axes, as demonstrated in Figs.~\ref{fig:thj_polarangs} (d)--(i).
However, the interference of $\mathcal{CP}$-odd effective couplings introduces a preferential direction of the charged lepton momentum along each of these axes.
The coefficients $C_{t\varphi}^I,\,C_{tW}^I$ and $C_{\varphi\widetilde{W}}$ determine whether leptons are produced preferentially in the forward or backward direction, and the linear dependence allows for the definition of a dedicated $\mathcal{CP}$-odd asymmetry about $\cos{\theta_e} = 0$, as shown in Sec.~\ref{sec:asym}.

\subsection{Top-pair production in association with a Higgs boson}
\label{sec:cpodd_tth}
Top-pair production in association with a Higgs boson has been widely explored in the context of the top-Yukawa coupling $\mathcal{CP}$ structure. In this analysis we will review the corresponding observables and explore whether \cp-odd effects in other interactions can play an important role in this process. 
At parton-level, from an EFT perspective, many observables rely on quadratic contributions to constrain the $\mathcal{CP}$-odd operators. Alternative approaches involve the construction of manifestly $\mathcal{CP}$-odd observables that can probe the $\mathcal{CP}$-odd interference directly at linear order \cite{Boudjema:2015nda,Goncalves:2018agy,Barman:2021yfh,Goncalves:2021dcu,Barman:2022pip,Azevedo:2022jnd}, or the use of machine learning techniques \cite{Ren:2019xhp,Barman:2021yfh,Bhardwaj:2021ujv,Bahl:2021dnc,Hall:2022bme,Butter:2022vkj,Ackerschott:2023nax,Esmail:2024gdc,Hammad:2025ewr}. In this work, we focus on the construction of \cp-sensitive observables sensitive to our subset of WCs. 

The sensitivity of this process to operators involving $W$ bosons is suppressed by the electroweak coupling. However, since we have considered the effects of the operator $\hat{O}_{tW}$ in $thj$ production, we also include it in $t\bar{t}h$ for completeness. In fact, due to the large uncertainties in $thj$ measurements, the constraints on $\hat{O}_{tW}$ are found to be stronger from $t\bar{t}h$ production \cite{CMS:2022hjj}, as discussed in Sec.~\ref{sec:fit}. Other operators, such as $\hat{O}_{tZ}$, which could generate effects similar to $\hat{O}_{tW}$ in $t\bar{t}h$ production, are omitted because they do not contribute to $thj$, and their impact on $t\bar{t}h$ is minimal. Furthermore, these operators can be more effectively constrained from other LHC processes \cite{Brivio:2019ius,Durieux:2019rbz,Ellis:2020unq,Ethier:2021bye,Miralles:2021dyw,Celada:2024mcf}.
Therefore, for the observables built from top-pair production in association with a Higgs boson, we have considered the effects of only five out of the seven operators considered: \{$\hat{O}_{t\varphi}$, $\hat{O}_{tG}$, $\hat{O}_{tW}$, $O_{\varphi G}$, $O_{\varphi\widetilde{G}}$ \}.
\begin{figure}[h!]
  \centering
  \begin{subfigure}[b]{0.42\textwidth}
    \includegraphics[width=1.\linewidth]{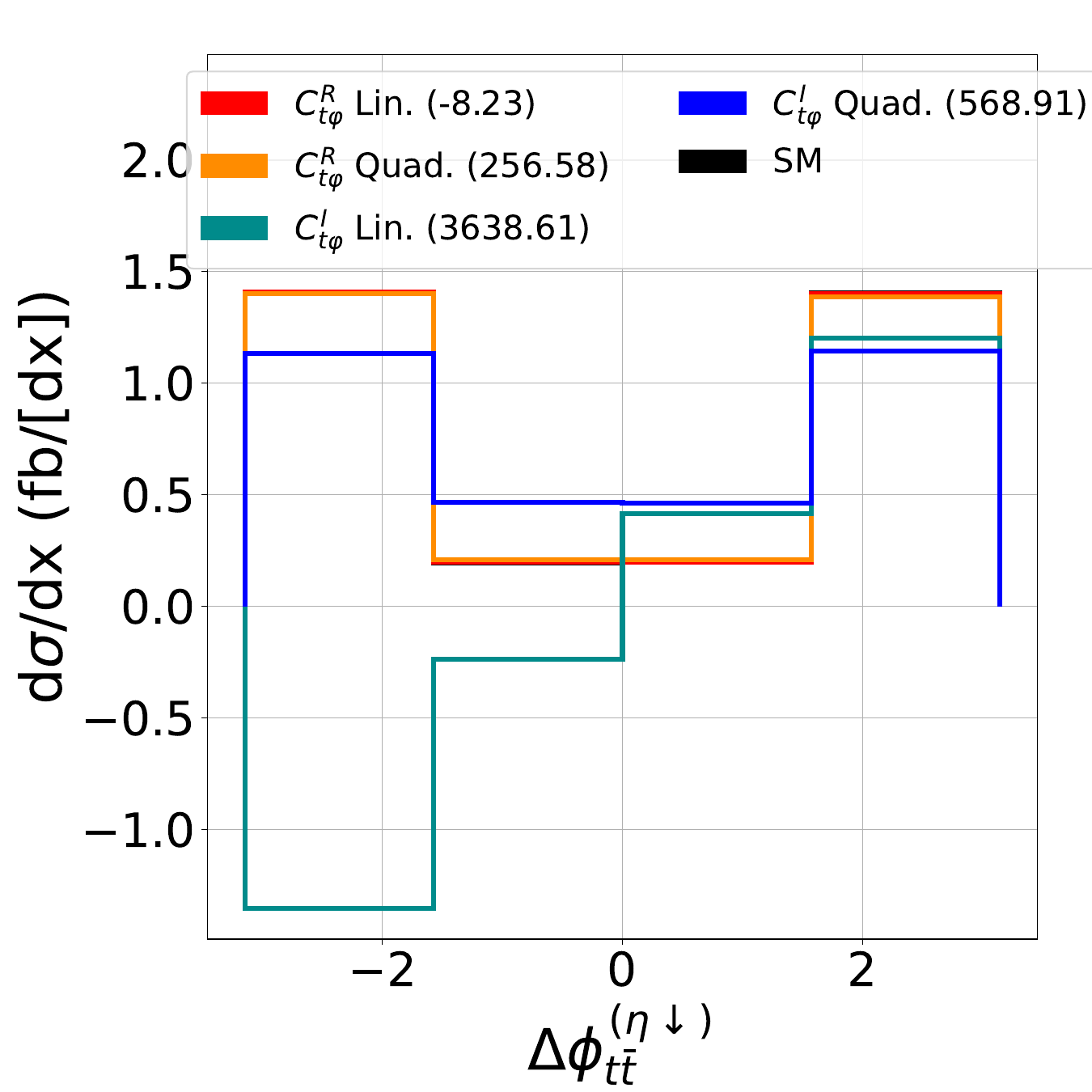}
    \caption{}
    \label{fig:sub1}
  \end{subfigure}%
  \begin{subfigure}[b]{0.42\textwidth}
    \includegraphics[width=1.\linewidth]{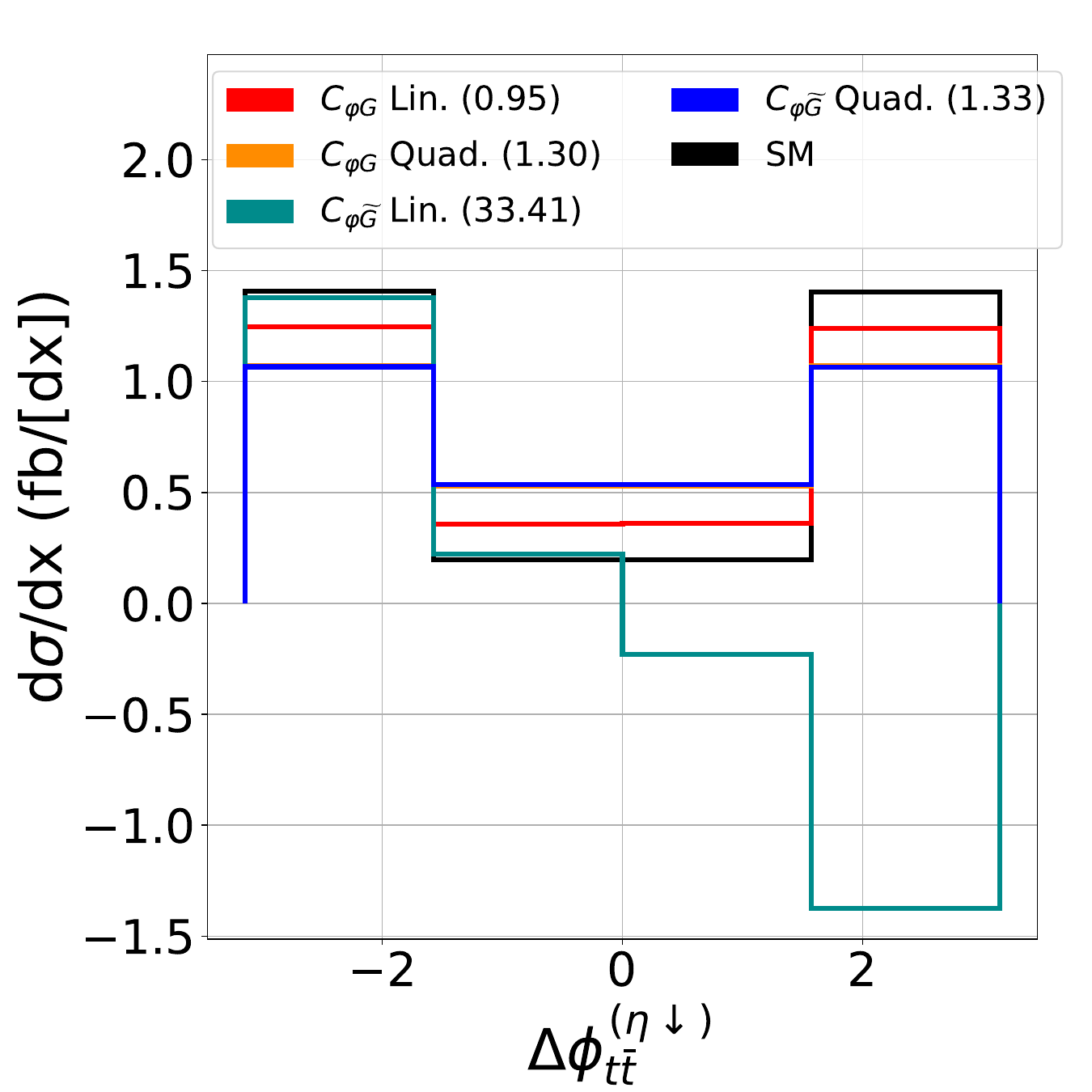}
    \caption{}
    \label{fig:sub1}
  \end{subfigure}
  \begin{subfigure}[b]{0.42\textwidth}
    \includegraphics[width=1.\linewidth]{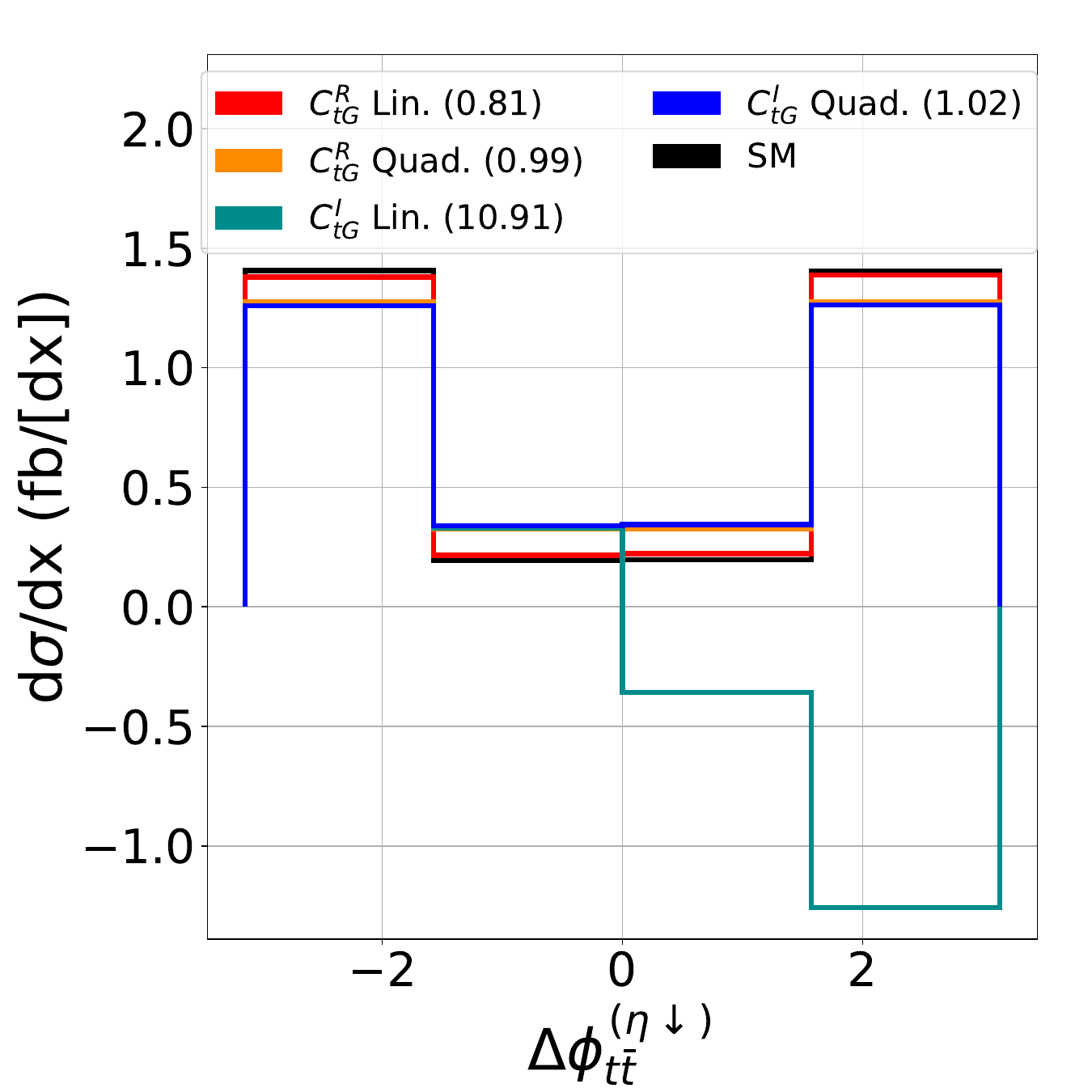}
    \caption{}
    \label{fig:sub1}
  \end{subfigure}%
  \begin{subfigure}[b]{0.42\textwidth}
    \includegraphics[width=1.\linewidth]{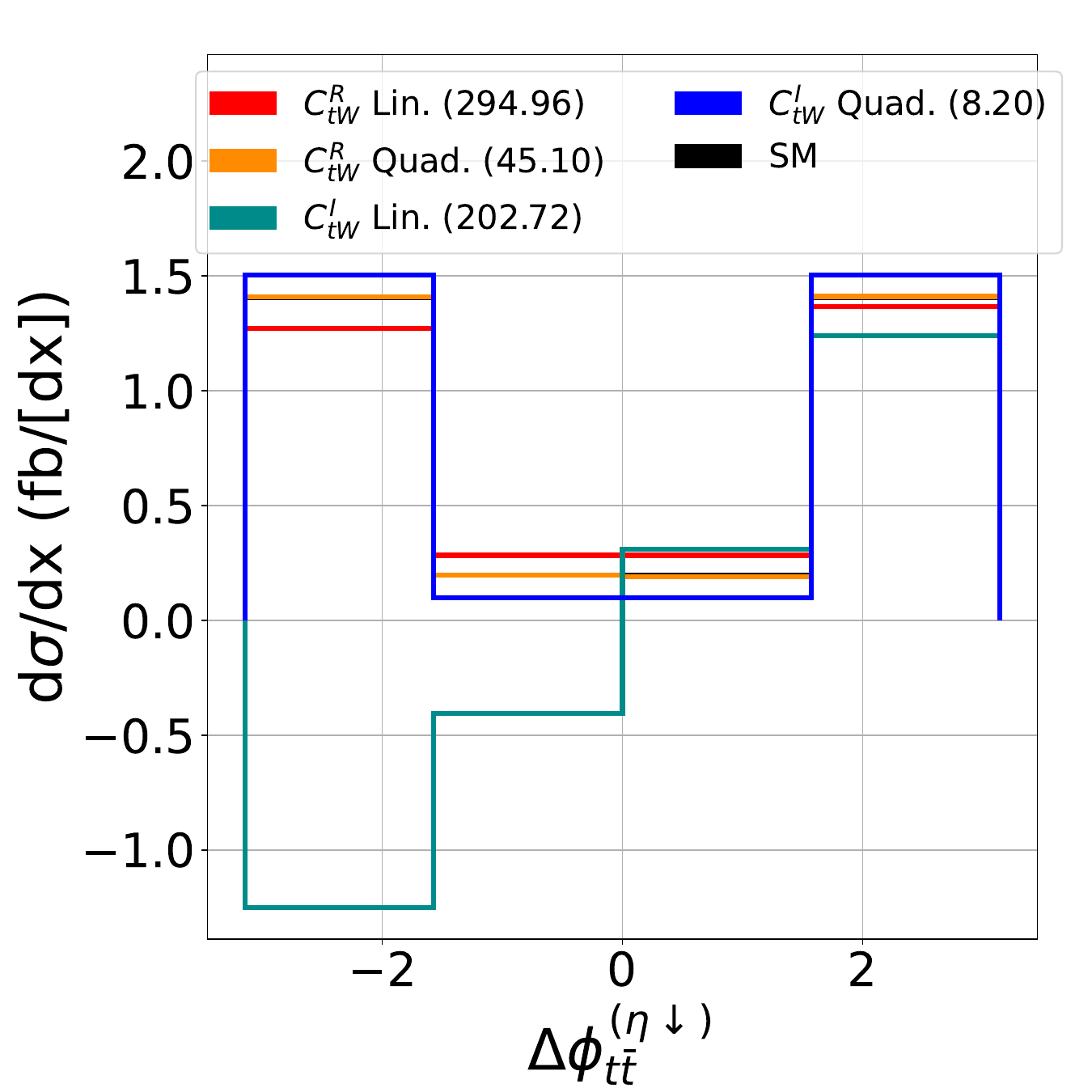}
    \caption{}
    \label{fig:sub1}
  \end{subfigure}%
  \caption{\label{fig:tth_dphi_tt}Differential cross sections of  $\Delta\phi_{tt}^{(\eta\downarrow)}$, for the operators $\hat{O}_{t\varphi}$ (top left), $O_{\varphi G/\widetilde{G}}$ (top right), $\hat{O}_{tG}$ (bottom left) and $\hat{O}_{tW}$ (right left). SMEFT predictions are scaled to match the SM curve area, and we show the multiplicative factor used in parenthesis in the label of each curve}. 
\end{figure}
\paragraph{Top-quark level observables}
The two-fermion operators generate identical kinematics for the top and antitop quarks. This ensures that, given the subset of operators considered here, angular observables constructed from top-quark four-momenta, ordered by their pseudorapdities, are invariant under charge conjugation while retaining their parity dependence. Therefore, these observables can be sensitive to the \cp-odd couplings at linear order. 

In this work, we have considered the pseudorapidity-ordered azimuthal angular separation between the top quarks, $\Delta\phi_{tt}^{\eta\downarrow} = (\phi(t_1)-\phi(t_2) : \eta(t_1)>\eta(t_2))$. 
This distribution is \cp-odd by construction and, as can be seen in Fig.~\ref{fig:tth_dphi_tt}, it can be used to discriminate the \cp-odd effects of $\hat{O}_{tW}$, $\hat{O}_{tG}$ and $O_{\varphi\widetilde{G}}$ by constructing an asymmetry observable with its distribution. However, the sensitivity to the \cp-odd effects of $\hat{O}_{t\varphi}$ is significantly smaller, making it less effective for probing this operator.

In the context of extended Higgs sectors, various observables have been proposed to discriminate non-SM top-Yukawa couplings~\cite{Gunion:1996xu}. The well-known $b_2$ and $b_4$ observables are \cp-even, but the quadratic effects of the \cp-odd couplings generate different shapes on their distributions. They are defined as~\cite{Gunion:1996xu}
\begin{align}
\label{eq:b2}
    b_2 &= \left({\hat{p}}_t \times \hat{z}\right) \cdot \left({\hat{p}}_{\bar{t}} \times \hat{z}\right) =  \sin{\theta_t}\sin{{\theta}_{\bar{t}}}\cos{\Delta\phi_{t\bar{t}}}, \\
\label{eq:b4}
    b_4 &= \left({\hat{p}}_t \cdot \hat{z}\right) \left({\hat{p}}_{\bar{t}} \cdot \hat{z}\right) = \cos{{\theta}_t} \cos{{\theta}_{\bar{t}}},
\end{align}
where $\hat{z}$, $\hat{p}_t$ and $\hat{p}_{\bar{t}}$ represent the 3-momentum of the beam axis, the top quark and the antitop quark, respectively. In our analysis we have included $b_4$ measured in the lab frame since is the one more sensitive to \cp-odd interactions, as shown in the ATLAS search for \cp-odd Higgs bosons~\cite{ATLAS:2023cbt}. The observable $b_2$ is highly correlated with $b_4$ and is therefore not included in our analysis. The dependence of $b_2$ and $b_4$ on the WC considered can be found in Fig.~\ref{fig:tth_b2_b4} of Appendix~\ref{app:extra_fig}.

\begin{figure}[h!]
  \centering
  \vspace{-1.2cm}
  \begin{subfigure}[b]{0.28\textwidth}
    \includegraphics[width=1.\linewidth]{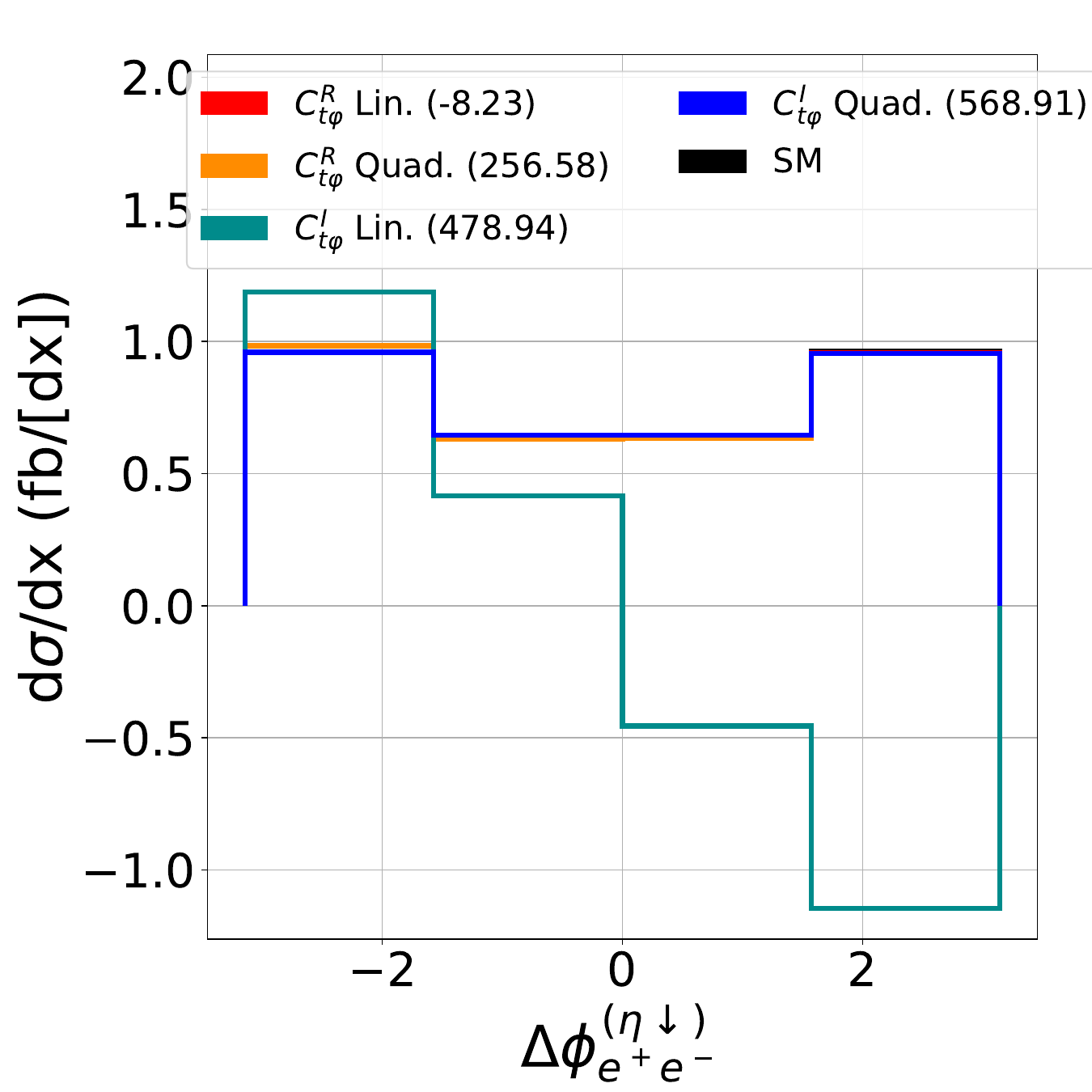}
    \caption{}
    \label{fig:sub1}
  \end{subfigure}%
  \begin{subfigure}[b]{0.28\textwidth}
    \includegraphics[width=1.\linewidth]{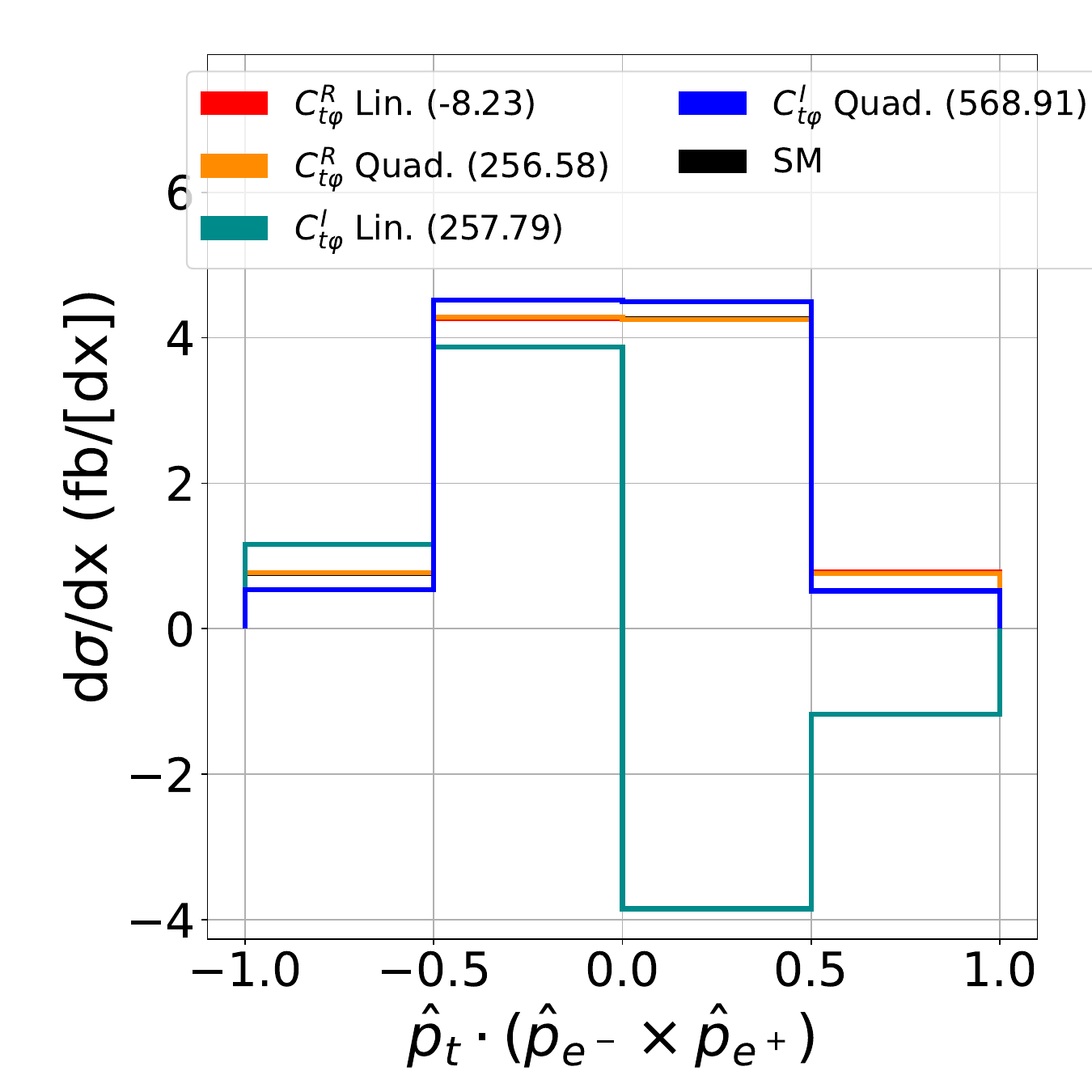}
    \caption{}
    \label{fig:sub2}
  \end{subfigure}%
    \begin{subfigure}[b]{0.28\textwidth}
    \includegraphics[width=1.\linewidth]{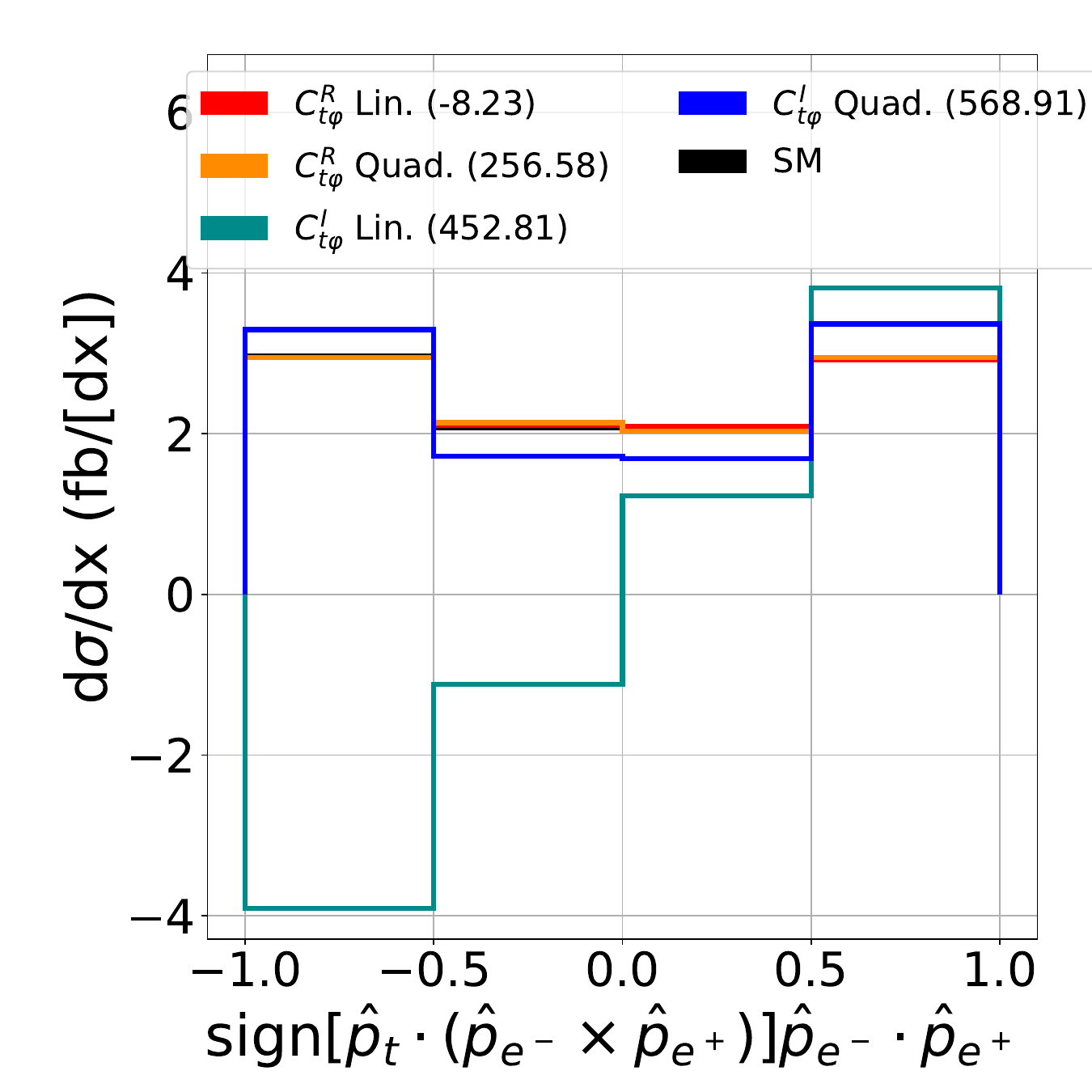}
    \caption{}
    \label{fig:sub2}
  \end{subfigure}
  \begin{subfigure}[b]{0.28\textwidth}
    \includegraphics[width=1.\linewidth]{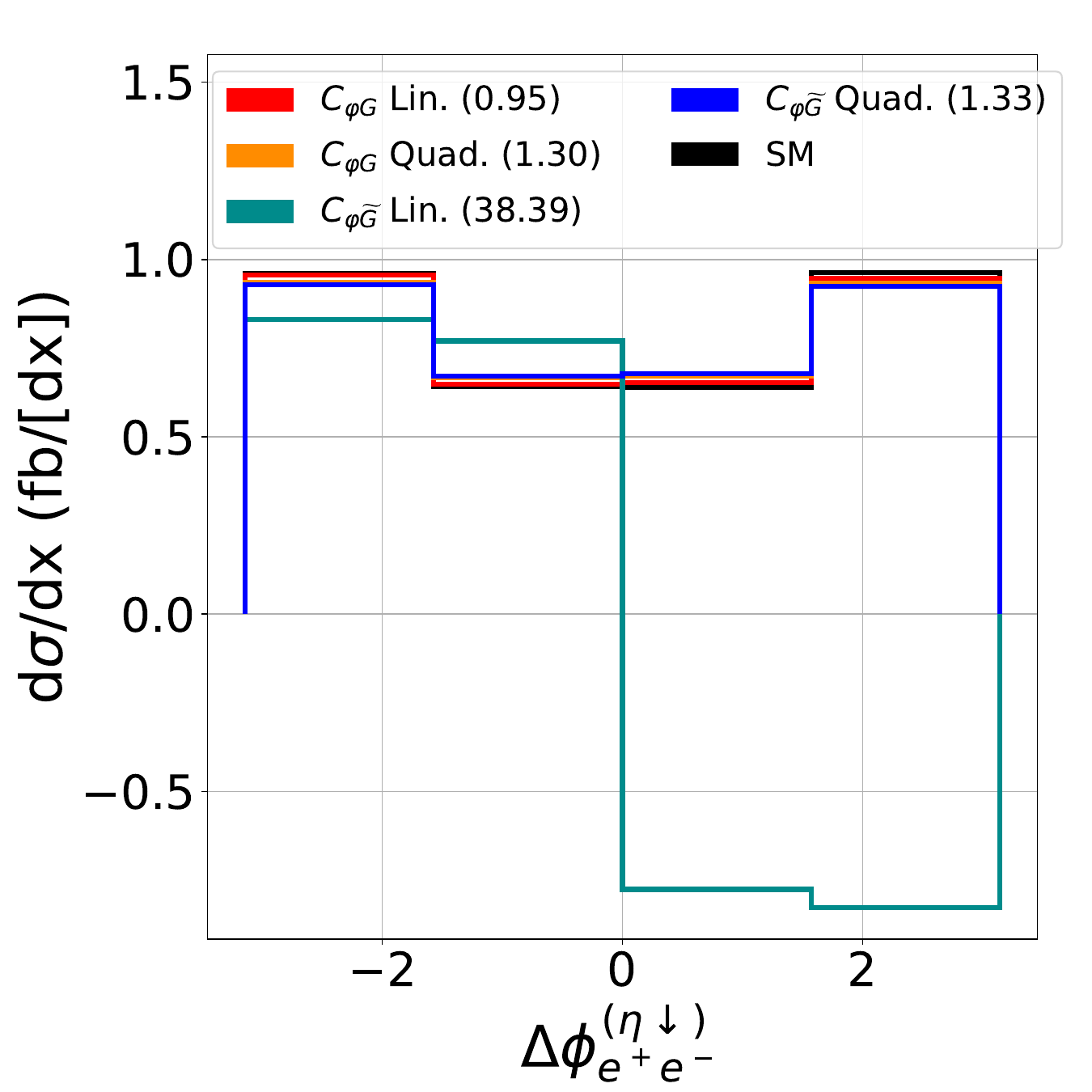}
    \caption{}
    \label{fig:sub1}
  \end{subfigure}%
  \begin{subfigure}[b]{0.28\textwidth}
    \includegraphics[width=1.\linewidth]{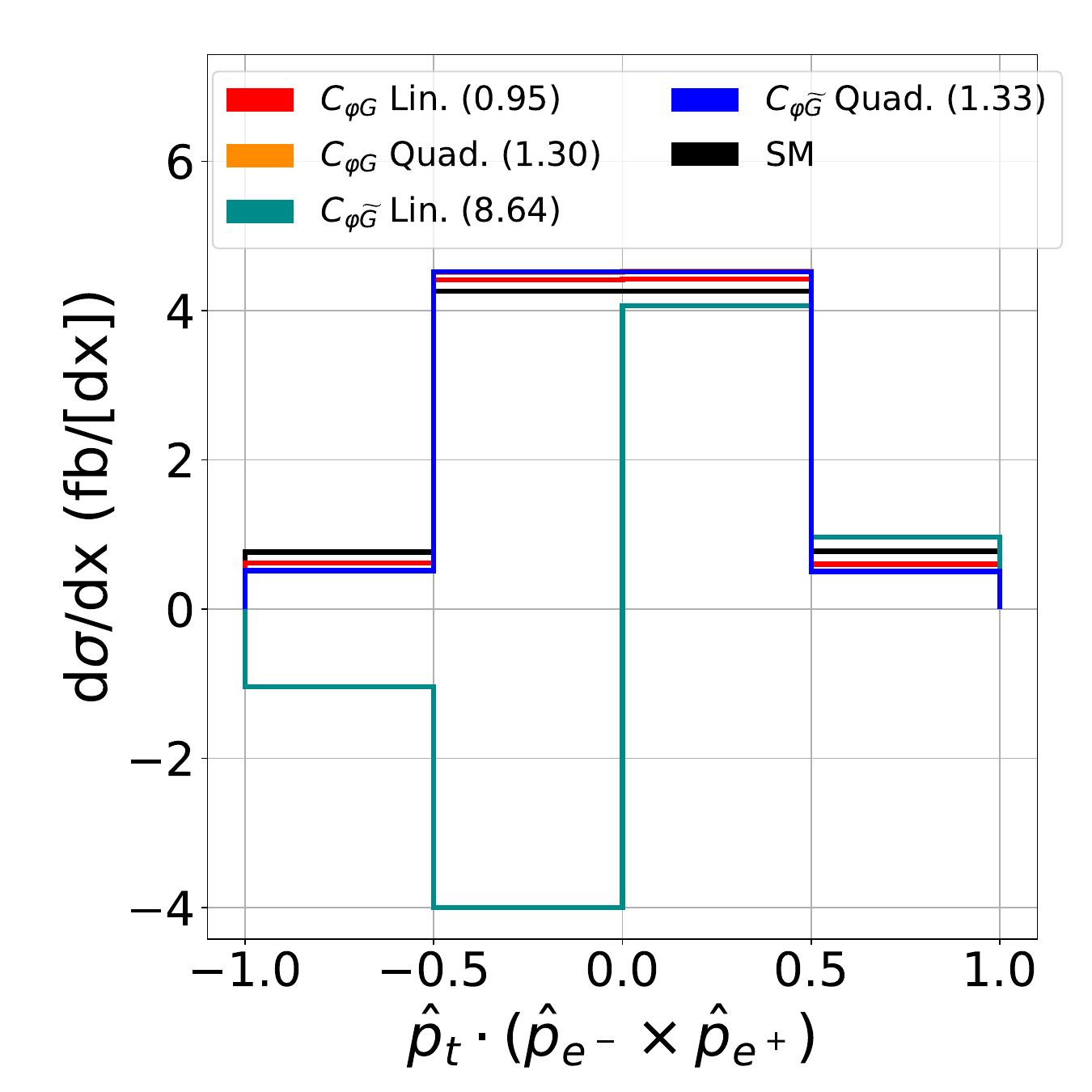}
    \caption{}
    \label{fig:sub2}
  \end{subfigure}
    \begin{subfigure}[b]{0.28\textwidth}
    \includegraphics[width=1.\linewidth]{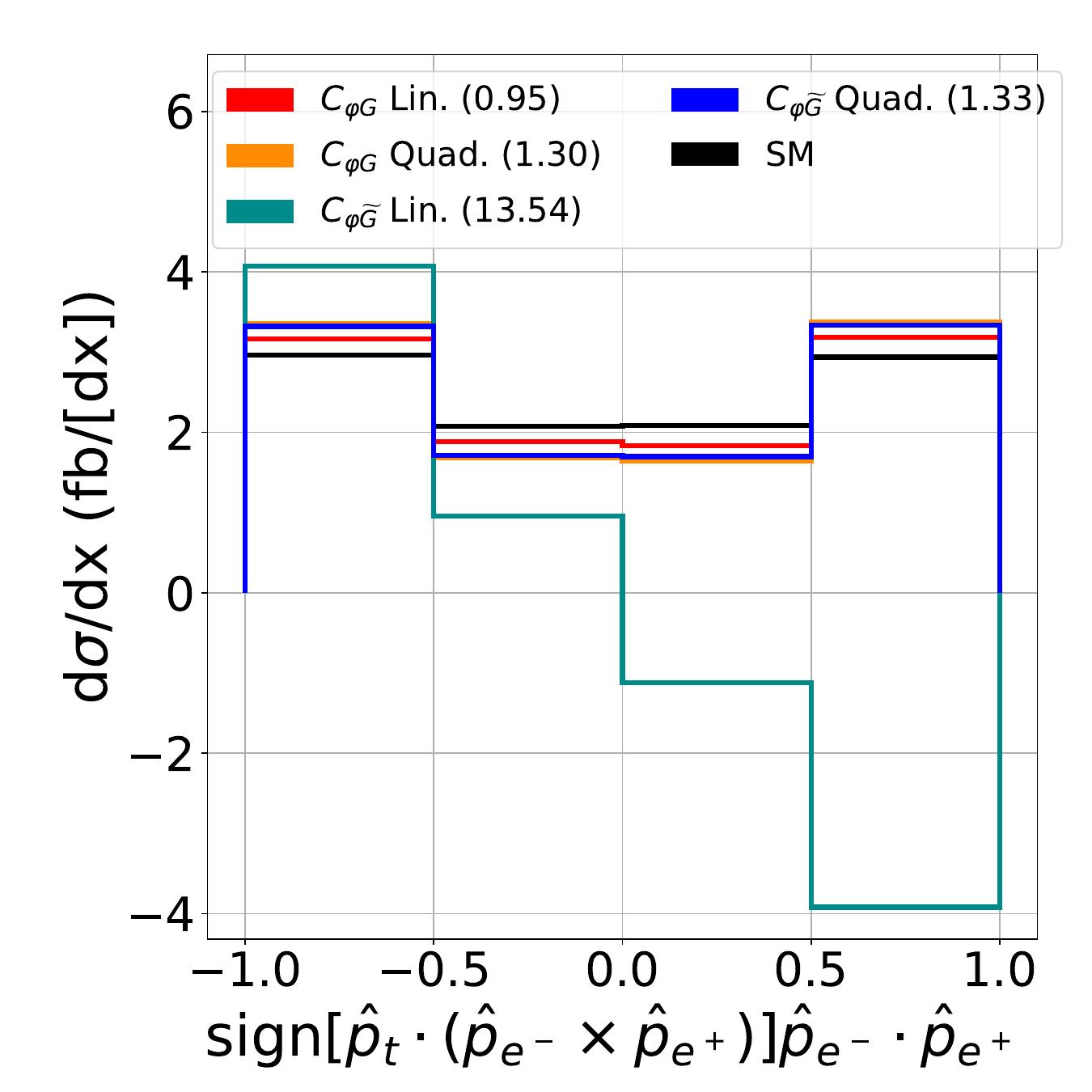}
    \caption{}
    \label{fig:sub2}
  \end{subfigure}
  \begin{subfigure}[b]{0.28\textwidth}
    \includegraphics[width=1.\linewidth]{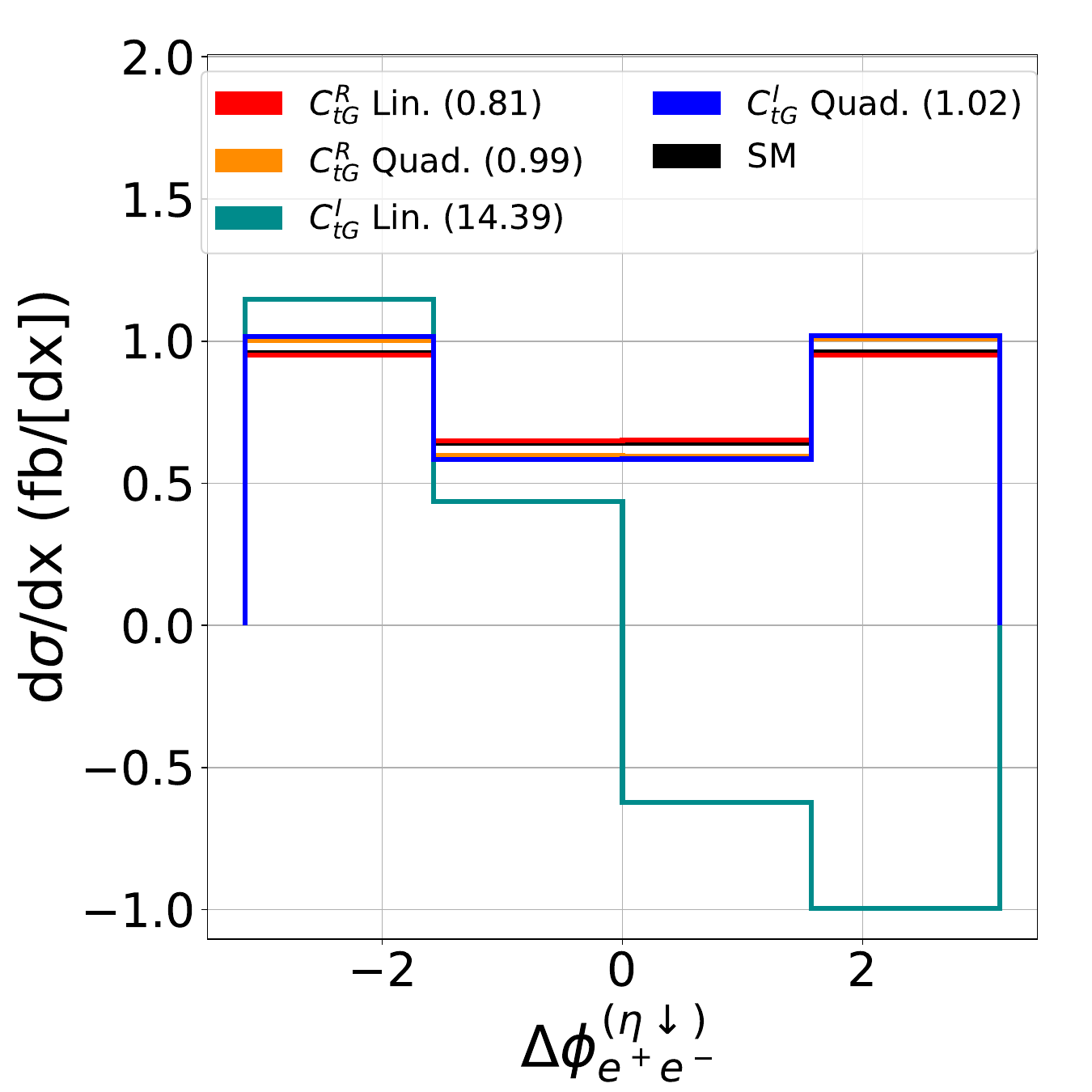}
    \caption{}
    \label{fig:sub1}
  \end{subfigure}%
  \begin{subfigure}[b]{0.28\textwidth}
    \includegraphics[width=1.\linewidth]{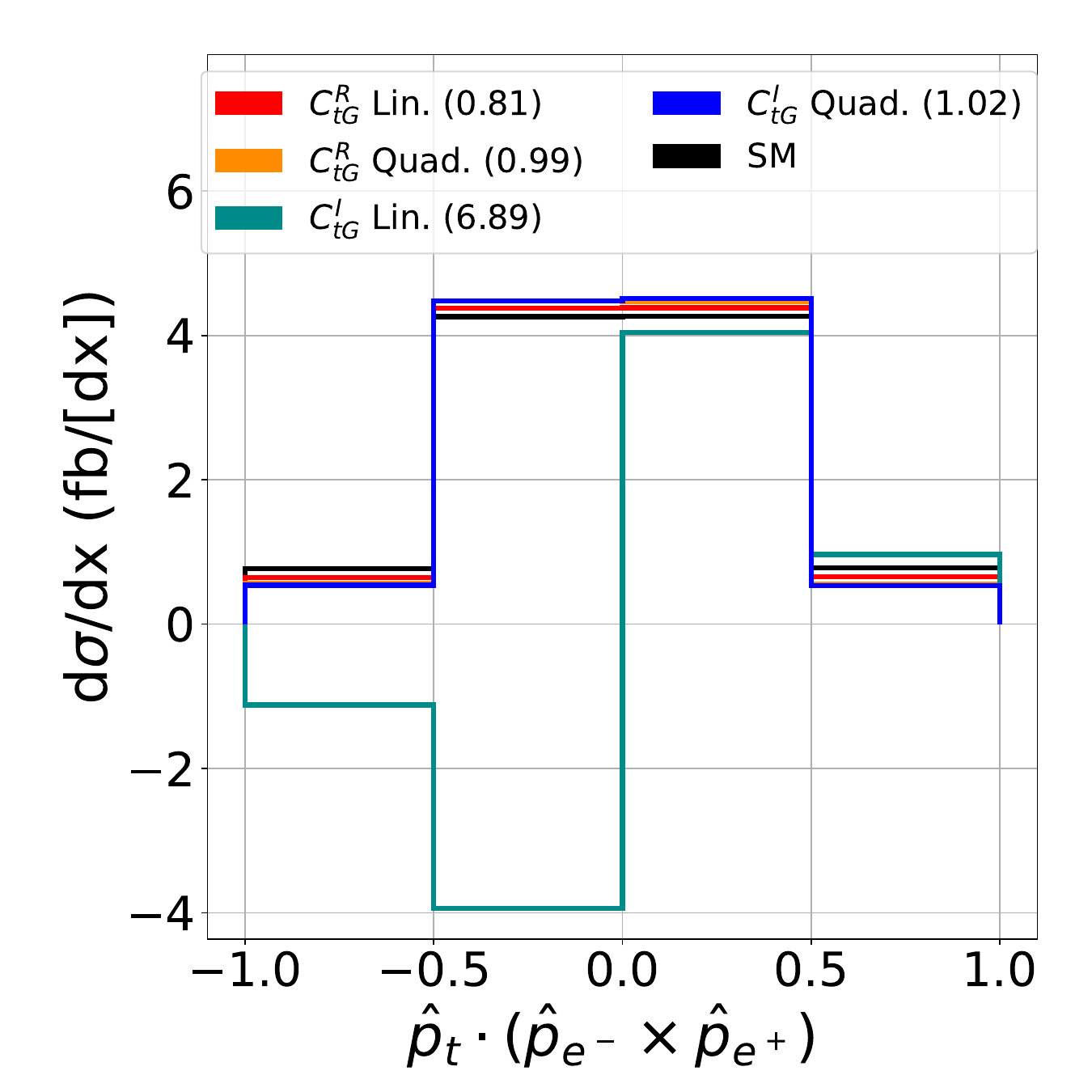}
    \caption{}
    \label{fig:sub2}
  \end{subfigure}%
    \begin{subfigure}[b]{0.28\textwidth}
    \includegraphics[width=1.\linewidth]{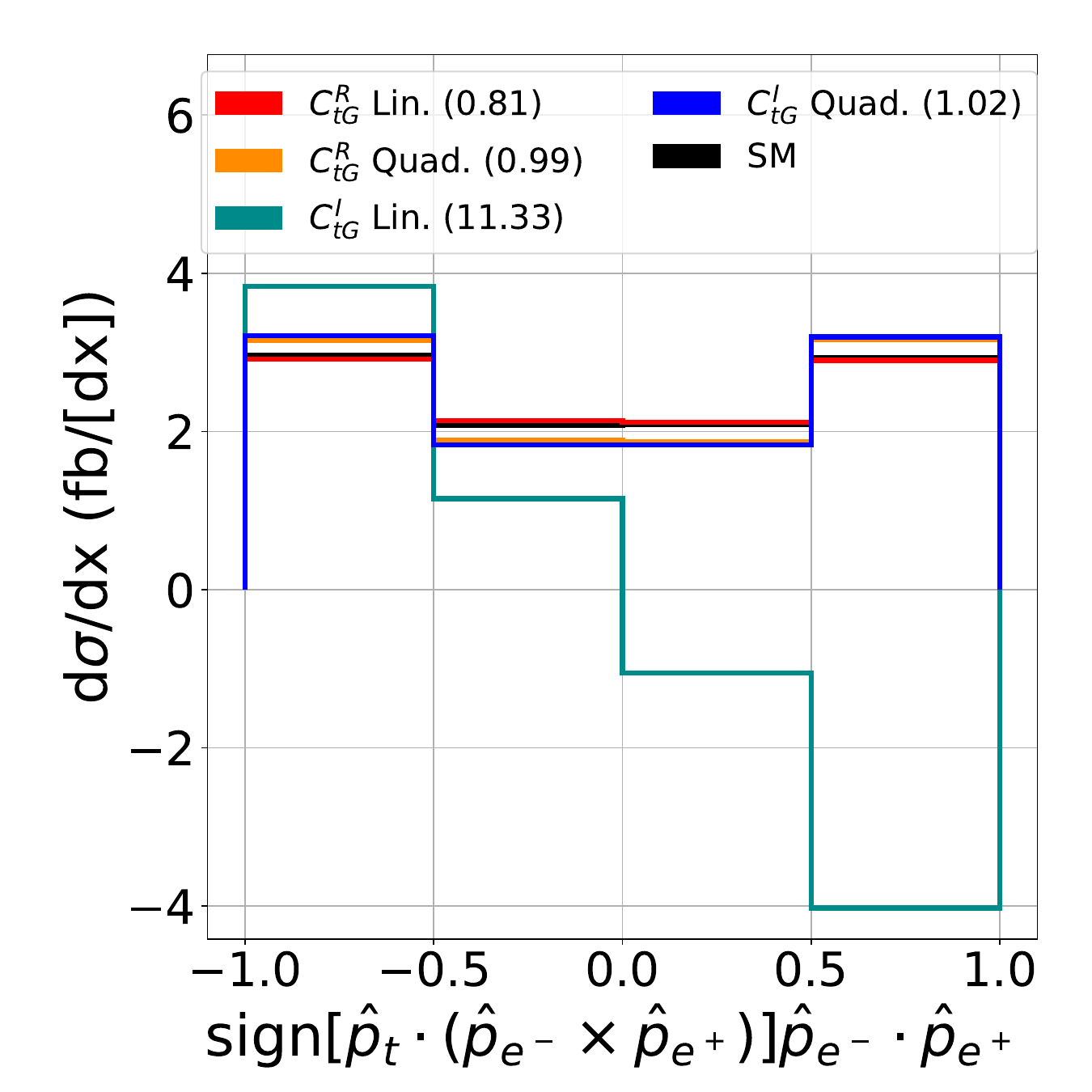}
    \caption{}
    \label{fig:sub2}
  \end{subfigure}
    \begin{subfigure}[b]{0.28\textwidth}
    \includegraphics[width=1.\linewidth]{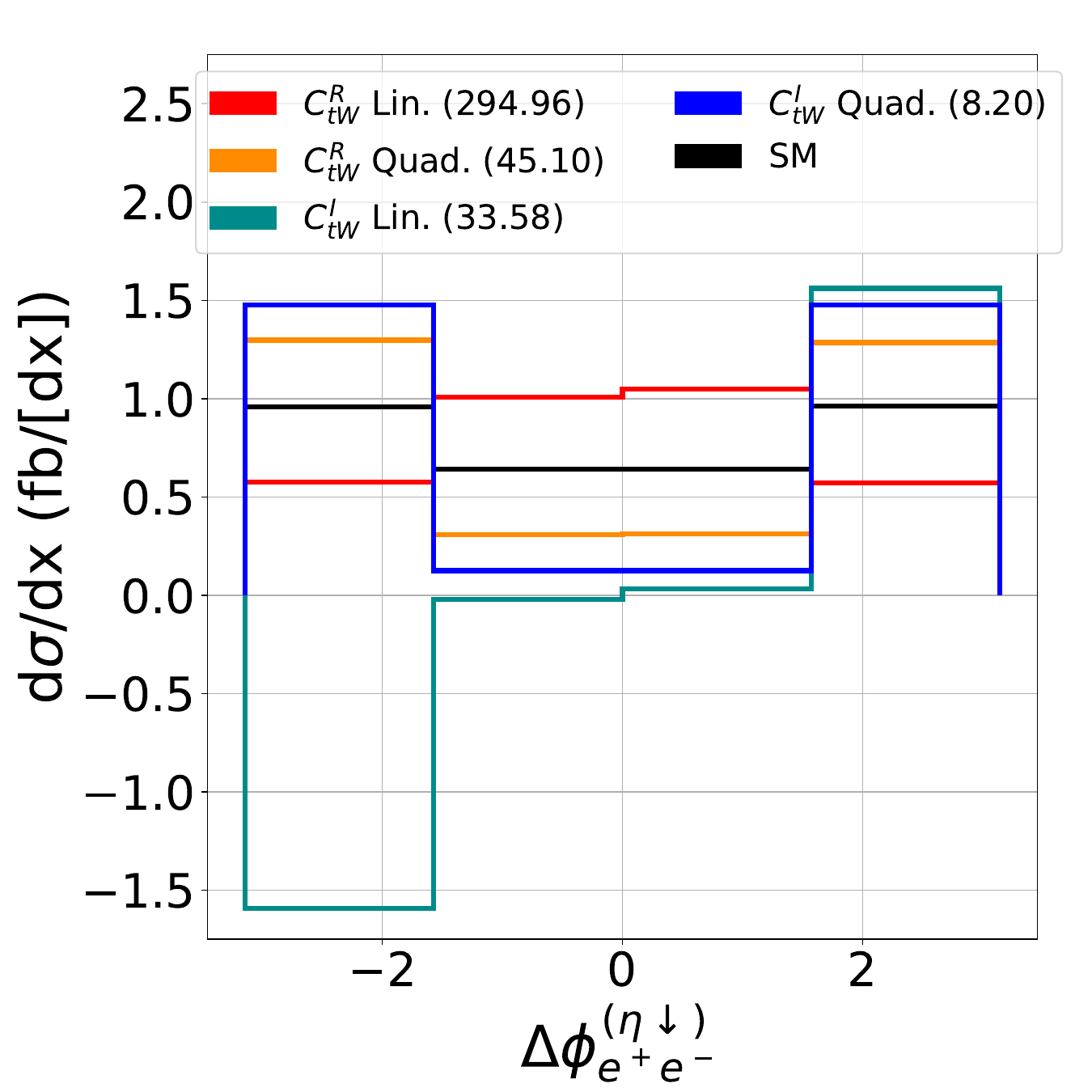}
    \caption{}
    \label{fig:sub1}
  \end{subfigure}%
  \begin{subfigure}[b]{0.28\textwidth}
    \includegraphics[width=1.\linewidth]{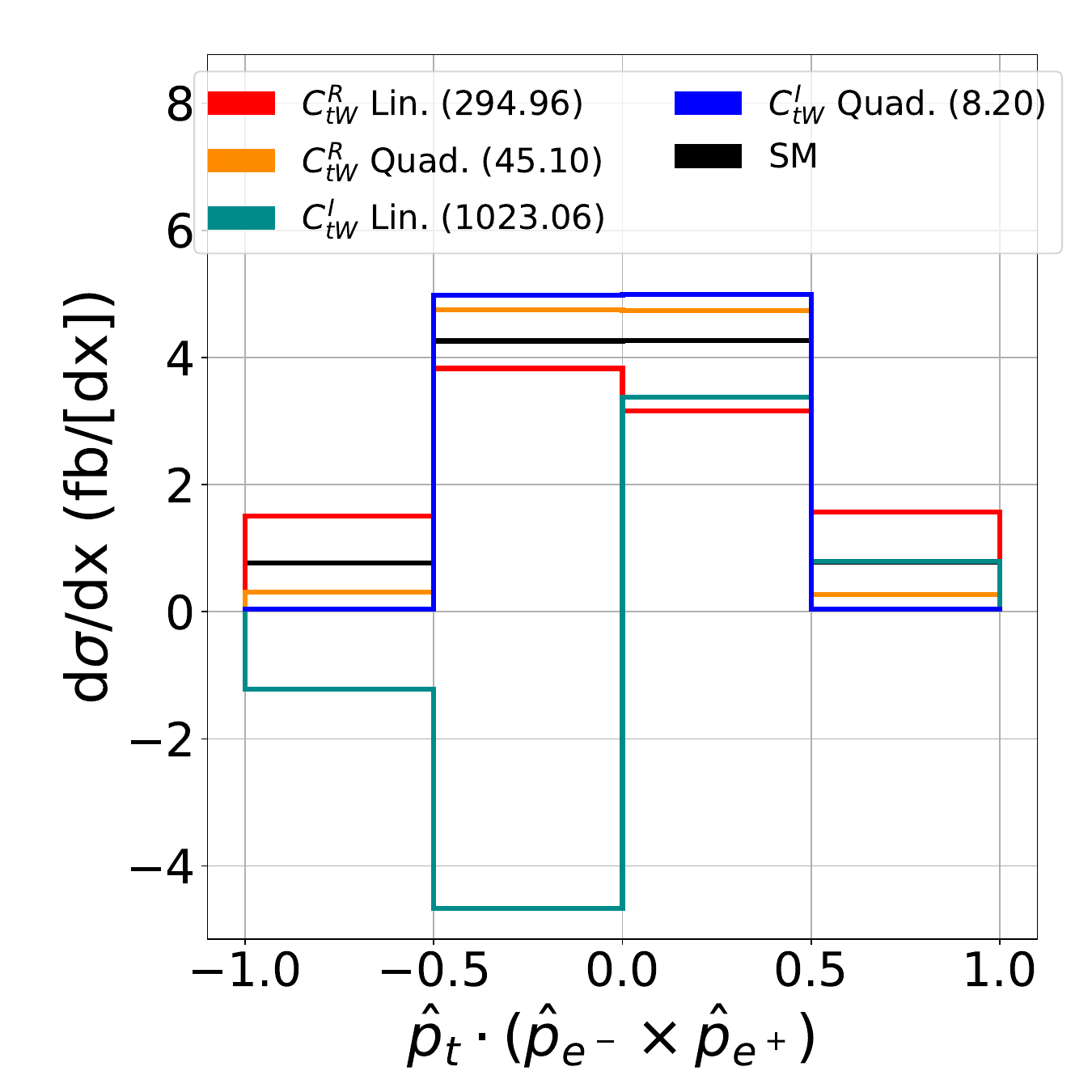}
    \caption{}
    \label{fig:sub2}
  \end{subfigure}%
    \begin{subfigure}[b]{0.28\textwidth}
    \includegraphics[width=1.\linewidth]{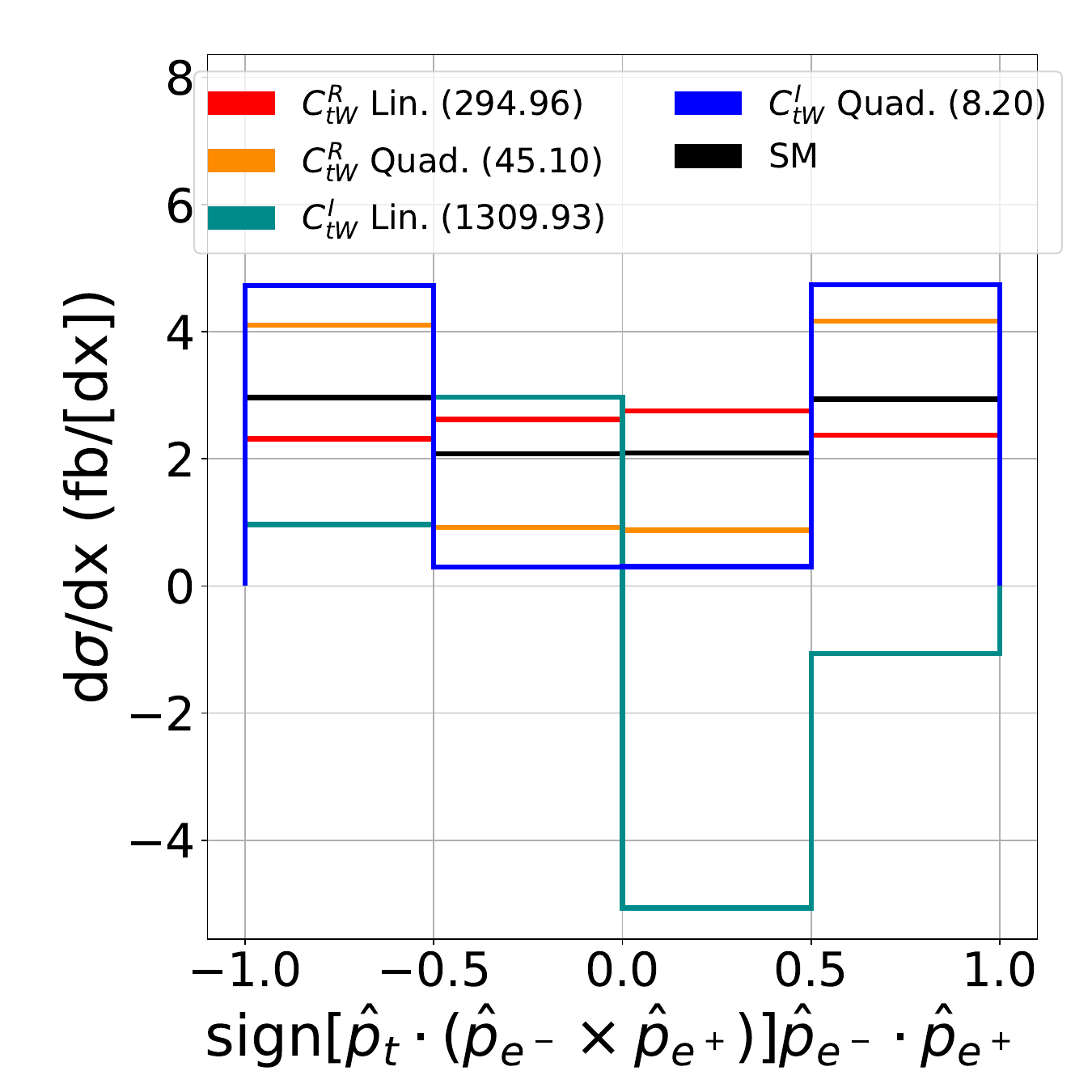}
    \caption{}
    \label{fig:sub2}
  \end{subfigure}
  \caption{\label{fig:tth_decay}Differential cross sections of the observables 
$\Delta\phi_{ee}^{\eta\downarrow}$ (lab), 
$\hat{p}_t \cdot (\hat{p}_{e^-} \times \hat{p}_{e^+})$ 
and $\mathrm{sgn}\left[\hat{p}_t \cdot (\hat{p}_{e^-} \times \hat{p}_{e^+})\right] \hat{p}_{e^-} \cdot \hat{p}_{e^+}$ (in the $t\bar{t}$ rest frame), 
for the operators $\hat{O}_{t\varphi}$ (top), $O_{\varphi G/\widetilde{G}}$ (centre top), $\hat{O}_{tG}$ (centre bottom) and $\hat{O}_{WG}$ (bottom). SMEFT predictions are scaled to match the SM curve area, and we show the multiplicative factor used in parenthesis in the label of each curve.}
\end{figure}

\paragraph{Decay level observables} Besides the parton-level observables, several particle-level observables can be defined to exploit additional information in the leptonic and semi-leptonic decays of the $t\bar{t}$ pair to probe top-quark spin correlations and thus the interactions entering the production of the top quarks. One promising observable is the azimuthal separation of the charged lepton pair in the $t\bar{t}$ rest frame, $\Delta\phi_{ll}^{t\bar{t}}$, which has been proven to be sensitive to the \cp-phase of an anomalous top-Yukawa coupling \cite{Azevedo:2022jnd}. 
In this work, instead, we have used the pseudorapidity-ordered azimuthal angular separation between the final electrons in the lab frame, $\Delta\phi_{ee}^{\eta\downarrow}$. The advantage of this observable is that it does not depend on the explicit reconstruction of the $t\bar{t}$ system, which is challenging when both top quarks decay to leptons, due to the presence of multiple neutrinos in the final state. Additionally, we have considered the triple product of the normalised 3-momenta of the top quark, the final electron, and the positron\footnote{We are only considering the electron channel decay of the top quarks in this work, as explained in Sec.~\ref{sec:methodology}.}, $\hat{p}_t \cdot (\hat{p}_{e^-} \times \hat{p}_{e^+})$, and the cosine of the angle formed by the final electron and positron scaled by the sign of the previous triple product, ${\mathrm{sgn}\left[\hat{p}_t \cdot (\hat{p}_{e^-} \times \hat{p}_{e^+})\right] \hat{p}_{e^-} \cdot \hat{p}_{e^+}}$, both in the $t\bar{t}$ rest frame. All three observables are sensitive to the \cp-odd couplings at linear level, as illustrated in Fig.~\ref{fig:tth_decay}. Furthermore, these particle-level observables show much better sensitivity to the \cp-odd effects of $\hat{O}_{t\varphi}$ due to the availability of additional spin information at particle-level. 

Experimentally the pseudorapidity-ordered angular separation among the charged leptons, $\Delta\phi_{ee}^{\eta\downarrow}$, is easily accessible since it can be constructed directly from the direction of the lepton pair in the lab frame. Furthermore, and despite the limited information used from the full event, the asymmetry built from $\Delta\phi_{ee}^{\eta\downarrow}$ has a sensitivity comparable in magnitude to that from observables requiring full event reconstruction, as we will see in Sec.~\ref{sec:asym}. Nevertheless, the triple product $\hat{p}_t \cdot (\hat{p}_{e^-} \times \hat{p}_{e^+})$, which requires full event reconstruction, is the one providing the best separation of positive and negative weights for an anomalous top-Yukawa \cp-phase.

\subsection{Asymmetries}
\label{sec:asym}

To maximise the linear sensitivity to the \cp-odd couplings, we have defined asymmetries for the \cp-odd observables presented in Secs.~\ref{sec:cpodd_th} and~\ref{sec:cpodd_tth}. For a given observable $X$, the associated asymmetry is defined as 
\begin{equation}
    A(X,x,c) = \frac{\sigma(X>x, c) - \sigma(X<x, c)}{\sigma(X>x, c) + \sigma(X<x, c)} \equiv \frac{\Delta(X,x,c)}{\sigma(c)}\;,
\end{equation}
where $\sigma(c)$ is the SMEFT cross section --- dependent on the value of the coefficient, $c$ --- and $\sigma(X>x, c)$ and $\sigma(X<x, c)$ are the fiducial rates in each of the phase-space regions, defined by the cut-off value, $x$, typically taken to be zero.
 
Interference and quadratic contributions, parameterised as
\begin{align}
\begin{split}
A(X,x,c) &= \frac{\Delta^{sm}(X,x) + C\Delta^{int}(X,x) + C^2 \Delta^{quad}(X,x)}{\sigma(C)}, \\ 
\Delta^i(X,x) &= \sigma^i(X>x) - \sigma^i(X<x),
\end{split}
\end{align}
are obtained for the subset of WCs considered in this work. Note that all of them except for the asymmetry constructed with $\cos{\theta^j_e}$ in $thj$  will only have a linear contribution in the numerator, given that they are \cp-odd by construction. Apart from the enhanced sensitivity to \cp-odd effects, any asymmetry observable benefits from a significant cancellation of systematic uncertainties and thus offers better prospects in the identification of new physics effects.

\begin{figure}[h]
  \centering
  \begin{subfigure}[b]{0.5\textwidth}
    \includegraphics[width=1.\linewidth]{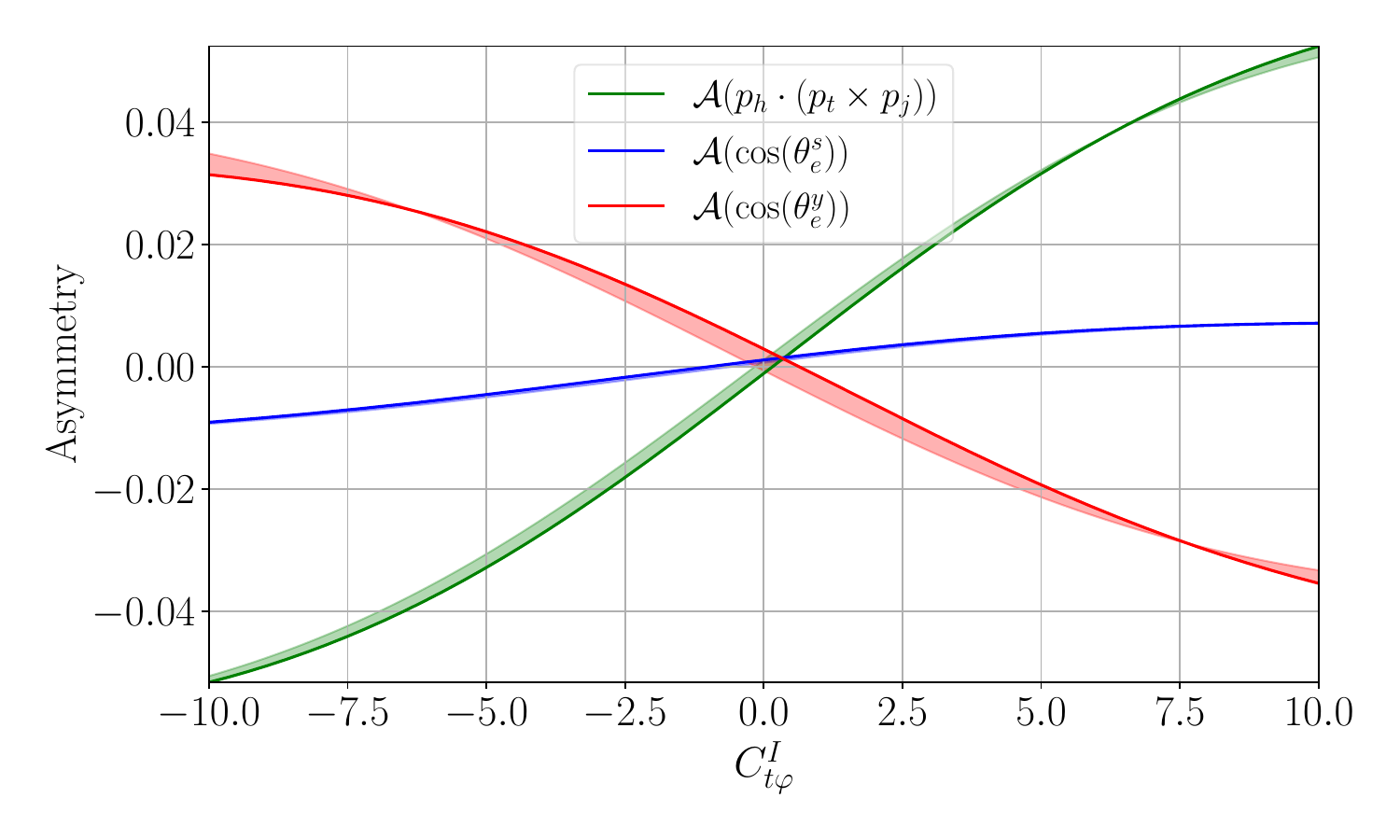}
    \caption{}
    \label{fig:sub1}
  \end{subfigure}%
  \begin{subfigure}[b]{0.5\textwidth}
    \includegraphics[width=1.\linewidth]{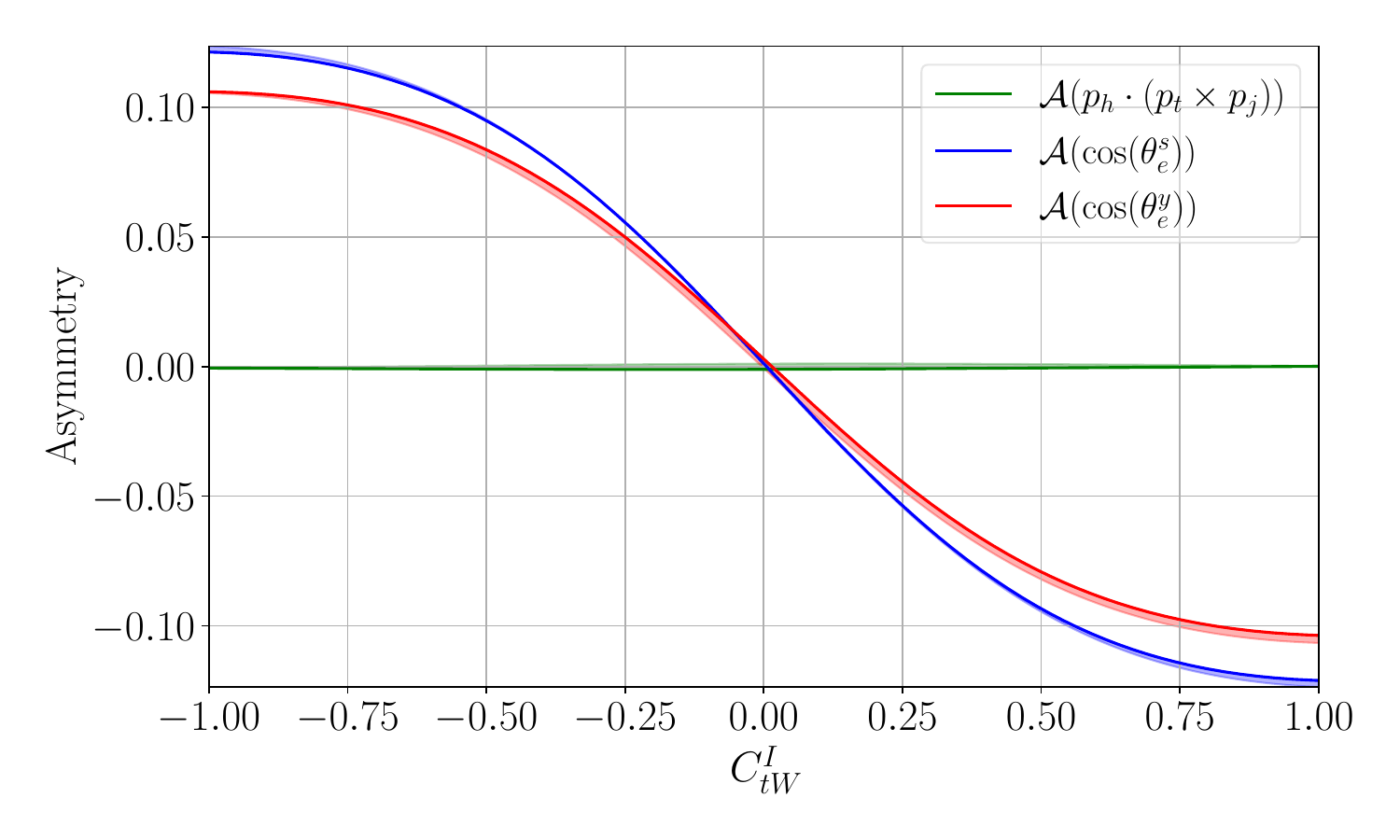}
    \caption{}
    \label{fig:sub2}
  \end{subfigure}

  \begin{subfigure}[b]{0.5\textwidth}
    \includegraphics[width=1.\linewidth]{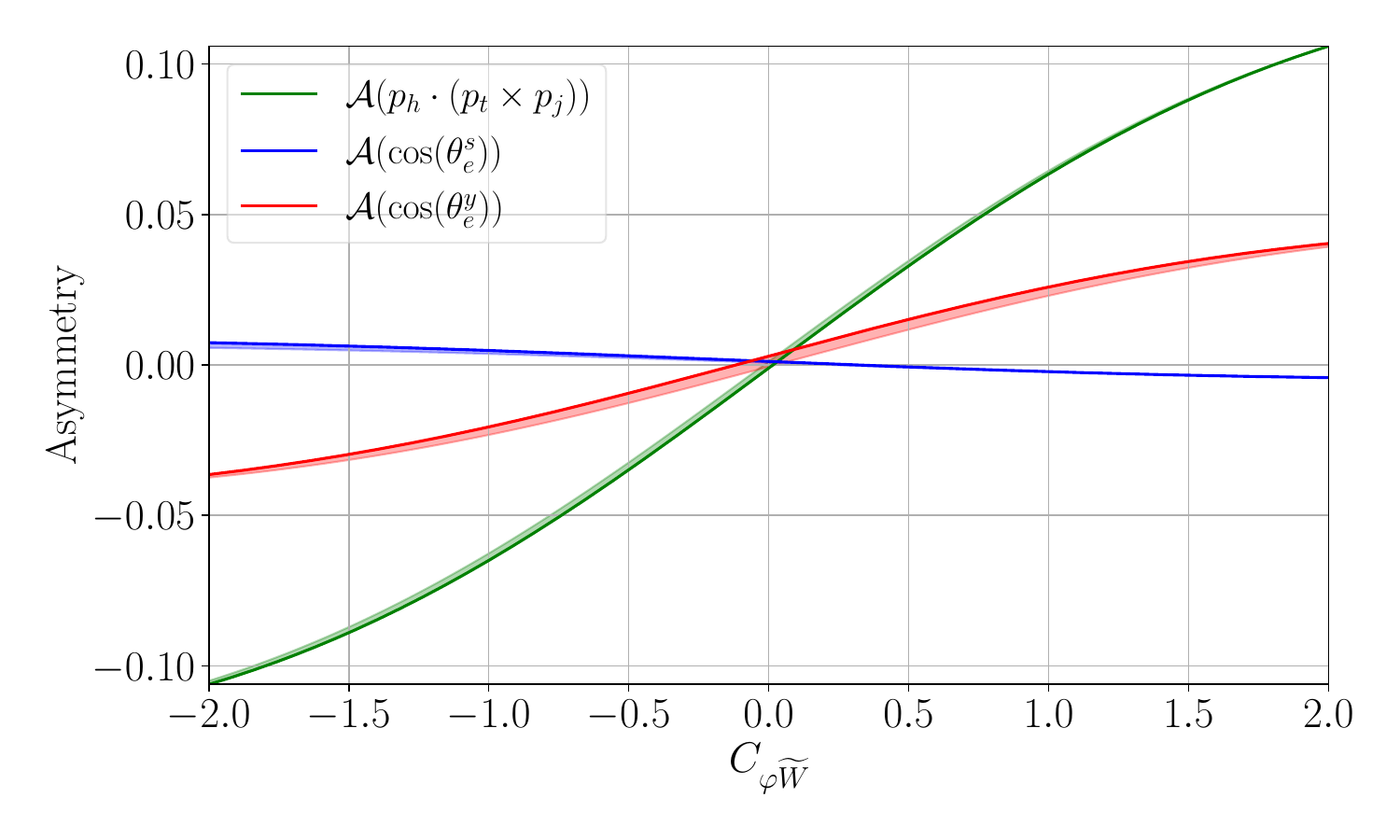}
    \caption{}
    \label{fig:sub1}
  \end{subfigure}%
  \begin{subfigure}[b]{0.5\textwidth}
    \includegraphics[width=1.\linewidth]{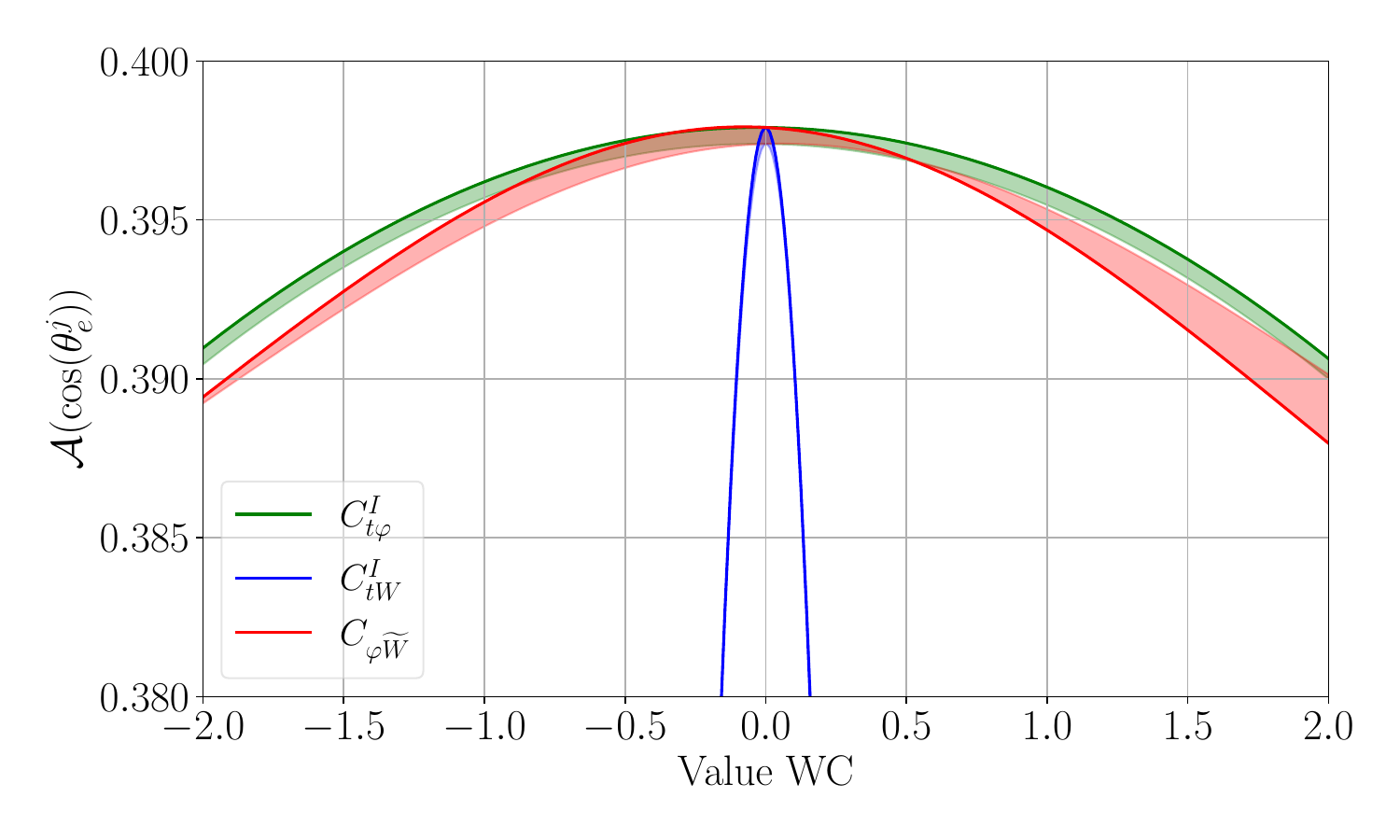}
    \caption{}
    \label{fig:sub2}
  \end{subfigure}
  \caption{\label{fig:thj_asym}Asymmetries in SMEFT $thj$ production, constructed from the observables $\hat{p}_h \cdot \left(\hat{p}_t \times \hat{p}_j\right)$, $\cos{\theta^y_e}$ and $\cos{\theta^s_e}$ (red, blue and green curves, respectively). The asymmetries as a function of Wilson coefficient values are shown for the operators $\hat{O}_{t\varphi},\hat{O}_{tW},O_{\varphi\widetilde{W}}$ ((a) to (c)). For better visibility, panel (d) shows   the asymmetry $A(\cos{\theta^j_e},0)$ for these three operators separately (red, blue and green curves, respectively). The shaded bands represent the estimation of the theoretical uncertainties obtained varying the renormalisation and factorisation scales by a factor 2 and $1/2$. The ranges of the WC in the plots are chosen to be within the values allowed by our prospects on LHC data. }
\end{figure}

\begin{figure}[h]
  \centering
  \begin{subfigure}[b]{0.5\textwidth}
    \includegraphics[width=1.\linewidth]{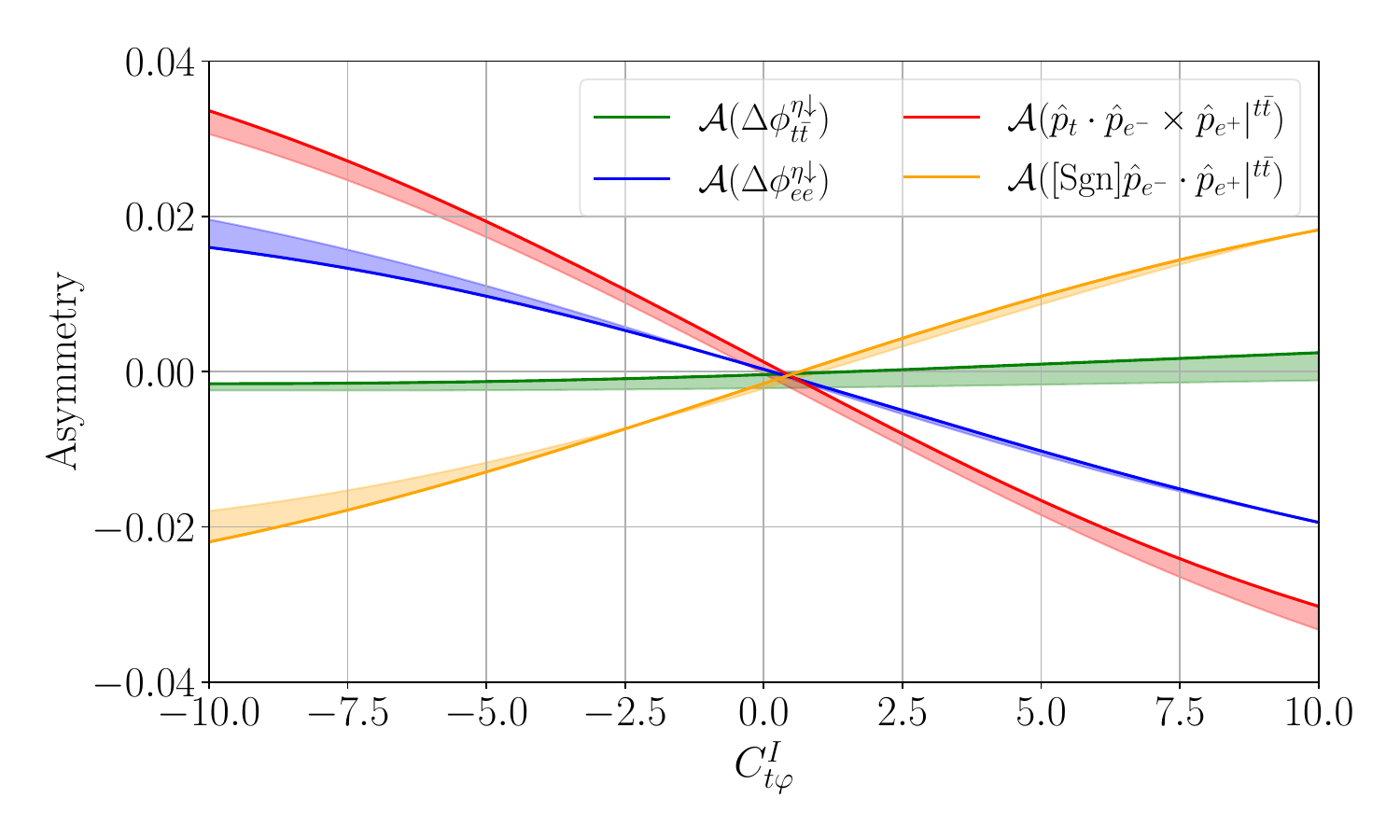}
    \caption{}
    \label{fig:sub1}
  \end{subfigure}%
  \begin{subfigure}[b]{0.5\textwidth}
    \includegraphics[width=1.\linewidth]{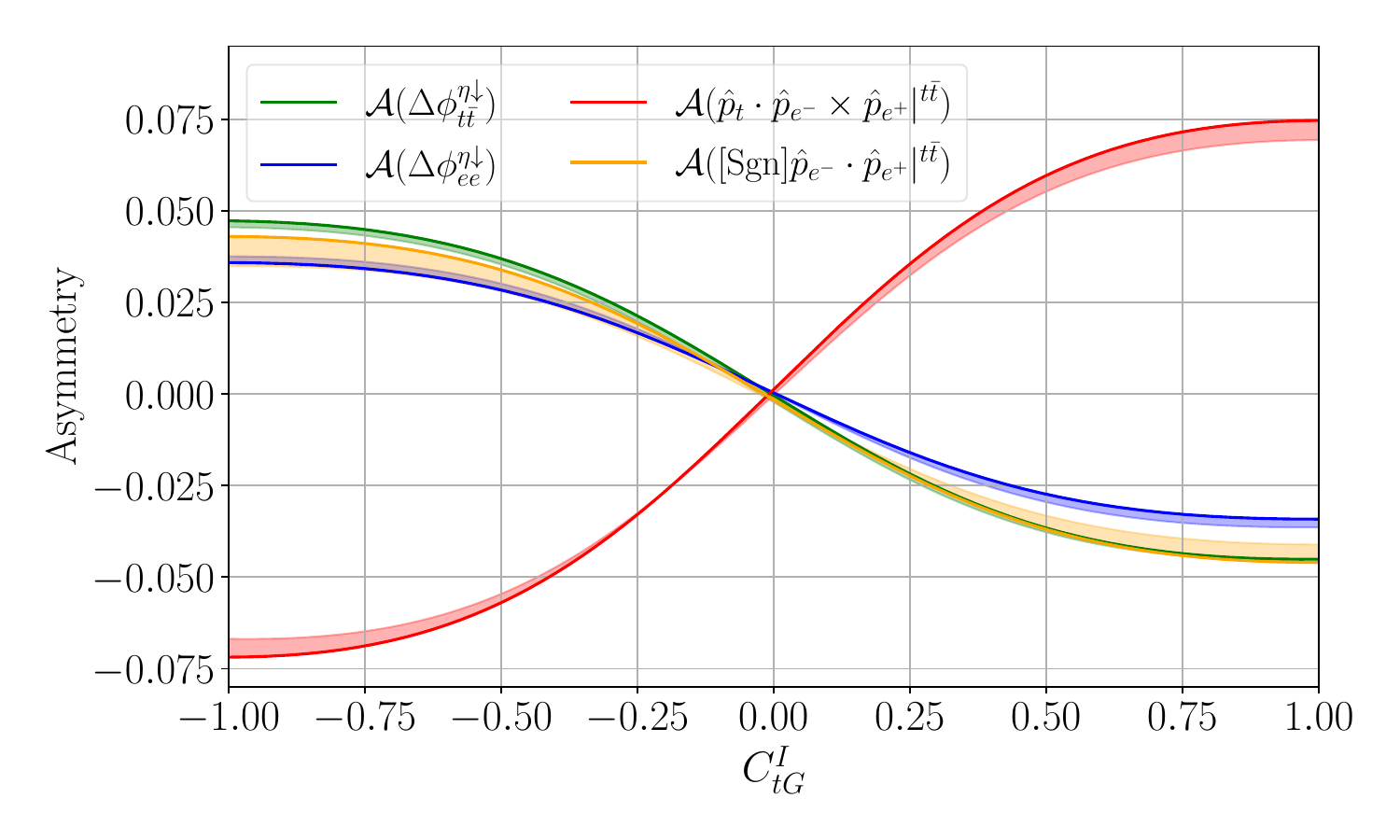}
    \caption{}
    \label{fig:sub2}
  \end{subfigure}

  \begin{subfigure}[b]{0.5\textwidth}
    \includegraphics[width=1.\linewidth]{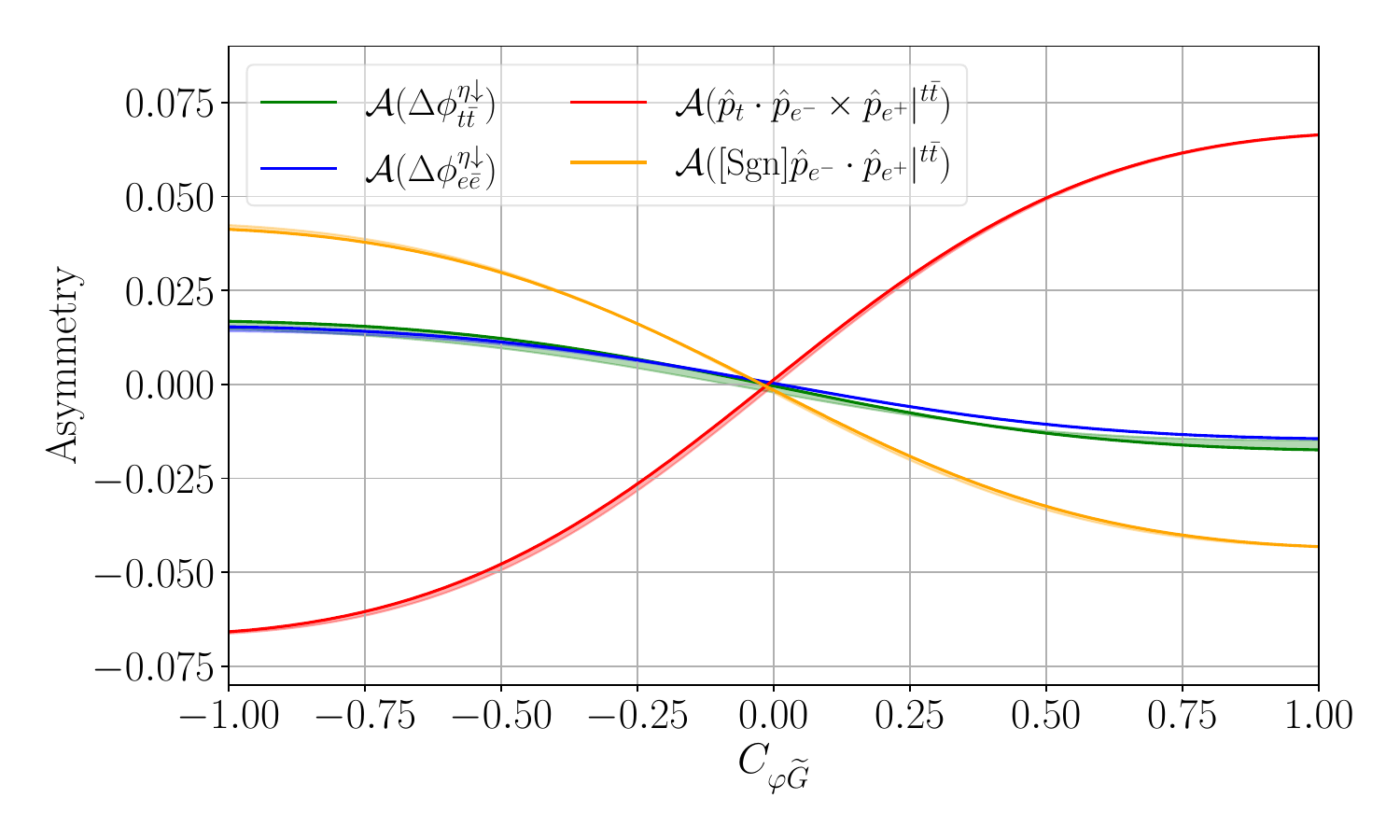}
    \caption{}
    \label{fig:sub1}
  \end{subfigure}%
\begin{subfigure}[b]{0.5\textwidth}
\includegraphics[width=1.\linewidth]{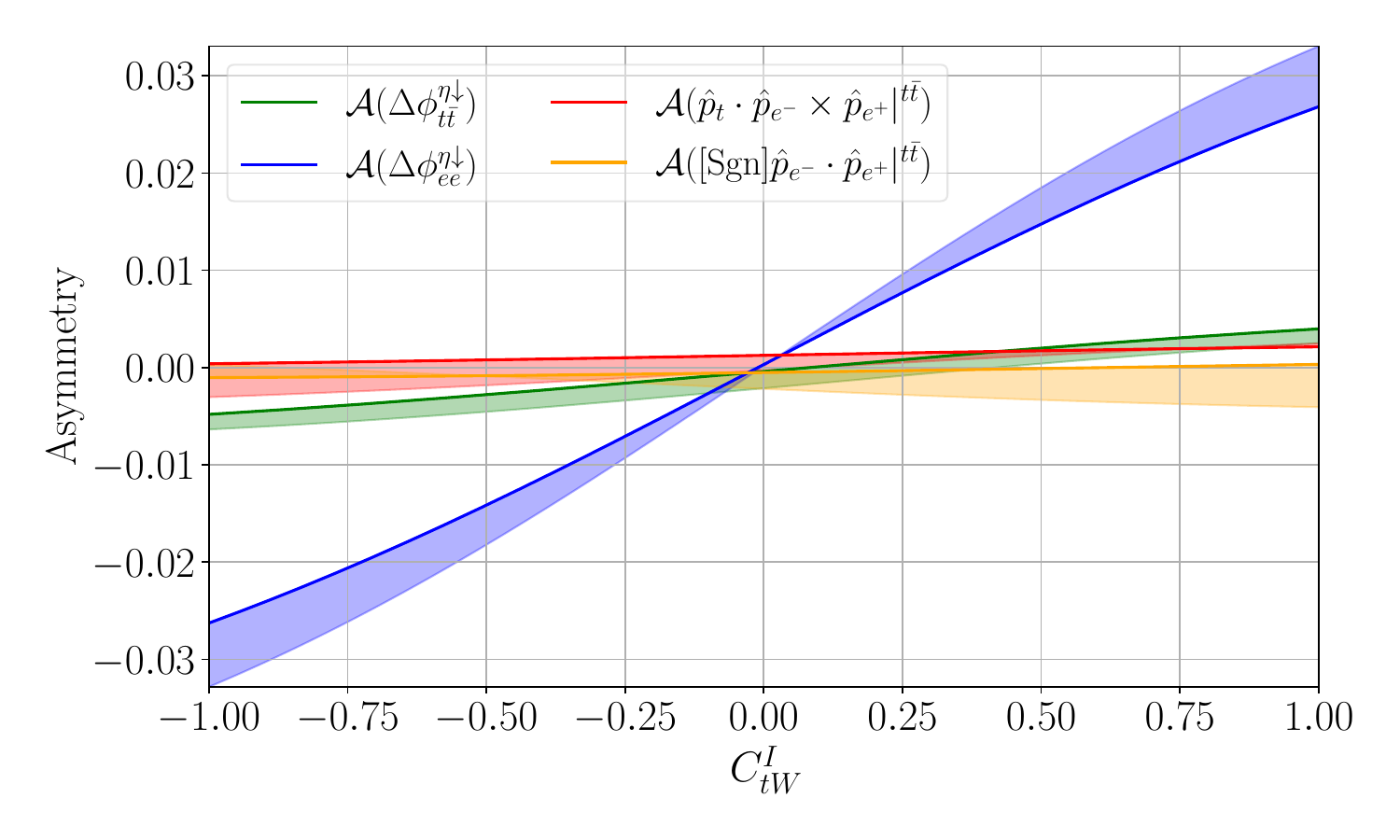}
\caption{}
\label{fig:sub1}
\end{subfigure}%
  \caption{\label{fig:tth_asym}Asymmetries in SMEFT $t\bar{t}h$ production, constructed from the observables pseudorapidity-ordered $\Delta\phi_{tt}^{(\eta\downarrow)}$, pseudorapidity-ordered $\Delta\phi_{ee}^{\eta\downarrow}$, $\hat{p}_t\cdot\left(\hat{p}_{e^-} \times \hat{p}_{e^+}\right)|^{t\bar{t}}$ and ${\mathrm{sgn}[\hat{p}_t\cdot \left(\hat{p}_{e^-} \times \hat{p}_{e^+}\right)](\hat{p}_{e^-} \cdot \hat{p}_{{e^+}})|^{t\bar{t}}}$ (yellow, red, blue and green curves, respectively). The asymmetries as a function of the coefficient value are shown for the ($\mathcal{CP}$-odd part of the) operators $\hat{O}_{t\varphi},\,\hat{O}_{tG},\,O_{\varphi\widetilde{G}}$, and $\hat{O}_{tW}$ ((a) to (d)).  The shaded bands represent the estimation of the theoretical uncertainties obtained varying the renormalisation and factorisation scales by a factor 2 and $1/2$. The ranges of the WC in the plots are chosen to be within the values allowed by our prospects on LHC data.}
\end{figure}

The asymmetries in $thj$ production are shown in Fig.~\ref{fig:thj_asym}. The asymmetry most sensitive to $O_{\varphi\widetilde{W}}$ and $\hat{O}_{t\varphi}$ is the one built from the triple product of outgoing partons, with enhanced sensitivity to $O_{\varphi\widetilde{W}}$. Nevertheless, this same asymmetry becomes irrelevant for $\hat{O}_{tW}$, as the triple product distribution switches sign three times (as shown in Fig.~\ref{fig:thj_tripleprods} (c)) diminishing any contribution to the asymmetry. Indeed, for this case measuring the binned distribution would be more suitable. For the observables considered in this work the best sensitivity to the \cp-odd couplings of $\hat{O}_{tW}$ at linear order is obtained by the \cp-odd polarisation angular distributions ($\cos{\theta^y_e}$ and $\cos{\theta^s_e}$). Among the two \cp-odd polarisation angular distributions, the one constructed from $\cos{\theta^s_e}$ exhibits stronger effects from $O_{\varphi\widetilde{W}}$ and $\hat{O}_{t\varphi}$. In contrast, the asymmetry extracted from $\cos{\theta^y_e}$ shows only mild effects from these operators. Finally, the asymmetry built from the \cp-even polarisation angular distribution ($\cos{\theta^j_e}$) shows, as expected, only quadratic dependencies, with particular sensitivity to $\hat{O}_{tW}$.

The corresponding asymmetries in $t\bar{t}h$ production are shown in Fig.~\ref{fig:tth_asym}. In this case the particle-level pseudorapidity-ordered azimuthal separation, $\Delta\phi_{ee}^{\eta\downarrow}$, clearly outperforms its parton-level counterpart, $\Delta\phi_{tt}^{\eta\downarrow}$. It shows significant sensitivity to $\hat{O}_{t\varphi}$ and $\hat{O}_{tW}$ (unlike the parton-level case) and provides similar sensitivity for $\hat{O}_{tG}$ and $O_{\varphi\widetilde{G}}$ as the parton-level case.  However, the most sensitive asymmetry, except for $\hat{O}_{tW}$, is the one constructed from the triple product of the final charged leptons and the top-quark 3-momentum, $\mbox{$\hat{p}_t \cdot \left(\hat{p}_{e^-} \times \hat{p}_{e^+}\right)\big|^{t\bar{t}}$}$, which requires full reconstruction of the top-quark momentum. A relevant intermediate approach is the asymmetry built from the cosine of the final charged leptons and the sign of this triple product, $\text{sgn}[\hat{p}_t\cdot \left(\hat{p}_{e^-} \times \hat{p}_{e^+}\right)](\hat{p}_{e^-} \cdot \hat{p}_{e^+})|^{t\bar{t}}$. In this case, only the reconstruction of the sign of the triple product is required. It outperforms the sensitivity of $\Delta\phi_{ee}^{\eta\downarrow}$ for $O_{\varphi\widetilde{G}}$ and demonstrates comparable sensitivity for $\hat{O}_{tG}$, $\hat{O}_{tW}$ and, $\hat{O}_{t\varphi}$.

Overall, the asymmetries for $thj$ can range from 5\% to 10\%, while those for $t\bar{t}h$ typically fall between 3\% and 6\%, for WC values within the ranges currently allowed by LHC data, as discussed in the next section. Although these values are non-negligible, the experimental uncertainties in these channels are often considerable, potentially compromising their impact on the global fit. This impact will be explored in detail in the next section.

\section{Constraints on Wilson coefficients}
\label{sec:constWC}
\subsection{Experimental observables}
In order to establish the experimental sensitivity to the relevant Wilson coefficients, we will perform a toy fit, using current data as well as future projections for the HL-LHC. 
The observables considered, with the appropriate binning, can be found in Tab.~\ref{tab:binning}. Besides the observables constructed to be sensitive to \cp-odd operators at linear order, there are other differential measurements that can be useful to constrain our subset of operators. In particular, in this work  we have also considered the angular separations between outgoing partons, defined as $\Delta R = \sqrt{(\Delta\phi)^2 + (\Delta \eta)^2}$,\footnote{Coordinates are defined as follows: the $z$-axis is aligned with the proton beam; $\phi$ is the azimuthal polar angle, transverse to the $z$-axis; $\eta=-\ln{\tan{\theta/2}}$, where $\theta$ is measured relative to the positive $z$-axis.} and some invariant mass distributions. 
In order to determine their constraining power, we have estimated their experimental uncertainties at the LHC Run 3 and in the future phase of the high luminosity LHC (HL-LHC).

\begin{table}[ht]
\centering
\begin{scriptsize}
\begin{tabular}{ccclccc}
\cline{2-3} \cline{6-7}
&\multicolumn{1}{|c|}{$thj$ obs.} & \multicolumn{1}{c|}{binning} &  & & \multicolumn{1}{|c|}{$t\bar{t}h$ obs.} & \multicolumn{1}{c|}{binning} \\ \cline{1-3} \cline{5-7} 
\multicolumn{1}{|c}{\multirow{4}{*}{\cp-odd}} &\multicolumn{1}{|c|}{$\hat{z}\cdot\hat{p}_t\times\hat{p}_j|_h$} & \multicolumn{1}{c|}{$[-1, -0.1, 0, 0.1, 1]$} & & \multicolumn{1}{|c}{\multirow{4}{*}{\cp-odd}} &  \multicolumn{1}{|c|}{$\Delta\phi_{t\bar{t}}^{\eta\downarrow}$} & \multicolumn{1}{c|}{} \\ \cline{2-3} \cline{6-6} 
\multicolumn{1}{|c}{}&\multicolumn{1}{|c|}{$\hat{p}_h\cdot\hat{p}_t\times\hat{p}_j$} & \multicolumn{1}{c|}{} & & \multicolumn{1}{|c}{} & \multicolumn{1}{|c|}{$\Delta\phi_{ee}^{\eta\downarrow}$} & \multicolumn{1}{c|}{Asymmetry} \\ \cline{2-2} \cline{6-6}
\multicolumn{1}{|c}{}&\multicolumn{1}{|c|}{$\cos\theta_e^y$} & \multicolumn{1}{c|}{Asymmetry} & & \multicolumn{1}{|c}{} & \multicolumn{1}{|c|}{$\hat{p}_t\cdot\hat{p}_{e^-}\times\hat{p}_{e^+}|^{t\bar{t}}$} & \multicolumn{1}{c|}{about 0}   \\ \cline{2-2} \cline{6-6}
\multicolumn{1}{|c}{} &\multicolumn{1}{|c|}{$\cos\theta_e^s$} & \multicolumn{1}{c|}{about 0} & & \multicolumn{1}{|c}{} &\multicolumn{1}{|c|}{[Sgn]$\hat{p}_{e^-}\cdot\hat{p}_{e^+}|^{t\bar{t}}$} & \multicolumn{1}{c|}{}\\ \cline{1-2} \cline{5-7}
\multicolumn{1}{|c}{}&\multicolumn{1}{|c|}{$\cos\theta_e^j$} & \multicolumn{1}{c|}{} & & \multicolumn{1}{|c}{\multirow{2}{*}{\cp-even}} & \multicolumn{1}{|c|}{$M(ht\bar{t})$ [GeV]} & \multicolumn{1}{c|}{$[450, 655, 860, 1270, 2500]$ } \\ \cline{2-3} \cline{6-7} 
\multicolumn{1}{|c}{\cp-even}&\multicolumn{1}{|c|}{$\Delta R(ht)$} & \multicolumn{1}{c|}{$[0, \pi, 8]$} & & \multicolumn{1}{|c}{} & \multicolumn{1}{|c|}{$b_4^{lab}$} & \multicolumn{1}{c|}{$[-1, -0.5, 0, 0.5, 1]$}  \\ \cline{2-3} \cline{5-7} 
\multicolumn{1}{|c}{}&\multicolumn{1}{|c|}{$M(ht)$ [GeV] } & \multicolumn{1}{c|}{$[200, 340, 424, 620, 1600]$} &  & & & \\ \cline{1-3} 
\end{tabular}
\caption{\label{tab:binning} Binning for the differential distributions of the considered observables. $\mathcal{CP}$-odd observables and $\cos\theta_e^j$ are interpreted as asymmetries, with the exception of the $thj$ observable $\hat{z}\cdot\hat{p}_t\times\hat{p}_j|^h$.}
\end{scriptsize}
\end{table}

The ATLAS and CMS collaborations of the LHC have published measurements of the $t\bar{t}h$ cross section \cite{ATLAS:2020ior, ATLAS:2021qou,CMS:2021kom,CMS:2021nnc,CMS:2023vtj} and the $\mathcal{CP}$ properties of the top-Yukawa coupling using $t\bar{t}h$ and $th$ events \cite{ATLAS:2020ior,CMS:2021nnc, ATLAS:2023cbt, CMS:2023vtj}.
Despite the lack of a direct observation of single top-Higgs associated production to date, upper bounds on the total rate are placed by both collaborations\footnote{Both collaborations include the $thq$ and $thW$ processes in their searches for single top-Higgs associated production.} \cite{CMS:2018jeh, ATLAS:2020ior, CMS:2021kom}. In order to estimate the expected statistical uncertainty at the LHC for our binned distributions, we have assumed the same acceptance and efficiency as the one obtained by the ATLAS collaboration for the inclusive cross section shown in Fig. 2 of Ref.~\cite{ATLAS:2022vkf}. In particular we take as reference
$$
\sigma_{t\bar{t}h}=369.6\pm86.0_{\rm{stat.}}\pm84.4_{\rm{syst.}} \,\rm{fb},\qquad\quad \sigma_{thj}=560.6\pm272.3_{\rm{stat.}}\pm201.7_{\rm{syst.}} \,\rm{fb}.
$$
Using these numbers together with the quoted SM predictions ($\sigma_{t\bar{t}h}=499.8\,\rm{fb}$ and $\sigma_{thj}=84.8\,\rm{fb}$)
we find that 480 and 8.2 events per million of $t\bar{t}h$ and $thj$ are properly reconstructed, respectively. For the systematic uncertainty we have simply assumed the same relative uncertainty as in the inclusive cross section. The LHC limits that we present are our prospects after full Run 3, so a total integrated luminosity of 300 fb$^{-1}$ has been assumed and, therefore, the statistical uncertainty has been scaled accordingly, leaving the systematic uncertainties at their current value. Finally, for the prospects of the HL-LHC we have assumed that, besides the appropriate scaling of the statistical uncertainty with the luminosity, the systematic uncertainties will be reduced by a factor of two by the operation time of the HL-LHC.

Since  we are combining observables coming from the same processes, it is crucial to determine their correlations. To obtain the correlations among the statistical uncertainties, we have assumed SM distributions\footnote{We have also verified that, when considering the effects of the WC, the correlation matrices remain very similar as long as the WC stay within reasonable values. The differences in the entries of the correlation matrix are typically less than 2\%.} and that bins with different events are completely uncorrelated. Thus, any correlation would arise from the events shared among the different binned observables. The systematic uncertainties are assumed to be 50\% correlated among the different bins\footnote{We have also tested that by setting the correlation to 0\% or 100\%. Similar results are obtained with a different of at most 20\% on the limits. We have also checked that assuming a correlation as high as 50\% between the $t\bar{t}h$ and $thj$ observables would only worsen our limits by at most 25\%.}.

\begin{figure}[h!]
    \includegraphics[width=0.5\textwidth]{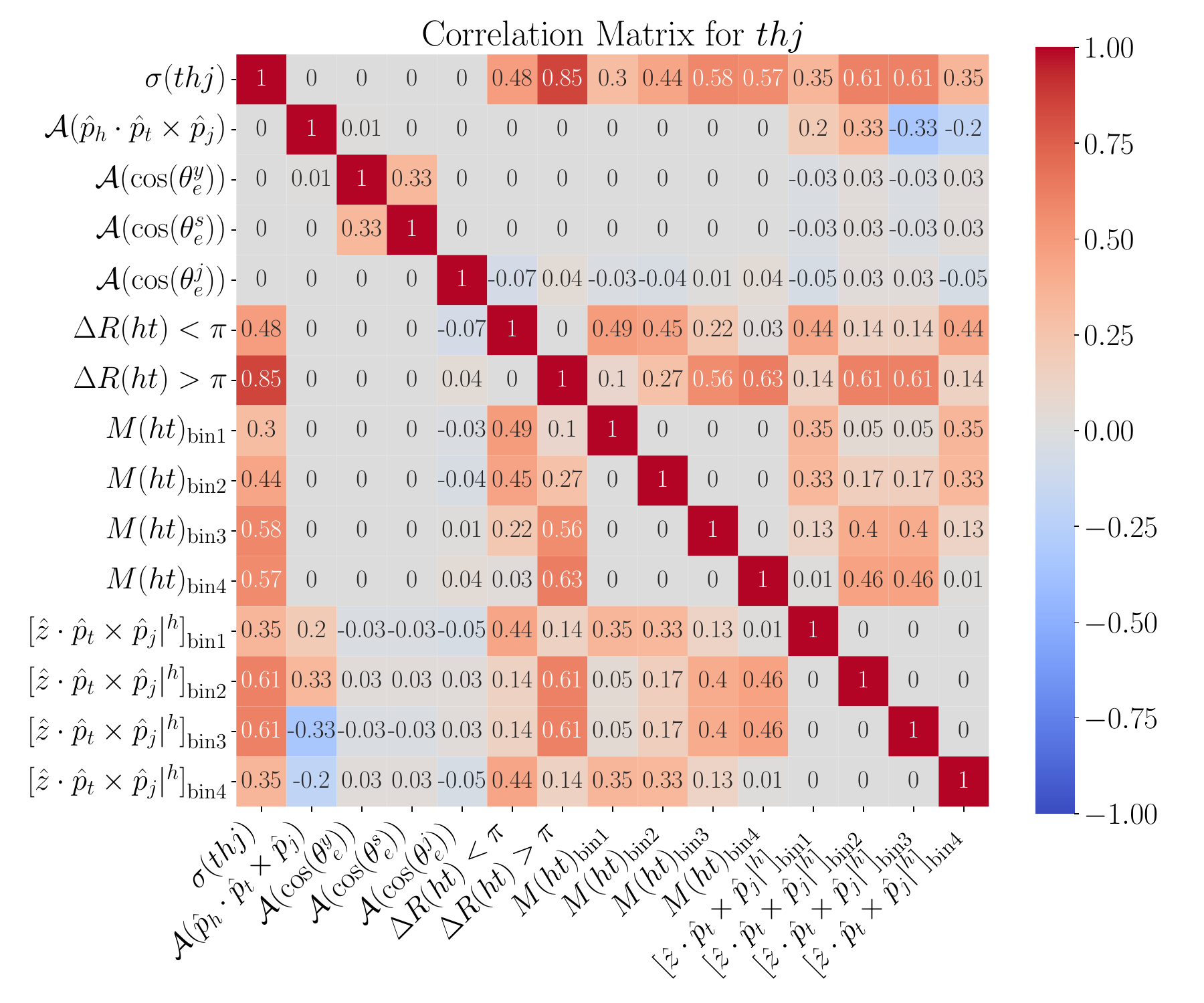} 
    \includegraphics[width=0.5\textwidth]{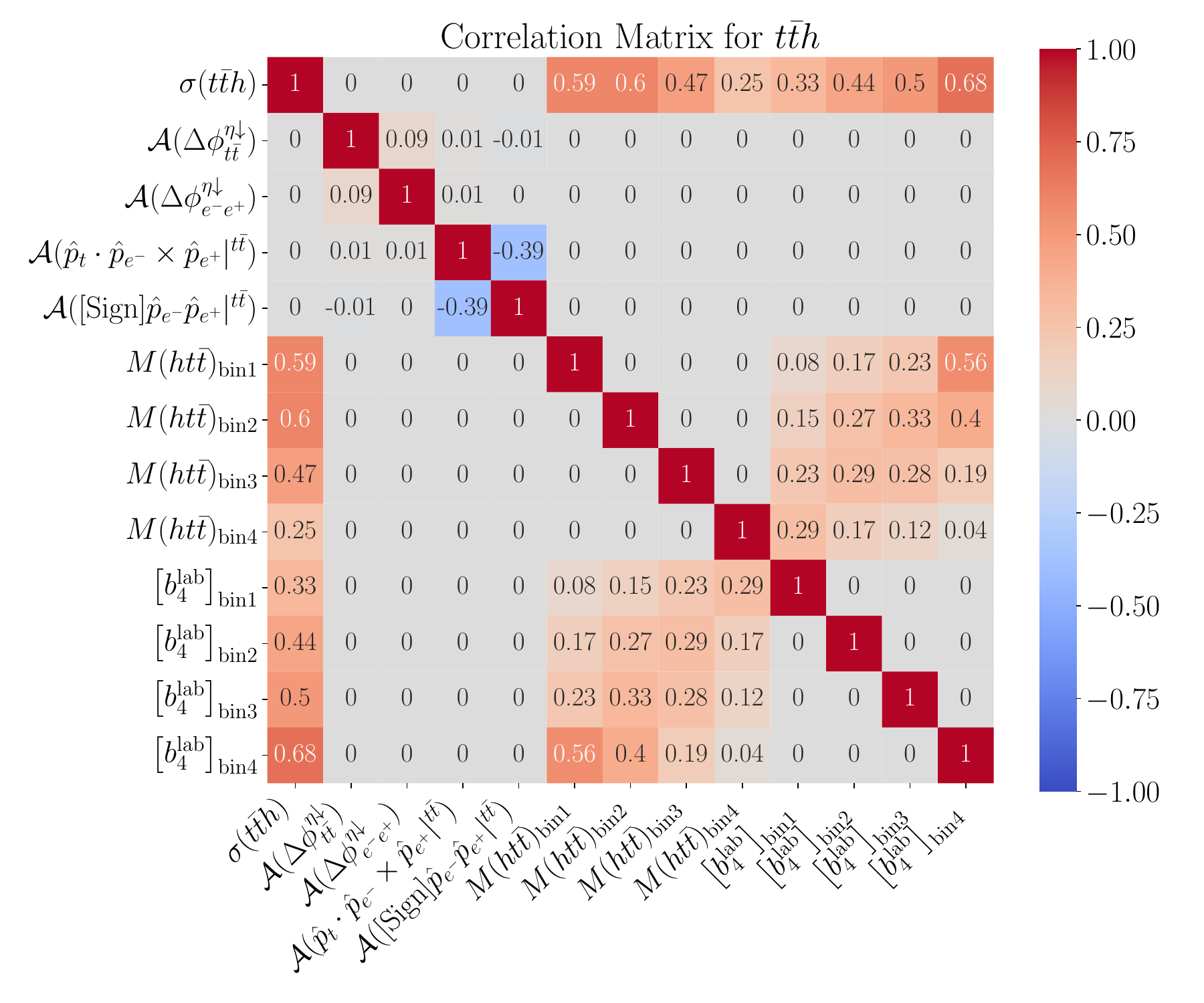} 
\caption{\label{fig:corr} 
Correlation matrices among the statistical uncertainties of the observables considered. In the left panel we show the correlation among the observables of the $thj$ channel and in the right panel the one for the observables of the $t\bar{t}h$ channel. The entries smaller than 5\textperthousand \, have been set to zero.
}
\end{figure}

The statistical correlation matrix among the observables considered is shown in Fig.~\ref{fig:corr}, in addition to the correlation with the total cross section. In general we observe small correlations of the asymmetries built from \cp-odd observables, with  several relevant exceptions.

In the case of $thj$ observables, the asymmetry of the triple product of the final particles in the lab frame is considerably correlated with the triple product of the beam axis, the (anti-)top quark and the jet in the Higgs rest frame. This correlation is expected since both quantities are defined relative to the plane formed by the final top-quark and jet. The other highly correlated asymmetries in this process are the ones defined with respect to the angles $\theta_e^y$ and $\theta_e^s$, which are highly correlated (33\%) since both measure projections of the electron direction of flight onto the plane normal to the jet momentum. Regarding the correlation among \cp-even quantities, it is worth noting the correlation among $\Delta R(ht)$ and $M(ht)$. In this case, low and high invariant masses are preferentially associated to angular separations smaller and larger than $\pi$, respectively. 

Concerning the $t\bar{t}h$ observables, we see a strong anti-correlation of the asymmetry of $\hat{p}_t\cdot \left(\hat{p}_{e^-} \times \hat{p}_{e^+}\right)|^{t\bar{t}}$ and $\text{sgn}[\hat{p}_t\cdot \left(\hat{p}_{e^-} \times \hat{p}_{e^+}\right)](\hat{p}_{e^-} \cdot \hat{p}_{e^+})|^{t\bar{t}}$, which is expected given that the former is proportional to the sine of the angle formed between the final charged leptons and the latter is proportional to the cosine of that same angle. With respect to the \cp-even observables, the lower bins of the invariant mass, $M(ht\bar{t})$, are more correlated with the higher bins of $b_4^{lab}$.

\subsection{Expected bounds at the LHC and beyond}
\label{sec:fit}

\begin{figure}[h!]
    \includegraphics[width=1.\textwidth]{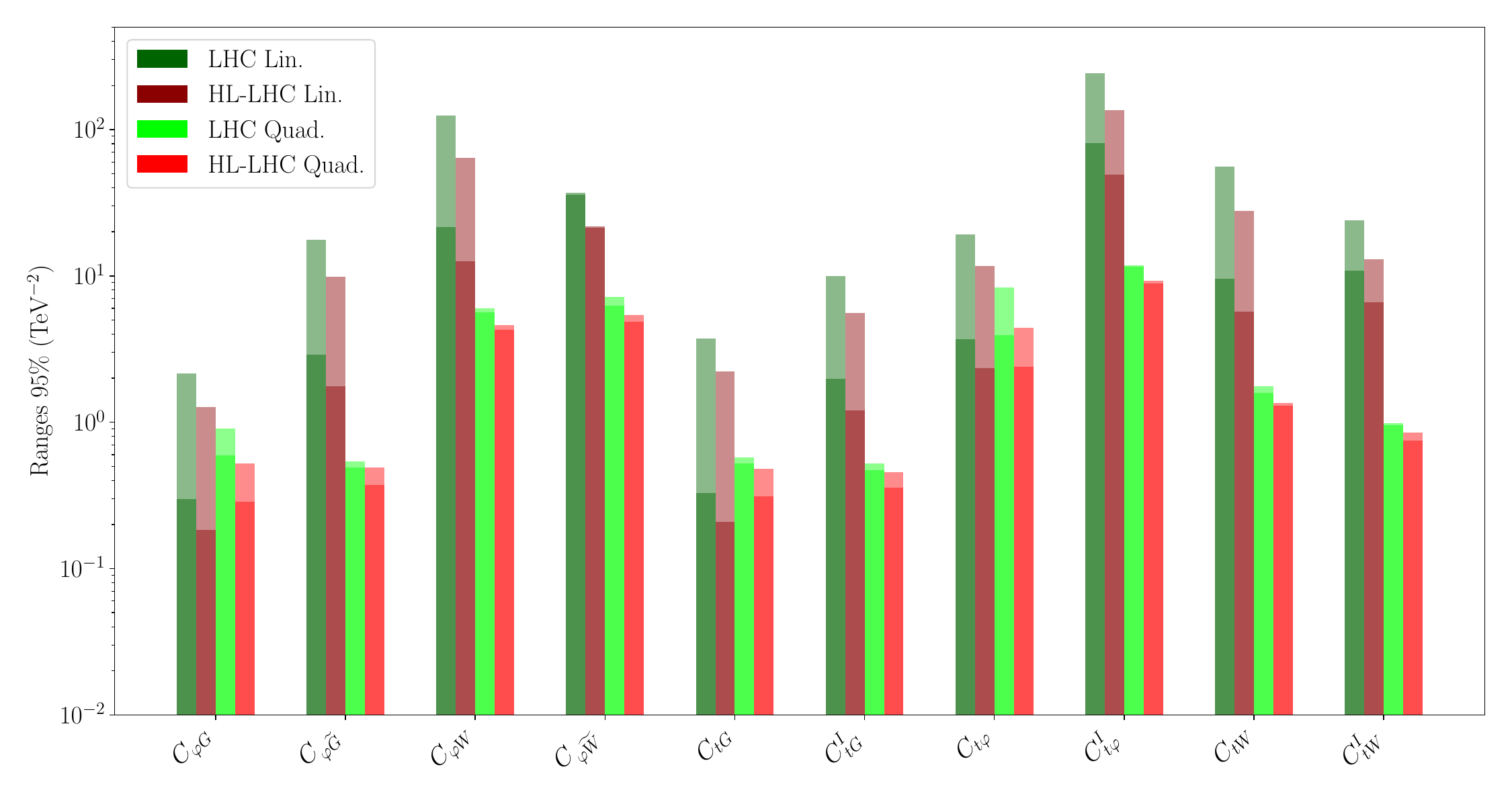}  
\caption{\label{fig:fit_comparison} 
Expected ranges at a 95\% probability for the WC considered in this work. The shadowed (solid) bars represent the marginalised (individual) limits. In green we show the limits expected for the LHC Run 3 and in red those for the HL-LHC. 
}
\end{figure}

Using the experimental information presented in the previous section, summarised in Tab.~\ref{tab:binning}, we perform a fit of the 10 degrees of freedom of our analysis:
$$
\{C_{\varphi G}, C_{\varphi \widetilde{G}}, C_{\varphi W}, C_{\varphi \widetilde{W}},\,\, C_{t\varphi},\,\, C^I_{t\varphi},\,\, C_{tG},\,\, C^I_{tG},\,\, C_{tW},\,\, C^I_{tW}\}.
$$
The fits are performed using the open source code \texttt{HEPfit} \cite{DeBlas:2019ehy} to which we have added the parametrisations of the observables considered in terms of the WC at linear and quadratic order. 
This code has previously been widely used to perform fits on the SM \cite{deBlas:2021wap}, specific SM extensions \cite{Coutinho:2024zyp,Karan:2023kyj,Eberhardt:2021ebh} and the SMEFT \cite{Durieux:2019rbz,Miralles:2021dyw,Cornet-Gomez:2025jot}. Given that  we are considering observables that have not been measured yet, we have set the central values to the SM predictions and estimated their uncertainties as described in the previous section.

In Fig.~\ref{fig:fit_comparison} we show the expected constraints on the subset of WC considered from a linear and quadratic fit at the LHC Run 3 and at the HL-LHC. The shadowed bars represent the marginalised limits for each WC obtained from the global fit, where all the coefficients have been varied at the same time. The solid bars, however, represent the individual limits obtained by performing a fit of one coefficient at a time. The difference between individual and marginalised fits is greater in the linear fit, where strong correlations (and cancellations) occur. There are some cases in which the individual linear fits provide better constraints than the quadratic one due to the appearance of several modes in the quadratic case but, in general, the quadratic fit provides much better constraints than the linear case.

\begin{figure}[h!]
    \includegraphics[width=1.\textwidth]{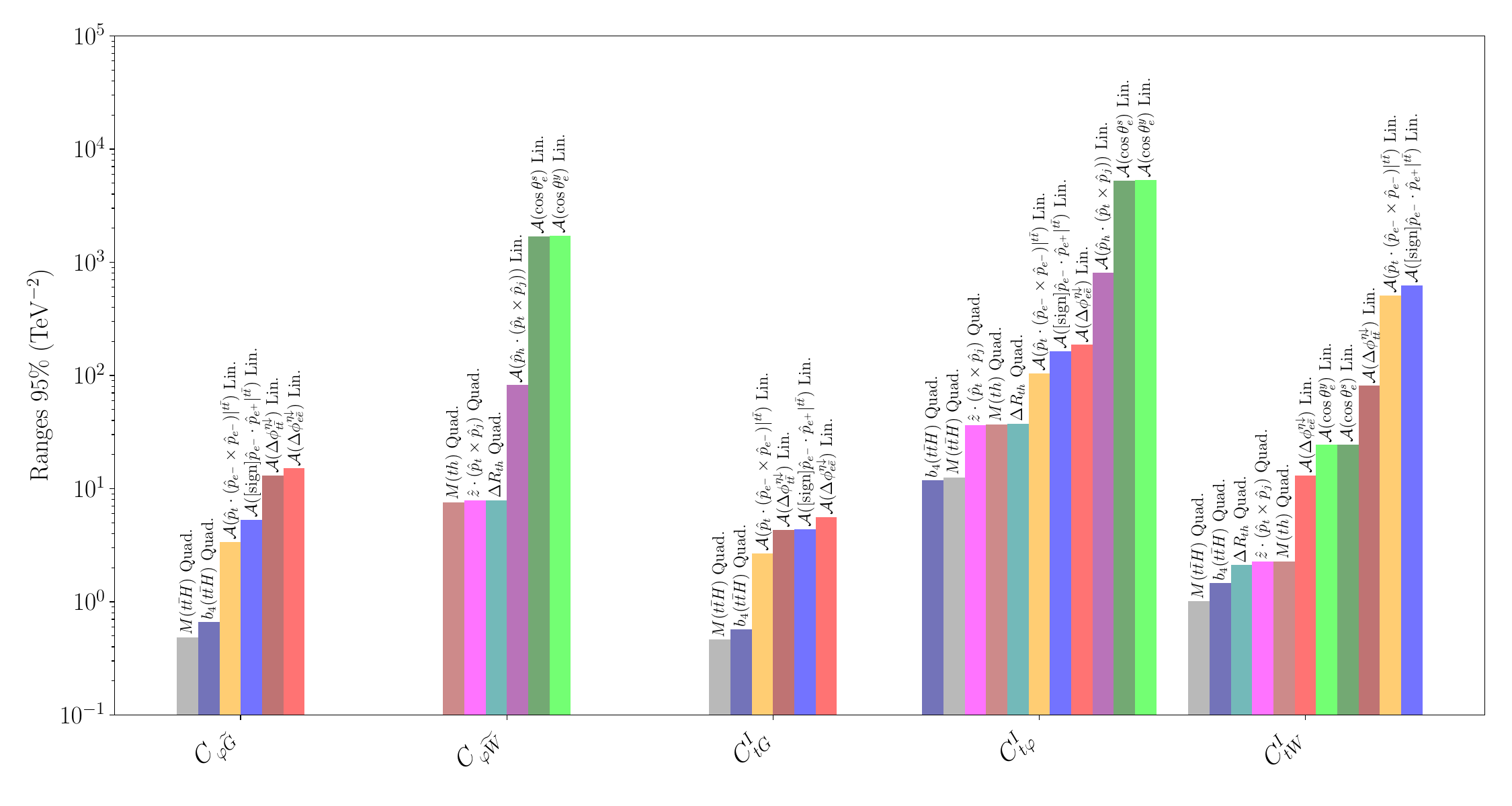}  
\caption{\label{fig:fit_sensitivity} 
Sensitivity of different observables for the \cp-even WC represented in terms of the allowed ranges at 95\%  probability, fitting one operator at a time with only one observable.  
}
\end{figure}

Indeed, we observe a huge dependence of the fit results on the quadratic terms, even in the HL-LHC. This shows that the precision expected on measuring these observables at the HL-LHC does not seem to be enough to make the proposed asymmetries  have a relevant impact on the fit, given that the asymmetries only enter at linear order in the \cp-odd couplings.  Therefore, even though the asymmetries are essential to constrain the  \cp-odd couplings at linear order, the effect of the linear terms is sub-leading compared to higher-order effects considering the precision that we will reach in the near/mid future. In general, one should consider the validity of the EFT when the quadratic pieces dominate, given that those terms are generally expected to be much more suppressed than the linear ones by the NP scale. In the case of the \cp-odd couplings, however, the suppression of the linear terms in the \cp-even observables is well justified due to the absence of \cp\,violation in the SM, making it natural for the quadratics to dominate. The difference between the individual constraints of the linear and quadratic fit of the \cp-even coupling of $O_{t\varphi}$ (the operator on which we focus within this work) is negligible, showing a well perturbative behaviour in this case. This is not the case for the other WC but there are other observables sensitive to those that could be added to specifically constrain them \cite{Bhardwaj:2021ujv,Barrue:2023ysk,Rossia:2024rfo,Thomas:2024dwd}. 
The aim of this work is not providing a global analysis of the \cp-odd couplings of the full SMEFT but to study the possible constraints on the effective top quark Yukawa, being also aware of the possible effects of other operators. 
It is worth noting that the dominance of quadratic terms in our global fit of dimension-six operators suggests that certain dimension-eight operators, whose leading contribution is suppressed by the NP scale at the same order as quadratic dimension-six terms, could influence the global analysis. This may be particularly relevant for the \cp-odd couplings, which are primarily constrained by \cp-even observables and contribute to these observables at the same order as the leading dimension-eight contributions, which were not considered in this analysis. Therefore, some caution is warranted when considering these limits.

\begin{figure}[b!]
\centering
    \includegraphics[width=0.7\textwidth]{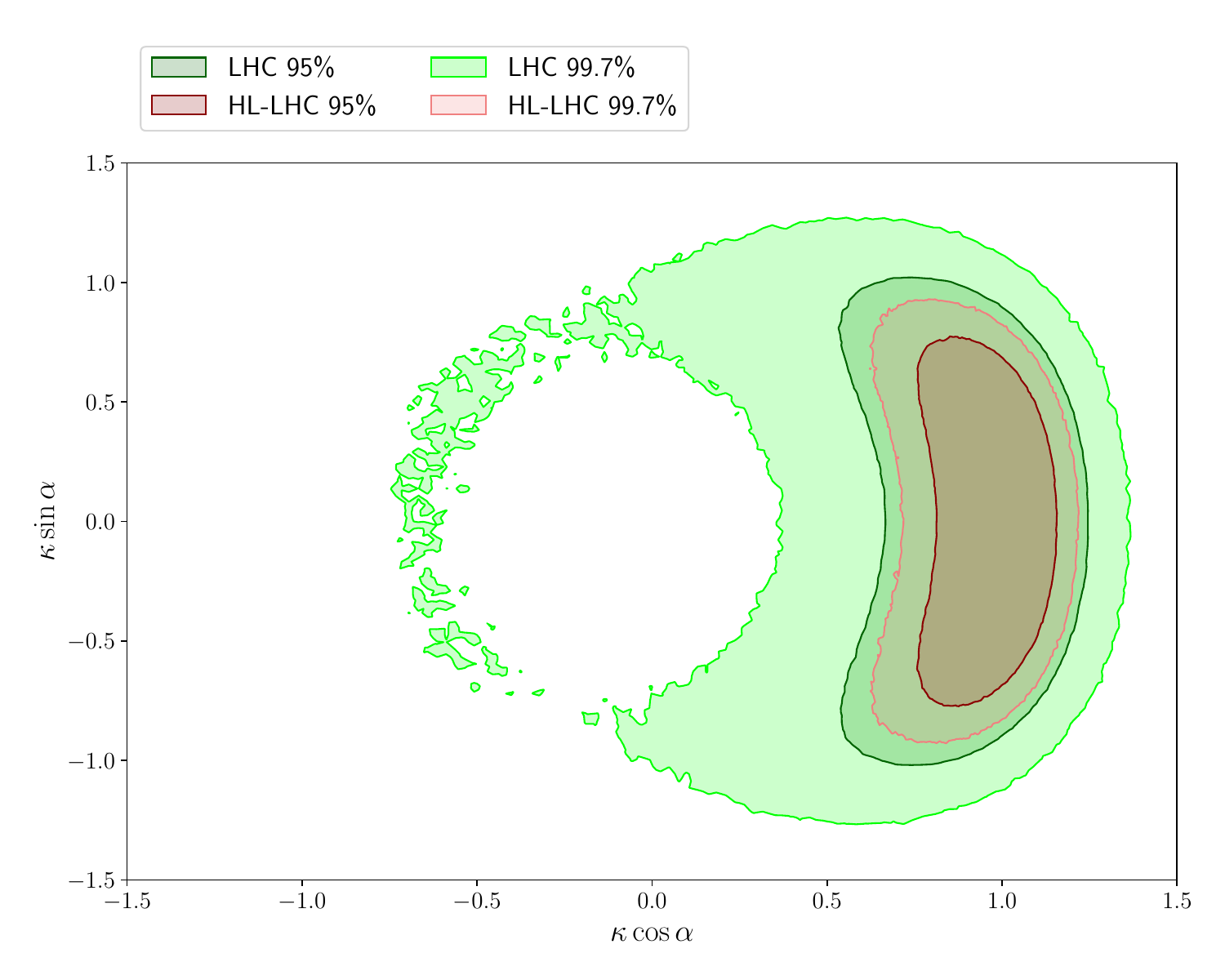}  
\caption{\label{fig:fit_yukawa} 
Expected ranges of the effective Yukawa couplings at the LHC Run 3 and at the HL-LHC.
}
\end{figure}

The breakdown of the sensitivity to the \cp-odd couplings for the observables of Tab.~\ref{tab:binning} is shown in Fig.~\ref{fig:fit_sensitivity}. In this figure we can see how the observables in $t\bar{t}h$ dominate the constraints on  $C_{t\varphi}^I$ thanks to the higher precision of this measurement. In general, the asymmetries provide limits around an order of magnitude worse than the \cp-even observables. Although $b_4$ is the observable most sensitive to $C_{t\varphi}^I$, the figure also shows that the differential mass distribution provides similar sensitivity. 

As mentioned in Sec.~\ref{sec:setup}, we can also reinterpret the constraints in terms of an effective Yukawa coupling which can be related to the WC $C_{t\varphi}^{(I)}$, as shown in Eq.~\eqref{eq:eff_yuk_lag}. 
In Fig.~\ref{fig:fit_yukawa} we show the results of a fit including only the effective Yukawa couplings. The constraints at 95\% probability are given  in Tab.~\ref{tab:yuk_values}. The results are compatible with those of a recent ATLAS study \cite{ATLAS:2023cbt}. We observe that the limits on the real part of the Yukawa coupling could be improved from LHC Run 3 to HL-LHC by a factor 2 and the imaginary part by a factor 1.5. 


\begin{table}[h]
    \centering
    \begin{tabular}{|c|c|c|c|}
        \hline
        & \textbf{ LHC (Run 3)} & \textbf{HL-LHC} \\ \hline
        $\kappa\cos\alpha$ & [0.61, 1.20] & [0.80, 1.12] \\ \hline
        $\kappa\sin\alpha$ & [-0.84, 0.84] & [-0.64, 0.64] \\ \hline
    \end{tabular}
    \caption{Predicted ranges within 95\% probability for $\kappa\cos\alpha$ and $\kappa\sin\alpha$ at LHC Run 3 and at HL-LHC from the fit including only the Yukawa couplings.}
    \label{tab:yuk_values}
\end{table}

\begin{figure}[h!]
    \includegraphics[width=0.5\textwidth]{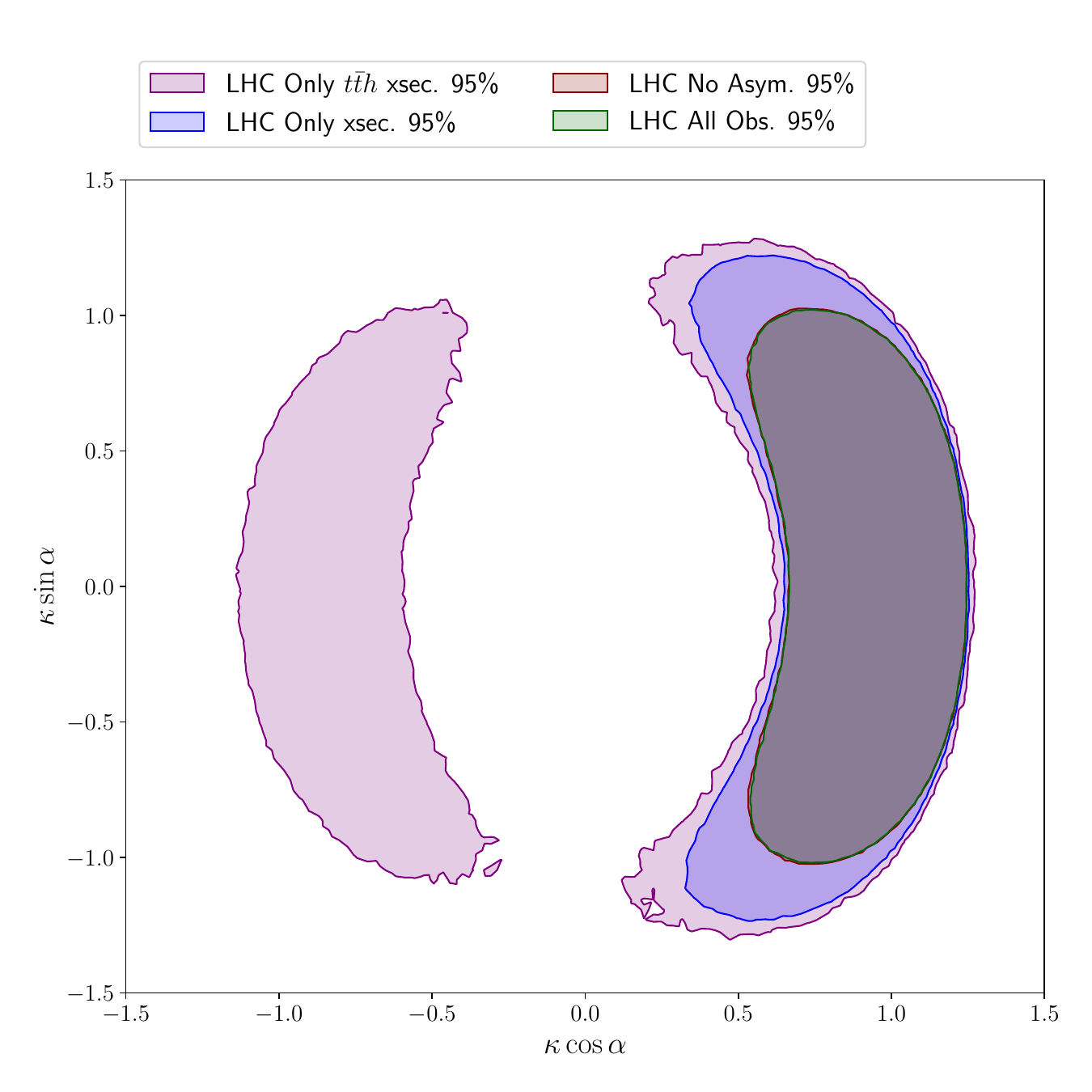} 
    \includegraphics[width=0.5\textwidth]{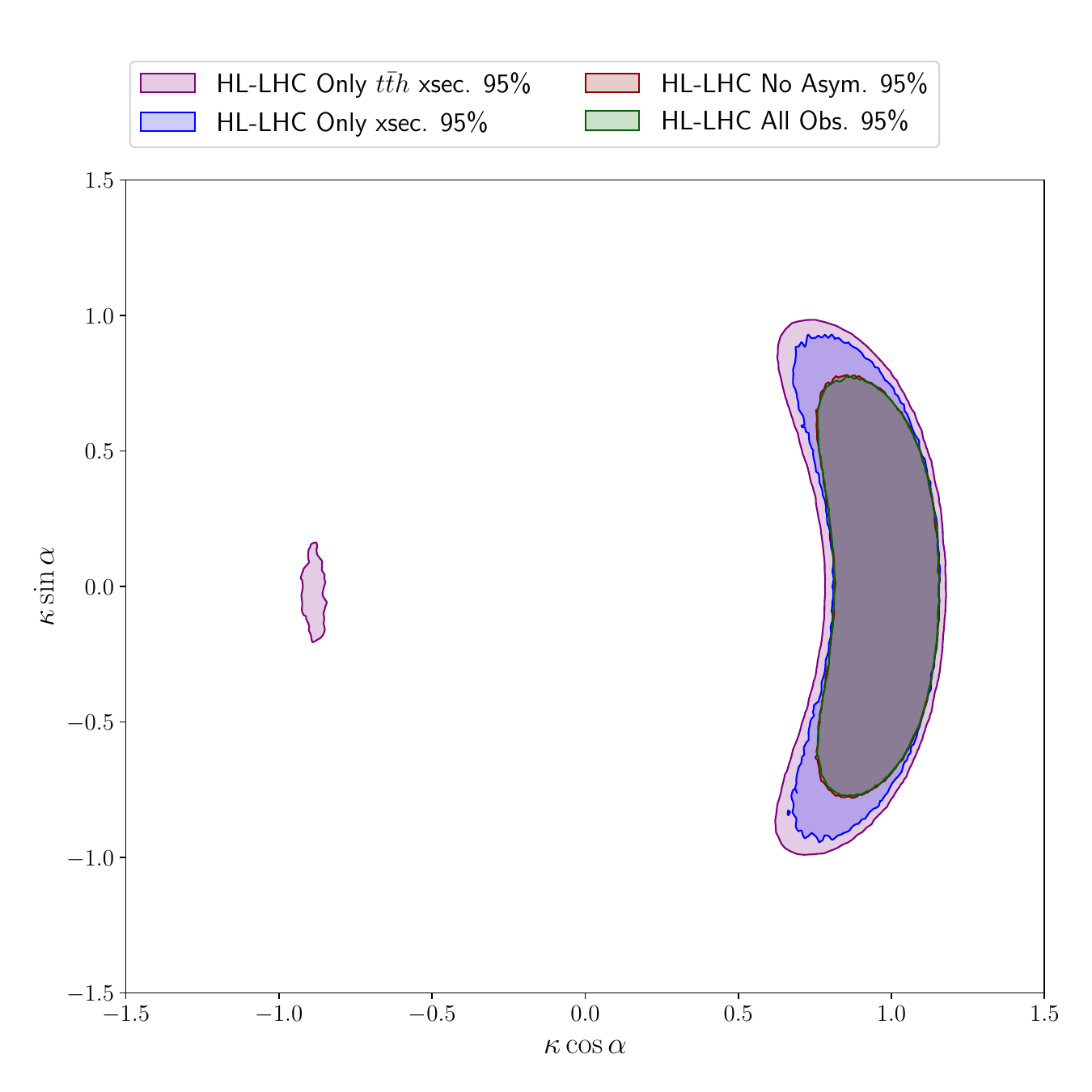} 
\caption{\label{fig:fit_yukawa_channels}
Constraining effect of the different sets of observables on the effective Yukawa couplings for the LHC Run 3 (left) and the HL-LHC (right). The label ``Only $t\bar{t}h$ xsec.'' refers to a fit in which only the differential cross section of $t\bar{t}h$ with respect to the invariant mass of $t\bar{t}h$ is added. The label ``Only xsec.'' includes also the invariant $thj$ differential cross section with respect to the invariant mass of the $th$ system. The label ``No Asym.'' removes the asymmetries of Tab.~\ref{tab:binning} from the fit. Finally, the label ``All Obs.'' refers to a fit including all the observables defined in Tab.~\ref{tab:binning}. 
}
\end{figure}

In Fig.~\ref{fig:fit_yukawa_channels} we can see the impact of each set of observables on the $\kappa\sin\alpha-\kappa\cos\alpha$ plane for the LHC Run 3 and the HL-LHC. The first remarkable feature is that, even though the differential cross section of $t\bar{t}h$ in $M(h t\bar{t})$ is one of the most sensitive observables, it is crucial to add the measurement of the differential cross section of $thj$ in $M(ht)$ in order to remove the degeneracy on the $\kappa\sin\alpha-\kappa\cos\alpha$ plane at 95\% probability. The inclusion of other quadratic observables like $b_4^{\rm{lab}}$, and differential distributions with respect to $\Delta R(ht)$ and $\hat{z} \cdot (\hat{p}_t \times \hat{p}_j)$ as shown in Tab.~\ref{tab:binning}, also help to shrink the allowed region. The asymmetries, however, do not have an effect once compared with the effect of the other observables. A similar behaviour is observed for the HL-LHC although the degeneracy on the $\kappa\sin\alpha-\kappa\cos\alpha$ is less pronounced than in the LHC case.

\section{Conclusions}
\label{sec:con}

In this work, we have studied the \cp-violating couplings in the top-Higgs sector using LHC processes. Specifically, we focus on two key processes: top-quark pair production in association with a Higgs boson and single top-Higgs associated production. To conduct a model-independent analysis, we parameterised the NP effects using SMEFT operators. Given our focus on these specific LHC processes, we restricted our analysis to the 10 Wilson coefficients relevant to these interactions: $\{C_{\varphi G},$ $C_{\varphi \widetilde{G}},$ $C_{\varphi W},$ $C_{\varphi \widetilde{W}},$ $C_{t\varphi},$ $C^I_{t\varphi},$ $C_{tG},$ $C^I_{tG},$ $C_{tW},$ $C^I_{tW}\}$, with particular emphasis on the top-Yukawa couplings $\{C_{t\varphi},\, C^I_{t\varphi}\}$. 

To enhance sensitivity to \cp-violating operators, we proposed and examined several observables that have been or could eventually be measured by the experimental collaborations. In the case of single top Higgs associated production, we studied several triple products of the momenta of the final and/or initial partons which constitute \cp-odd observables by construction. Besides these observables we also examined observables sensitive to the top-quark polarisation. These observables are angular distributions of the top-quark decay products, in our case the final electron coming from the decay of the $W$ produced from the top-quark decay. In the top-pair production in association with a Higgs boson, we studied both parton- and particle-level observables. Besides the well-studied \cp-even observable $b_4$ we also proposed pure \cp-odd observables built from the azimuthal separation of the leptons as well as triple products of the final state particles.

We then projected the uncertainties associated with these observables at the final stage of the Run 3 of the LHC and in the HL-LHC, assuming integrated luminosities of 300 fb$^{-1}$ and 3000 fb$^{-1}$, respectively.
We have performed a global fit considering the linear and the linear plus quadratic terms of the SMEFT parametrisation. Given the small number of processes considered, we observe a significant difference between the marginalised and individual limits in the linear global fit due to cancellations among the contributions of the different operators. This is not the case for the quadratic fit, which also provides more stringent constraints, especially for the marginalised constraints. Indeed, we expect the difference between the marginalised and the individual limits to be reduced when adding other observables since in this work we have only considered the observables most relevant to constrain the top-Yukawa couplings from LHC data.

In the quadratic fit, the pure \cp-odd observables proposed will not be measured with enough precision at the HL-LHC to compete with other \cp-even observables. However, for the linear case these observables are essential to constrain the \cp-odd couplings, for which the \cp-even observables offer no sensitivity at linear order.

Finally, by reinterpreting the SMEFT limits in terms of an effective Yukawa coupling, we anticipate a factor of two improvement in the constraints on the imaginary Yukawa coupling at the HL-LHC compared to the current bounds established by ATLAS \cite{ATLAS:2023cbt}. In this analysis, the combination of $t\bar{t}h$ and $thj$ processes --- despite the limited precision expected for $thj$ --- is crucial to resolve degeneracies in the parameter space of the real and imaginary Yukawa couplings. Purely \cp-odd observables, however, are not relevant for this scenario and the constraints come mainly from quadratic terms in \cp-even observables.

Our work establishes a proof of concept for incorporating \cp-violating effects in the top-Higgs sector within broader global fits of the SMEFT framework. 
By integrating this analysis alongside constraints from other high-precision probes, such as flavor, electroweak, and Higgs observables, we can progress toward a more comprehensive and interconnected understanding of \cp\,violation in the SMEFT. Notably, the inclusion of low-energy observables, such as lepton EDMs, which were not studied here, could provide further insights for the top-Higgs sector. At present, most global fits within the SMEFT framework assume \cp~conservation in new physics sectors, an assumption that future analyses should reconsider and relax. Additionally, advancements in experimental techniques, including enhanced reconstruction methods and machine learning tools, could significantly improve the precision of the measurements beyond our initial conservative estimates. Such improvements are essential for isolating subtle \cp-violating signals from background effects, thereby advancing the robustness and reach of \cp\,violation studies in the high-luminosity era of the LHC.

\acknowledgments

V.M. would like to thank Giuseppe Ventura and Valentina Vecchio for valuable discussions during this work. The authors would also like to thank Simone Tentori for spotting a typo in the first version of this draft.
The work of V.M. and E.V. is supported by the European Research Council (ERC) under the European
Union’s Horizon 2020 research and innovation programme (Grant agreement No. 949451)
and a Royal Society University Research Fellowship through grant URF/R1/201553. The work of Y.P. and J.W.  is supported by the ERC under the European
Union’s Horizon 2020 research and innovation programme (Grant agreement No. 817719).

\appendix

\section{SMEFT operator basis}
\label{app:operators}

The SMEFT operators that we have used in our analysis are defined as
\begin{gather}
\nonumber
    O_{\varphi G} = \left(\pdp\right)G^{\mu\nu}_{\sss A} G_{\mu\nu}^{\sss A}, \qquad
    O_{\varphi \widetilde{G}} = \left(\pdp\right)\widetilde{G}^{\mu\nu}_{\sss A} G_{\mu\nu}^{\sss A}, \\
    O_{\varphi W} = \left(\pdp\right)W^{\mu\nu}_{\sss I} W_{\mu\nu}^{\sss I}, \qquad
    O_{\varphi \widetilde{W}} = \left(\pdp\right)\widetilde{W}^{\mu\nu}_{\sss I} W_{\mu\nu}^{\sss I}, \\
\nonumber
    \hat{O}_{t\varphi} = \left(\pdp\right)\bar{Q} t \tilde{\varphi}, \qquad
    \hat{O}_{tG} = g_{\sss S} \big(\bar{Q}\sigma^{\mu\nu} T_{\sss A} t\big) \tilde{\varphi} G^A_{\mu\nu}, 
    \qquad
    \hat{O}_{tW} = i\big(\bar{Q} \sigma^{\mu\nu} \tau_{\sss I} t\big) \tilde{\varphi} W^I_{\mu\nu},
\end{gather}
where $\varphi$ ($\widetilde{\varphi}$) represents the (charge-conjugate) SM Higgs doublet, $Q$ is the bottom- and top-quark left-handed doublet, $t$ is the right-handed top-quark, $G^{\mu\nu}_A$ and $W_{\mu\nu}^I$ stand for the $SU(3)_C$ and $SU(2)_L$ field strength tensors, and $T_A$ and $\tau_I$ are the generators of the $SU(3)_C$ and $SU(2)_L$ groups, respectively. 

The other operators mentioned in the text but not included in the analysis are

\begin{gather}
\nonumber
    O_{\varphi} = \left(\pdp\right)^3, \quad
    O_{\varphi\Box} = \left(\pdp\right)\Box(\pdp), \quad
    O_{\varphi D} = \left(\varphi^{\dagger} D^\mu \varphi\right)^*\left(\varphi^{\dagger} D_\mu \varphi\right), \\\nonumber
    O_G = f^{ABC} G^{A\nu}_{\ \ \mu} G^{B\rho}_{\ \ \nu} G^{C\mu}_{\ \ \rho}, \qquad\qquad
    O_W = \epsilon^{IJK} W^{I\nu}_{\ \ \mu} W^{J\rho}_{\ \ \nu} W^{K\mu}_{\ \ \rho}, \\
    O_{\widetilde{G}} = f^{ABC} \widetilde{G}^{A\nu}_{\ \ \mu} G^{B\rho}_{\ \ \nu} G^{C\mu}_{\ \ \rho}, \qquad\qquad
    O_{\widetilde{W}} = \epsilon^{IJK} \widetilde{W}^{I\nu}_{\ \ \mu} W^{J\rho}_{\ \ \nu} W^{K\mu}_{\ \ \rho}, \\\nonumber
    O_{\varphi B} = \left(\pdp\right) B^{\mu\nu} B_{\mu\nu}, \qquad\qquad
    O_{\varphi WB} = \left(\pdp\right) W^{\mu\nu}_{\sss I} B_{\mu\nu}, \\\nonumber
    O_{\varphi \widetilde{B}} = \left(\pdp\right) \widetilde{B}^{\mu\nu} B_{\mu\nu}, \qquad\qquad
    O_{\varphi \widetilde{W}B} = \left(\pdp\right) \widetilde{W}^{\mu\nu}_{\sss I} B_{\mu\nu},
\end{gather}
with $f^{ABC}$ and $\epsilon^{IJK}$ the structure constants of the $SU(3)_C$ and $SU(2)_L$ groups, respectively.

\section{Additional figures}
\label{app:extra_fig}

Here we present the dependence on the WC of some additional observables presented in the main text.

\begin{figure}[h!]
  \centering
  \begin{subfigure}[b]{0.33\textwidth}
    \includegraphics[width=1.\linewidth]{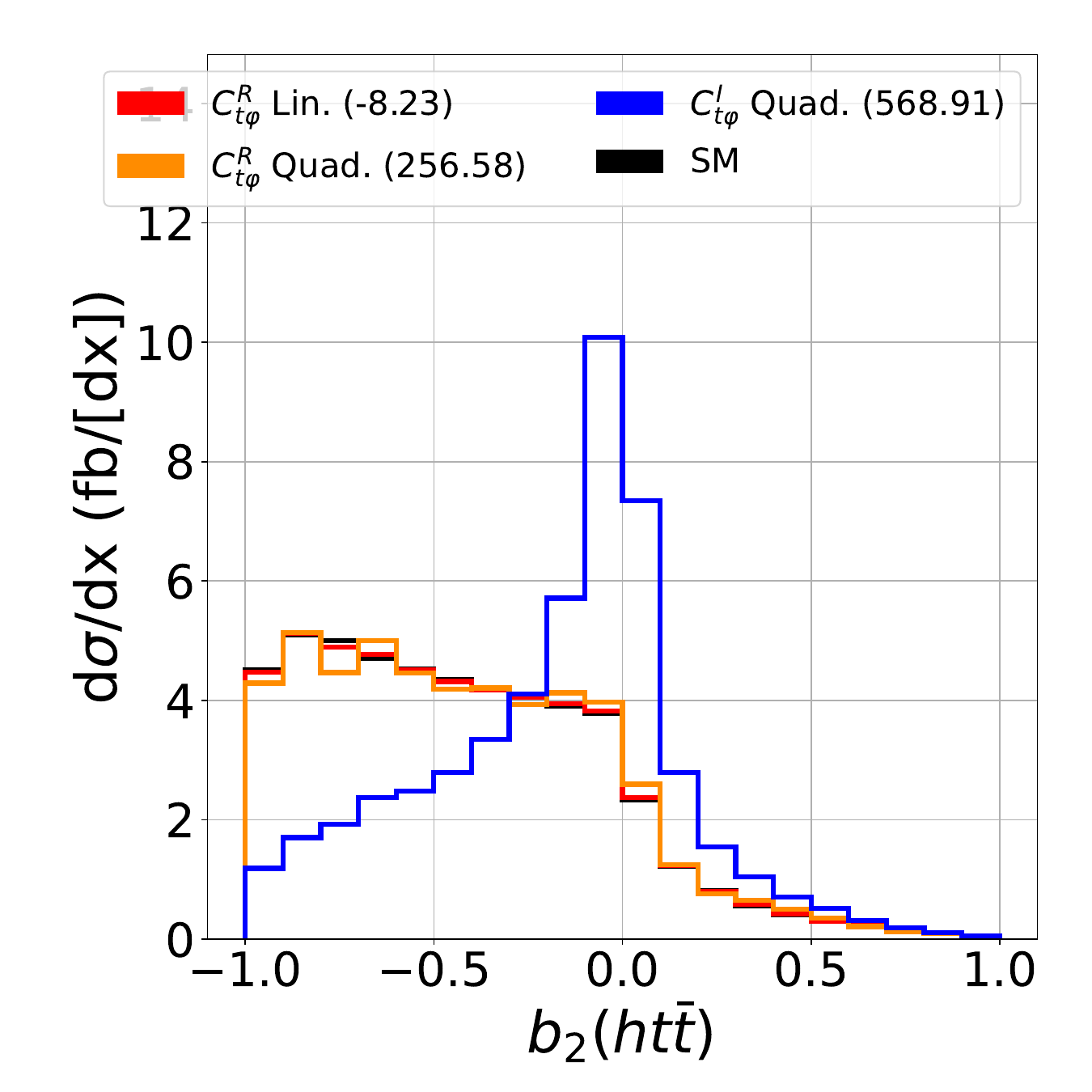}
    \caption{}
    \label{fig:sub1}
  \end{subfigure}%
  \begin{subfigure}[b]{0.33\textwidth}
    \includegraphics[width=1.\linewidth]{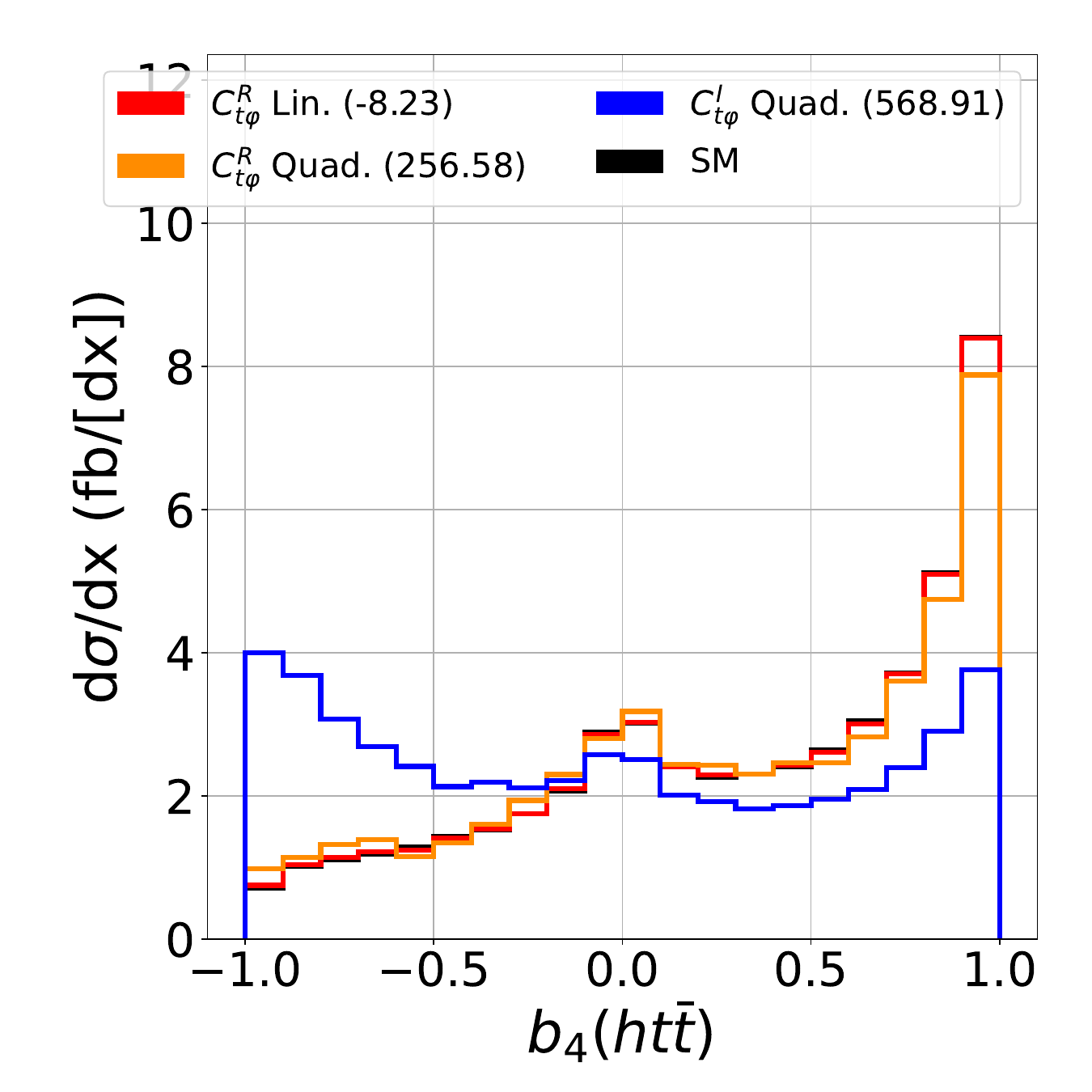}
    \caption{}
    \label{fig:sub1}
  \end{subfigure}
  \begin{subfigure}[b]{0.33\textwidth}
    \includegraphics[width=1.\linewidth]{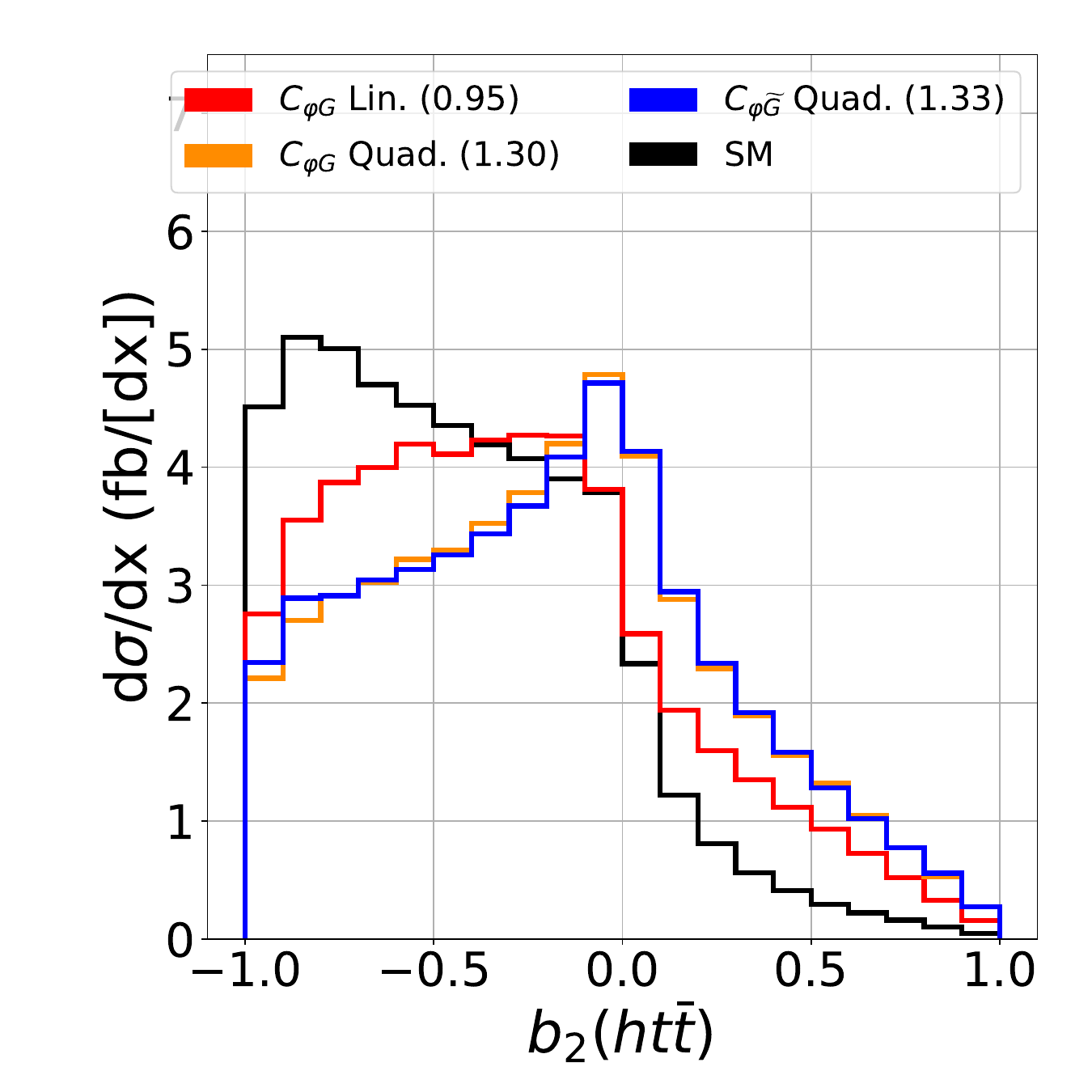}
    \caption{}
    \label{fig:sub2}
  \end{subfigure}%
  \begin{subfigure}[b]{0.33\textwidth}
    \includegraphics[width=1.\linewidth]{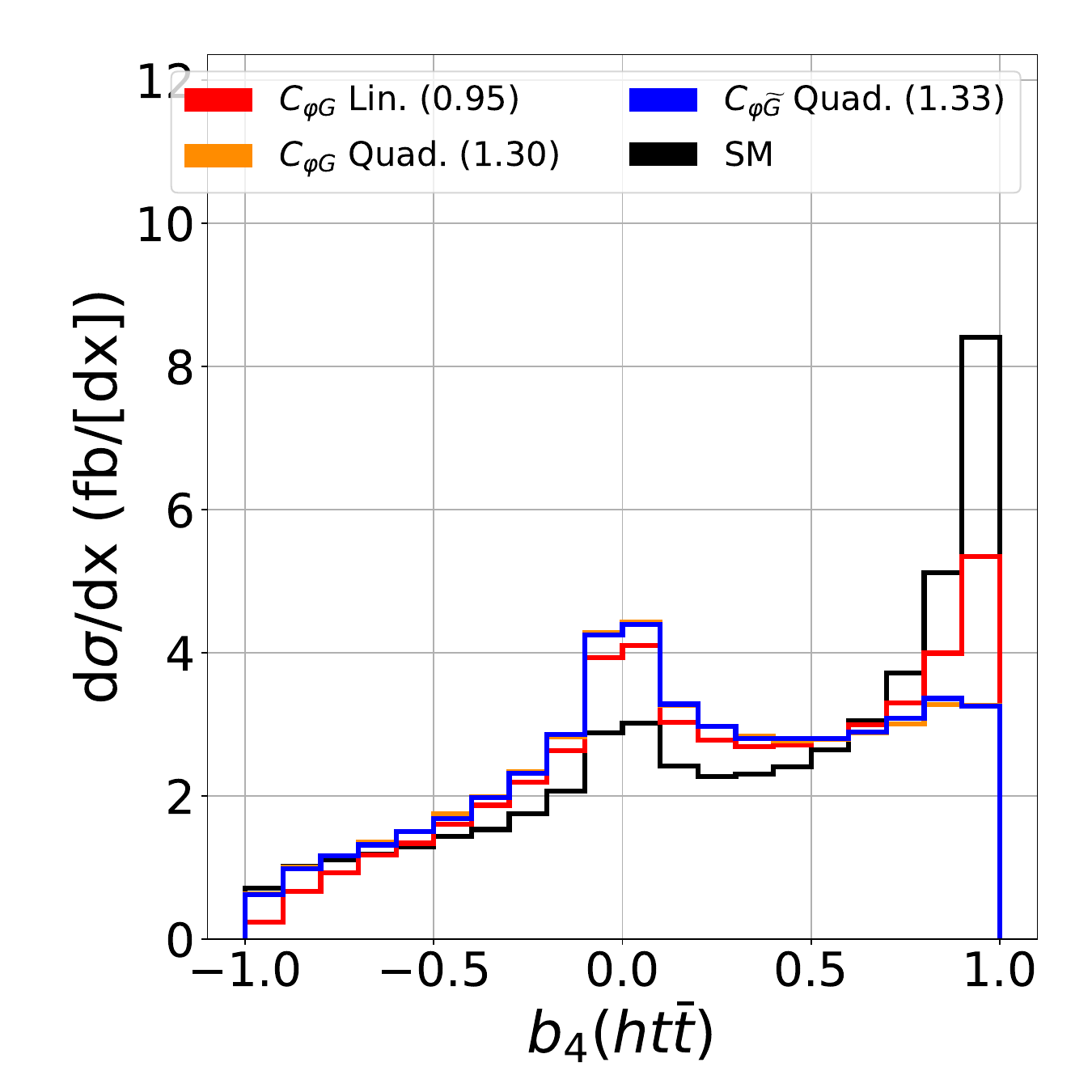}
    \caption{}
    \label{fig:sub2}
  \end{subfigure}
    \begin{subfigure}[b]{0.33\textwidth}
    \includegraphics[width=1.\linewidth]{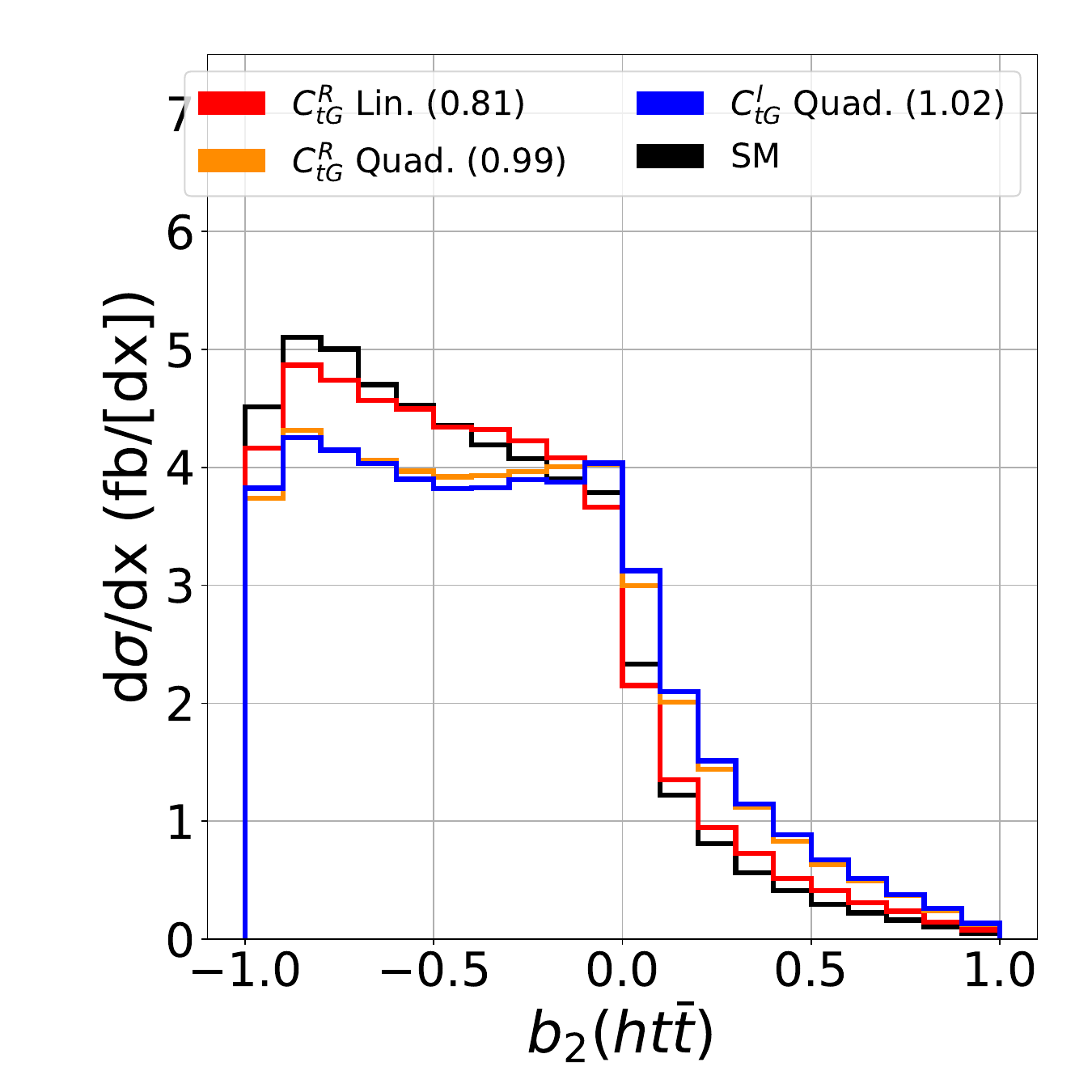}
    \caption{}
    \label{fig:sub2}
  \end{subfigure}%
    \begin{subfigure}[b]{0.33\textwidth}
    \includegraphics[width=1.\linewidth]{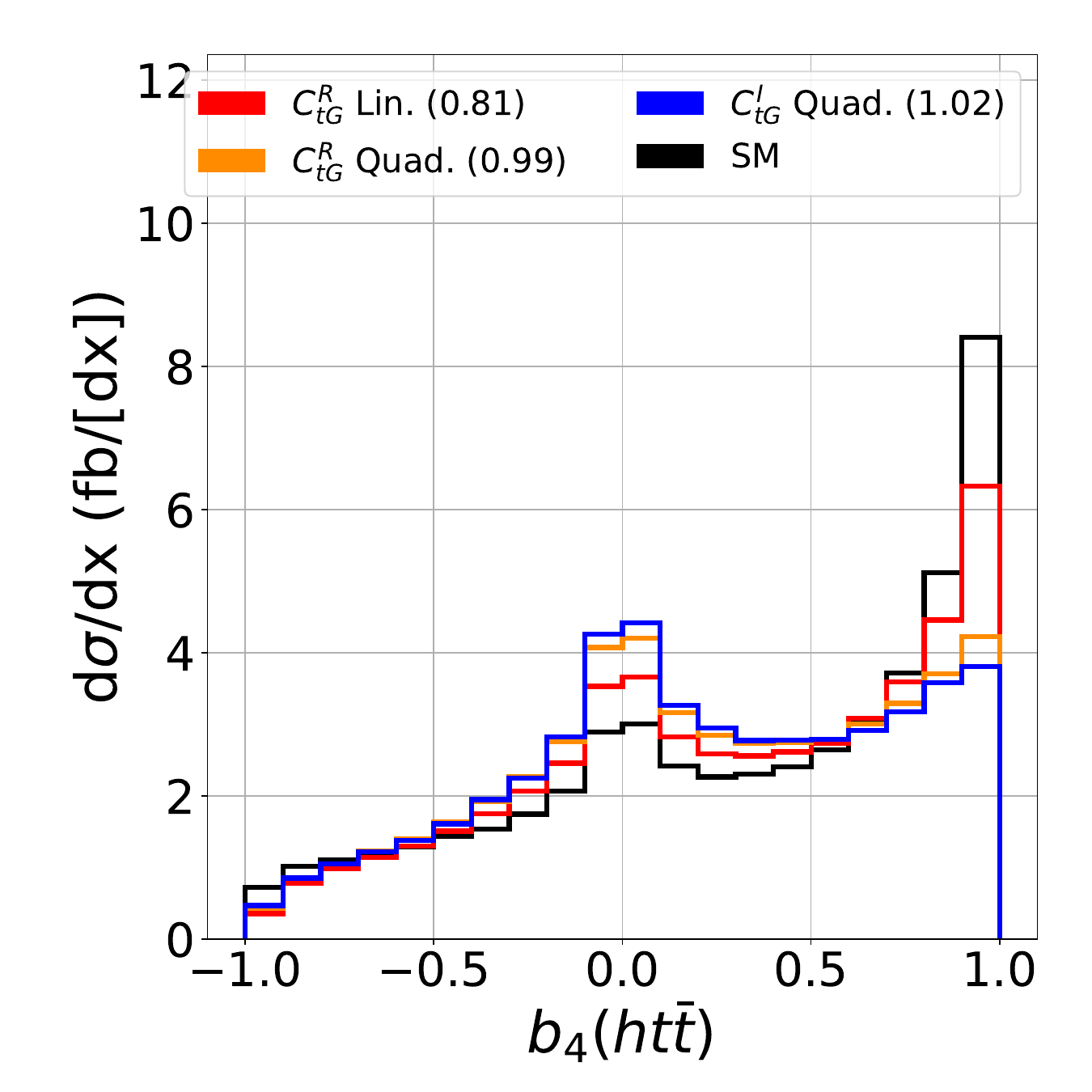}
    \caption{}
    \label{fig:sub2}
  \end{subfigure}
  \begin{subfigure}[b]{0.33\textwidth}
    \includegraphics[width=1.\linewidth]{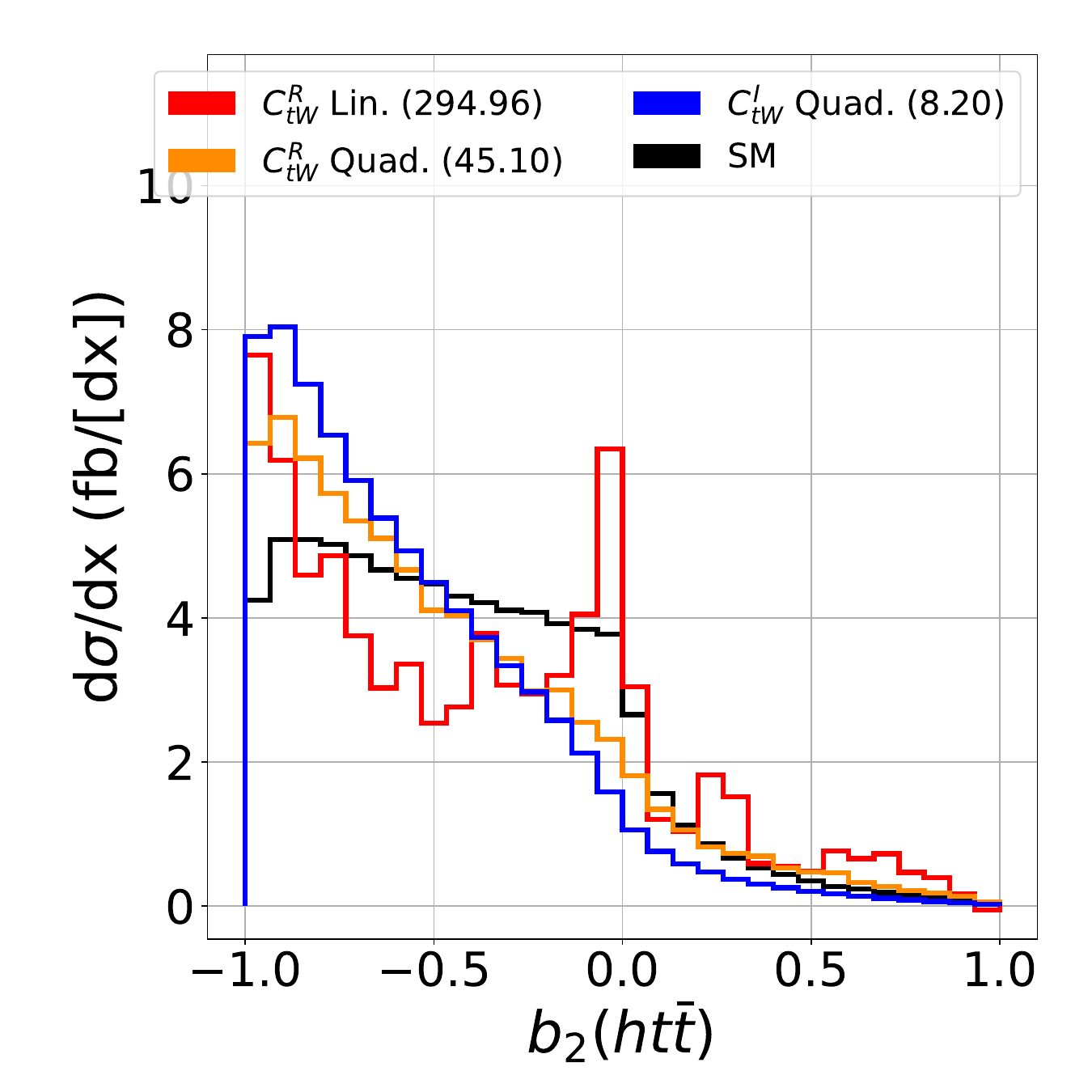}
    \caption{}
    \label{fig:sub2}
  \end{subfigure}%
    \begin{subfigure}[b]{0.33\textwidth}
    \includegraphics[width=1.\linewidth]{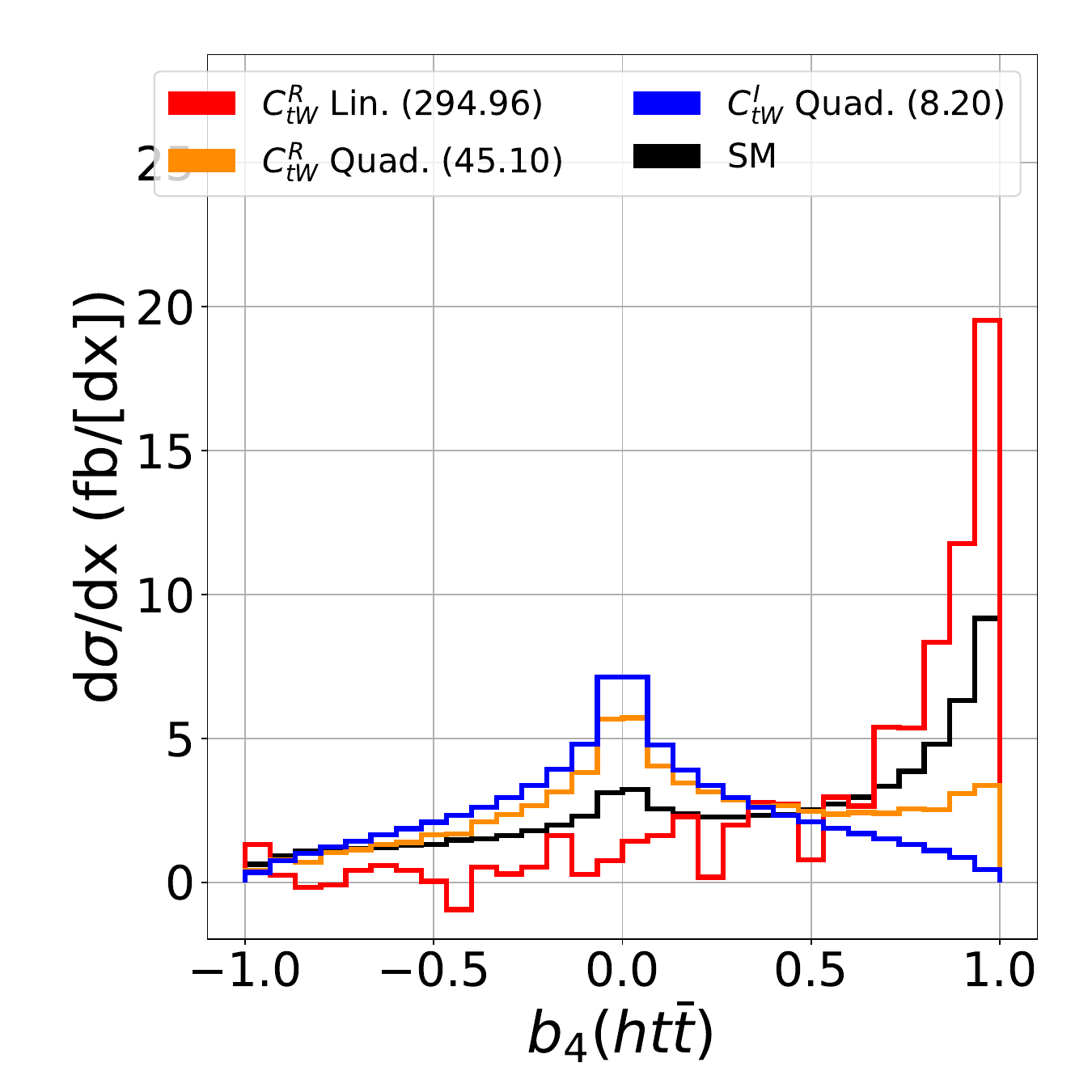}
    \caption{}
    \label{fig:sub2}
  \end{subfigure}

  \caption{\label{fig:tth_b2_b4} Differential cross sections of the observables $b_2$ (Eq.~\ref{eq:b2}) and $b_4$ (Eq.~\ref{eq:b4}) 
for the operators $\hat{O}_{t\varphi}$ (up), $O_{\varphi G/\widetilde{G}}$ (centre up), $\hat{O}_{tG}$ (centre down) and $\hat{O}_{tW}$ (down). }
\end{figure}
\FloatBarrier

\bibliographystyle{JHEP}
\bibliography{biblio}

\end{document}